\documentclass[aps,prd,preprint,showpacs,preprintnumbers,nofootinbib,superscriptaddress]{revtex4}
\usepackage{graphicx}
\usepackage{amsmath}
\usepackage{amsfonts}
\usepackage{amssymb}
\usepackage{feynmf}

\allowdisplaybreaks

\def\siml{{\ \lower-1.2pt\vbox{\hbox{\rlap{$<$}\lower6pt\vbox{\hbox{$\sim$}}}}\ }}
\def\simg{{\ \lower-1.2pt\vbox{\hbox{\rlap{$>$}\lower6pt\vbox{\hbox{$\sim$}}}}\ }}

\def \br {\mathbf{r}}

\def \bx {\mathbf{x}}

\def \bq {\mathbf{q}}

\def \cf {C_F}
\def \crr {C_R}
\def \nc {N}
\def \ca {C_A}

\def \bfnabla {\boldsymbol{\nabla}}
\def \bk {\mathbf{k}}
\def \mbk {\vert\bk\vert}
\def \bp {\mathbf{p}}

\def \mbq {\vert\bq\vert}
\def \mbp {\vert\bp\vert}

\def \trt {\tilde{\mathrm{Tr}}}
\def \als {\alpha_{\mathrm{s}}}

\def \m2   {\mu^{2 \epsilon}}

\newcommand{\MS}{{\overline{\rm MS}}}
 
\begin{document}
\title{Polyakov loop and  correlator of Polyakov loops at next-to-next-to-leading order}
\preprint{TUM-EFT 2/09}\preprint{INT-PUB-10-056}
\author{Nora Brambilla}

\affiliation{Physik-Department, Technische Universit\"at M\"unchen,
James-Franck-Str. 1, 85748 Garching, Germany}

\author{Jacopo Ghiglieri}
\affiliation{Physik-Department, Technische Universit\"at M\"unchen,
James-Franck-Str. 1, 85748 Garching, Germany}
\affiliation{Excellence Cluster Universe, Technische Universit\"at M\"unchen, 
Boltzmannstr. 2, 85748, Garching, Germany}

\author {P\'eter Petreczky}
\affiliation{Physics Department \\ 
Brookhaven National Laboratory, Upton, NY 11973, USA}

\author{Antonio Vairo}
\affiliation{Physik-Department, Technische Universit\"at M\"unchen,
James-Franck-Str. 1, 85748 Garching, Germany}

\begin{abstract}
We study the Polyakov loop and the correlator of two Polyakov loops at finite
temperature in the weak-coupling regime. 
We calculate the Polyakov loop at order $g^4$.
The calculation of the correlator of two Polyakov loops is performed at distances 
shorter than the inverse of the temperature and for electric screening masses 
larger than the Coulomb potential. In this regime, it is accurate up
to order $g^6$.  We also evaluate the Polyakov-loop correlator in an effective 
field theory framework that takes advantage of the hierarchy of energy scales 
in the problem and makes explicit the bound-state dynamics.
In the effective field theory framework, we show that the Polyakov-loop correlator is
at leading order in the multipole expansion the sum of a colour-singlet and a colour-octet 
quark-antiquark correlator, which are gauge invariant, 
and compute the corresponding colour-singlet and colour-octet free energies.
\end{abstract}

\date{\today}


\maketitle

\section{Introduction}
The Polyakov loop and the correlator of two Polyakov loops are 
the order parameters of the deconfinement phase transition in SU$(\nc)$ gauge
theories \cite{Kuti:1980gh,McLerran:1981pb}. The phase transition is signaled
by a non-vanishing expectation value of the Polyakov loop and a qualitative
change in the large-distance behaviour of the correlation function (from
confining to exponentially screened) \cite{McLerran:1981pb}.
In the deconfined phase, these quantities  provide information about 
the electric screening and can be calculated at sufficiently high temperatures $T$
in perturbation theory. For the correlation function of
Polyakov loops, the validity of the perturbative expansion is limited to
distances $r$ smaller than the magnetic screening length $r\ll 1/(g^2 T)$ 
\cite{Rebhan:1994mx,Arnold:1995bh}. 

From a phenomenological perspective, the Polyakov-loop correlator is
interesting because it provides an insight into the in-medium modifications of the 
quark-antiquark interaction. 
Indeed, in-medium modified heavy-quark potentials, inspired also by the behaviour of the 
Polyakov-loop correlator, have been used since long time in 
potential models (see e.g. Ref.~\cite{Mocsy:2007jz}). 
However, although the spectral decomposition of the Polyakov-loop correlator 
is known, its relation with the heavy-quark potential is still a matter of debate 
and in need of a clarifying analysis \cite{Philipsen:2008qx}. 
The issue has become particularly relevant since recently an 
in-medium modified heavy-quark potential has been derived rigorously 
from QCD \cite{Laine:2006ns,Beraudo:2007ky,Brambilla:2008cx,Escobedo:2008sy,Brambilla:2010vq}.
One of the aims of the paper is to discuss, in the weak-coupling regime, the relation  
between the Polyakov-loop correlator and these recent findings.

The Polyakov-loop correlator is a gauge-invariant quantity, 
hence it is well suited for lattice calculations.
In fact, the correlator of two Polyakov loops has be calculated on the lattice for the pure gauge
theory \cite{Kaczmarek:1999mm,Petreczky:2001pd,Digal:2003jc,Bazavov:2008rw} 
as well as for full QCD \cite{Karsch:2000kv,Petreczky:2004pz}
(for a review see Ref.~\cite{Petreczky:2005bd}). 
Surprisingly, not much is known instead about the correlator in perturbation theory. 
The correlator is known at leading order (LO)
since long time \cite{McLerran:1981pb,Gross:1980br}; beyond leading order, 
it was computed only for distances of the same order as the electric screening 
length in Ref.~\cite{Nadkarni:1986cz}.

The purpose of the paper is to evaluate the (connected) Polyakov-loop correlator 
up to order $g^6$ at short distances, $rT \ll 1$. 
This corresponds to a next-to-next-to-leading order (NNLO) calculation, 
if we count the order $g^4$ as LO and the order $g^5$ as next-to-leading order (NLO).
We also revisit the calculation of the expectation value of the Polyakov loop at order $g^4$, 
which corresponds also to a NNLO calculation, if we count $1$ as the leading-order 
result and $g^3$ as the NLO one. We will find a result that differs from the long-time accepted 
result of Gava and Jengo \cite{Gava:1981qd}. Finally, we will add on the discussion about 
the relation between the Polyakov-loop correlator and the in-medium heavy-quark potential.

The paper is organized as follows. In the next section, we 
discuss the gluon propagator in static gauge at one-loop level. 
Section \ref{sec_ploop} contains the calculation of the Polyakov loop at NNLO, 
while in section \ref{sec_corr} we calculate the Polyakov-loop correlator. 
In Sec.~\ref{sec_EFT}, we rederive the Polyakov-loop correlator in an effective 
field theory language. There, we also define a singlet and an octet free energy 
that we compute. Finally, section \ref{sec_concl} contains the conclusion and outlook.

\section{The static gauge and the self energy\label{self_energy}}
The Polyakov loop and the Polyakov-loop correlator are gauge-invariant quantities.
We may exploit the gauge freedom by choosing the most suitable gauge.
A convenient gauge choice is the \emph{static gauge} \cite{D'Hoker:1981us},
defined as\footnote{
We will work in Euclidean space-time and $0$ will label the Euclidean-time component.
} 
\begin{equation}
\partial_0A^0(x)=0.
\end{equation}
The reason for using the static gauge is that in this gauge the Polyakov line has 
a very simple form
\begin{equation}
\label{ploop}
L(\bx) = {\rm P} \exp\left(ig\int_0^{1/T} d\tau A^0(\bx,\tau)\right)=\exp\left(\frac{igA^0(\bx)}{T}\right),
\end{equation}
where $\rm P$ stands for the path-ordering prescription.
The spatial part of the gluon propagator reads
\begin{equation}
\label{propnonstatic}
D_{ij}(\omega_n,\bk)=
\frac{1}{k^2}\left(\delta_{ij}+\frac{k_ik_j}{\omega_n^2}\right)(1-\delta_{n0})
+\frac{1}{\bk^2}\left(\delta_{ij}-(1-\xi)\frac{k_ik_j}{\bk^2}\right)\delta_{n0},
\end{equation}
where $\omega_n=2\pi T n$, $n \in\mathbb{Z}$, are the bosonic Matsubara frequencies and
$k^2=\omega_n^2+\bk^2$. Throughout the paper italic letters will refer to Euclidean 
four-vectors and bold letters to the spatial components.
The parameter $\xi$ is a residual gauge-fixing parameter.
We call \emph{non-static modes} those propagating with nonzero Matsubara
frequencies and conversely we employ the term \emph{static mode} for the zero mode. 
The first term in the r.h.s. of Eq. \eqref{propnonstatic}, proportional to $(1-\delta_{n0})$,
is then the non-static part, whereas the second, proportional to $\delta_{n0}$,  
is the static part. The temporal part of the gluon propagator reads
\begin{equation}
\label{propstatic}
D_{00}(\omega_n,\bk)=\frac{\delta_{n0}}{\bk^2},
\end{equation}	
which is purely static.  Note that the gauge-fixing parameter affects only 
the static part of the spatial gluon propagator.
The complete set of Feynman rules in this gauge has been discussed in Refs. 
\cite{D'Hoker:1981us,Curci:1982fd,Curci:1984rd}. Feynman rules are listed 
in appendix \ref{app_rules} together with our Feynman diagram conventions.
We will adopt the static gauge in all the calculations of the paper, if not otherwise specified.

A necessary ingredient for the calculation of the Polyakov-loop expectation value and the Polyakov-loop
correlator at NNLO is the temporal component of the gluon self energy at LO. 
In the static gauge, due to the static nature of the temporal propagator in Eq. \eqref{propstatic} only 
$\Pi_{00}(\bk) \equiv \Pi_{00}(0,\bk)$ enters. Furthermore, at LO  
static and non-static modes do not mix in $\Pi_{00}(\bk)$, which can thus be conveniently split into
\begin{equation}
\Pi_{00}(\bk)=\Pi_{00}^{\mathrm{NS}}(\bk)+\Pi_{00}^{\mathrm{S}}(\bk)+\Pi_{00}^{\mathrm{F}}(\bk),
\end{equation}
where the three terms correspond to the contribution of the non-static gluons, 
the static gluons and the fermion loops respectively.

\begin{figure}[ht]
\begin{center}
\includegraphics[width=10cm]{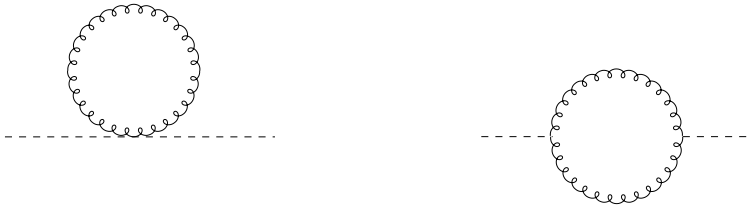}
\end{center}
\caption{Diagrams contributing to the non-static part of the gluon
  self-energy in the gluonic sector. 
  Dashed lines are temporal gluons, curly lines are spatial non-static gluons.}
\label{fig:ns}
\end{figure}

\begin{enumerate}
\item{\emph{$\Pi_{00}^{\mathrm{NS}}(\bk)$}}\\
In the gluonic sector, the non-static part of the self-energy receives contributions 
only from the two diagrams shown in Fig.~\ref{fig:ns}. Using the Feynman rules of 
appendix \ref{app_rules},  
it can be written in terms of five dimensionally-regularized master sum integrals 
\begin{equation}
\Pi_{00}^{\mathrm{NS}}(\bk)=
-2g^2\ca\left(\frac{d-1}{2}I_0 - (d-1) I_1 + I_2+\frac{1}{2}I_3 +\frac{1}{4}I_4\right),
\label{defp00}
\end{equation}
where $\ca = N =3$ is the number of colours, $d=3-2\epsilon$ is the number of dimensions, 
\begin{equation}
\label{defmaster}
I_0=\int_p^\prime\frac{1}{p^2},\quad 
I_1=\int_p^\prime\frac{\bp^2}{p^2q^2},\quad 
I_2=\int_p^\prime\frac{\bk^2}{p^2q^2},\quad 
I_3=\int_p^\prime\frac{\bk^2}{\bp^2 p^2},\quad 
I_4=\int_p^\prime\frac{\bk^4}{p^2q^2\omega_n^2},
\end{equation}
$q=k-p$, $\displaystyle \int_p^\prime$ is a shorthand notation for the non-static, $n\ne0$, sum integral:
\begin{equation}
\label{shorthand}
\int_p^\prime \equiv T\sum_{n\ne0} \m2  \int\frac{d^dp}{(2\pi)^d},
\end{equation}
and $\mu$ is the scale in dimensional regularization.
The result \eqref{defp00} can be conveniently cast in a sum of a vacuum part, a matter part, 
a part made of the subtracted zero modes and a part that we may call \emph{singular}, 
because it is singular for $T\to0$; the singular part is a peculiar 
feature of the static gauge. We then have
\begin{eqnarray}
\Pi_{00}^{\mathrm{NS}}(\bk)&=&
\Pi_{00}^{\mathrm{NS}}(\bk)_{\mathrm{vac}}+\Pi_{00}^{\mathrm{NS}}(\bk)_{\mathrm{mat}}
+\Pi_{00}^{\mathrm{NS}}(\bk)_{\mathrm{zero}}+\Pi_{00}^{\mathrm{NS}}(\bk)_{\mathrm{sing}},
\label{pi00nonstatic}
\end{eqnarray}
\begin{eqnarray}
\Pi_{00}^{\mathrm{NS}}(\bk)_{\mathrm{vac}}&=&
-\frac{g^2\bk^2}{(4\pi)^2}C_A\left[\frac{11}{3}
\left(\frac{1}{\epsilon}-\gamma_E+ \ln (4 \pi) -\ln\frac{\bk^2}{\mu^2}\right)
+\frac{31}{9}\right],
\label{pi00vacuum}\\
\nonumber
\Pi_{00}^{\mathrm{NS}}(\bk)_{\mathrm{mat}}&=&
g^2\ca\left\{\int_0^\infty d\mbp\frac{\mbp n_{\mathrm{B}}(\mbp)}{\pi^2}\left[1-\frac{\bk^2}{2\bp^2}
\right.\right.
\\
&& \left.\left. \hspace{3.8cm}
+\left(\frac{\mbp}{\mbk}-\frac{\mbk}{2\mbp}+\frac{\mbk^3}{8\mbp^3}\right)
\ln\left\vert\frac{\mbk+2\mbp}{\mbk-2\mbp}\right\vert\right]\right\}\!,
\label{pi00matter}\\
\Pi_{00}^{\mathrm{NS}}(\bk)_{\mathrm{zero}}&=&
g^2\ca\frac{T\mbk^{1-2\epsilon} \m2 }{4}\left[1+\epsilon(-1-\gamma_E + \ln (16\pi))\right], 
\label{pi00zero}\\
\Pi_{00}^{\mathrm{NS}}(\bk)_{\mathrm{sing}}&=&-g^2\ca\frac{\mbk^3}{192T},
\label{pi00sing}
\end{eqnarray}
where $\gamma_E$ is the Euler constant and 
$n_\mathrm{B}(k)=1/\left(e^{k/T}-1\right)$ is the Bose--Einstein distribution. 
We refer the reader to appendix \ref{app_pi00} for details on the derivation of these equations. 
The vacuum part \eqref{pi00vacuum} agrees with the static gauge computation in \cite{Curci:1982fd}. 
Furthermore, the vacuum part and the matter part are identical 
to the $k^0\to0$ limit of their Coulomb gauge counterparts, computed respectively 
in \cite{Duncan:1975kt,Appelquist:1977es} and \cite{Kajantie:1982xx,Heinz:1986kz}. 
$\Pi_{00}^{\mathrm{NS}}(\bk)_{\mathrm{zero}}$ consists of the subtracted zero modes.  
In the $\epsilon\to 0$ limit, it is $T\mbk/4$; we have kept the order $\epsilon$ corrections, 
because, in the Polyakov-loop correlator calculation of Sec.~\ref{sec_corr},
we will need to evaluate the Fourier transform of $\mbk^{1-2\epsilon}/\mbk^4$, coming 
from a self-energy insertion in a temporal-gluon propagator, which is divergent. \\ 

\item{\emph{$\Pi_{00}^{\mathrm{F}}(\bk)$}}\\
At leading order in the coupling, $\Pi_{00}^{\mathrm{F}}(\bk)$ may be written in terms 
of three di\-men\-sionally-regularized master sum integrals \cite{Heinz:1986kz} 
\begin{equation}
\Pi^{\mathrm{F}}_{00}(\bk) = 2g^2n_f\left(-\tilde{I}_0+2\tilde{I}_1+\frac{1}{2}\tilde{I}_2\right),
\label{pi00fermion}
\end{equation}
where
\begin{eqnarray} 
\nonumber
\tilde{I}_0  &=& T\sum_{n=-\infty}^{+\infty}\m2 \int\frac{d^dp}{(2\pi)^d}\frac{1}{p^2},  \quad
\tilde{I}_1   = T\sum_{n=-\infty}^{+\infty}\m2 \int\frac{d^dp}{(2\pi)^d}\frac{\tilde{\omega}_n^2}{p^2q^2}, 
\\
\tilde{I}_2 &=& T\sum_{n=-\infty}^{+\infty}\m2 \int\frac{d^dp}{(2\pi)^d}\frac{\bk^2}{p^2q^2},
\label{deffermionmaster}
\end{eqnarray}
$q=p+k$ and $\tilde{\omega}_n=(2n+1)\pi T$ are the fermionic Matsubara frequencies and
$n_f$ is the number of massless quarks contributing to the fermion loops.
 Since no fermionic Matsubara frequency vanishes, fermions are purely non-static.
The fermionic contribution can be cast into a sum of a vacuum and a matter part: 
$\Pi^{\mathrm{F}}_{00}(\bk)=\Pi_{00}^\mathrm{F}(\bk)_{\mathrm{vac}}+\Pi_{00}^\mathrm{F}(\bk)_{\mathrm{mat}}$. 
After the Matsubara frequencies summation, the matter part can be read from \cite{Kapusta:2006pm}
\begin{equation}
\label{matterfermion}
\Pi_{00}^\mathrm{F}(\bk)_{\mathrm{mat}}=
\frac{g^2}{2\pi^2}n_f\int_{0}^{\infty}d\mbp\,\mbp 
n_\mathrm{F}( \mbp)\left[2+\frac{4\bp^2-\bk^2}{2\mbp\mbk}
\ln\left\vert\frac{\mbk+2\mbp}{\mbk-2\mbp}\right\vert\right],
\end{equation} 
where $n_\mathrm{F}(k)=1/\left(e^{k/T}+1\right)$ is the Fermi--Dirac distribution.
The vacuum part is given by
\begin{equation}
\label{vacuumfermion}		
\Pi_{00}^\mathrm{F}(\bk)_{\mathrm{vac}}=
\frac{2}{3}\frac{g^2\bk^2}{(4\pi)^2}n_f
\left[\frac{1}{\epsilon}-\gamma_E +\ln (4\pi)-\ln\frac{\bk^2}{\mu^2}+\frac{5}{3}\right].
\end{equation}

\item{\emph{$\Pi_{00}^{\mathrm{NS}}(\bk) + \Pi_{00}^{\mathrm{F}}(\bk)$}}\\
Let us now consider the sum $\Pi_{00}^{\mathrm{NS}}(\bk) + \Pi_{00}^{\mathrm{F}}(\bk)$.
The divergences in the vacuum parts \eqref{pi00vacuum} and \eqref{vacuumfermion} are of 
ultraviolet origin and are accounted for by the charge renormalization.
In the $\MS$ scheme, the renormalized sum of vacuum parts reads 
\begin{equation}
\Pi_{00}^\mathrm{NS}(\bk)_{\mathrm{vac}}+\Pi_{00}^\mathrm{F}(\bk)_{\mathrm{vac}}
=-\frac{g^2\bk^2}{(4\pi)^2}\left[\beta_0\ln\frac{ \mu^2}{\bk^2}+\frac{31}{9}\ca-\frac{10}{9}n_f\right],
\label{vacuum}
\end{equation}
where $\beta_0=11\ca/3-2n_f/3$. 

Simple analytical expressions can be obtained for the renormalized 
sum $\Pi_{00}^{\mathrm{NS}}(\bk) + \Pi_{00}^{\mathrm{F}}(\bk)$ in the two limiting cases $\mbk\ll T$ and $\mbk\gg T$. 
In the former case, we have 
\begin{eqnarray}
\nonumber
\left(\Pi_{00}^{\mathrm{NS}}+\Pi_{00}^{\mathrm{F}}\right)(\mbk\ll T)
&=&\frac{g^2T^2}{3}\left(\ca+\frac{n_f}{2}\right)
\\
\nonumber
&-& \frac{g^2\bk^2}{(4\pi)^2}\left[\frac{11}{3} \ca\left( -\ln\frac{(4\pi
    T)^2}{\mu^2}+1+2 \gamma_E \right)\right.
\\
\nonumber
&&\left. \hspace{1.3cm} -\frac{2}{3}n_f\left(-\ln\frac{(4\pi T)^2}{\mu^2}-1+2 \gamma_E   + 4 \ln 2
\right)\right]
\\
&+& g^2\bk^2\mathcal{O}\left(\frac{\bk^2}{T^2}\right),
\label{kllt}
\end{eqnarray}
where the leading-order term is momentum independent and can be 
identified with the (square of the) Debye mass $m_D$, 
\begin{equation}
m_D^2 \equiv \frac{g^2T^2}{3}\left(\nc+\frac{n_f}{2}\right),
\label{fulldebye}
\end{equation}
which provides, in the weak-coupling regime, the inverse of an electric 
screening length.
We note that Eq. \eqref{kllt} presents a logarithm of
the renormalization scale over the temperature rather than over the
momentum: this happens because in the limit $\mbk\ll T$ the matter part produces
a term proportional to $\bk^2\beta_0\ln(T^2/\bk^2)$ that combines
with the logarithm in the renormalized vacuum part \eqref{vacuum} to
cancel its momentum dependence.

In the opposite limit $\mbk\gg T$, we have
\begin{eqnarray}
\nonumber
\left(\Pi_{00}^{\mathrm{NS}}+\Pi_{00}^{\mathrm{F}}\right)(\mbk\gg T)
&=&\Pi_{00}^\mathrm{NS}(\bk)_{\mathrm{vac}}+\Pi_{00}^\mathrm{F}(\bk)_{\mathrm{vac}}
+g^2 C_A \left(-\frac{T^2}{18} -\frac{\mbk^3}{192T}\right)
\\
&&+g^2\ca \frac{T\mbk^{1-2\epsilon} \m2 }{4}\left[1+\epsilon(-1-\gamma_E + \ln (16\pi))\right]
\nonumber\\
&&+g^2T^2\mathcal{O}\left(\frac{T^2}{\bk^2}\right).
\label{kggt}
\end{eqnarray}
We observe that, in this limit and at the considered order, fermions
enter only through their contribution to the vacuum part. It should
be also noted that, while the $-g^2\ca T^2/18$ term appears also in Coulomb
gauge \cite{Kajantie:1982xx,Brambilla:2008cx}, the term proportional to $\mbk^3$ 
is instead a peculiar feature of the static gauge. 
The terms proportional to $\epsilon T \mbk^{1-2\epsilon}$, which appear in the second line,
come from the subtracted zero modes and contribute only when plugged into divergent amplitudes. 
Details on the derivation of these expressions can be found in appendix \ref{app_exp}.

\begin{figure}[ht]
\begin{center}
\includegraphics{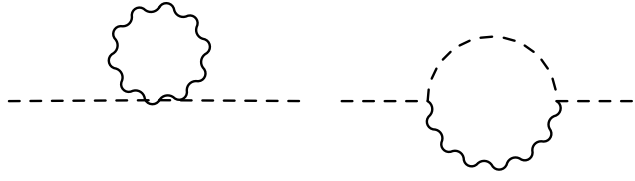}
\end{center}
\caption{Diagrams contributing to the static part of the self-energy: the
  dashed lines are temporal-gluon propagators,  the wavy lines are static
  spatial-gluon propagators. Loops made of two static spatial-gluon propagators  
  and of ghosts vanish.}
\label{fig:static}
\end{figure}

\item{\emph{$\Pi_{00}^{\mathrm{S}}(\bk)$}}\\
The diagrams contributing to the static part of the gluon self energy are shown
in Fig.~\ref{fig:static}. 
They are not sensitive to the scale $T$, since, by definition, 
static gluon propagators are just made of zero modes, however they 
are to the scale $m_D$. 
Hence, when evaluating the static contribution, it is important
to keep in mind that, if the incoming momentum is of the order of the Debye mass,
then insertions of gluon self-energies of the type of Eq. \eqref{kllt} into the 
temporal-gluon propagator need to be resummed modifying the temporal-gluon propagator into
\begin{equation}
\label{propscreened}
D_{00}(\omega_n,\bk)=\frac{\delta_{n0}}{\bk^2+m_D^2}.
\end{equation}
The static part of the gluon self energy with resummed propagators 
reads, for all values of the gauge-fixing parameter $\xi$, 
\begin{eqnarray}
\Pi_{00}^{\mathrm{S}}(\bk)&=&
g^2C_AT \m2  \int\frac{d^dp}{(2\pi)^d}\left(\frac{1}{\bp^2+m_D^2}+\frac{d-2}{\bp^2}\right.
\nonumber\\
&&\left.+\frac{2(m_D^2-\bk^2)}{\bp^2(\bq^2+m_D^2)}
+(\xi-1)(\bk^2+m_D^2)\frac{\bp^2+2\bp\cdot\bk}{\bp^4(\bq^2+m_D^2)}\right),
\label{staticvacpol}
\end{eqnarray}
where $q=k+p$. The result agrees with Ref.~\cite{Rebhan:1993az,Rebhan:1994mx}.  
Note that Eq. \eqref{staticvacpol} applies for all gauges sharing 
the same static propagator, among which the static and the covariant gauges. 
The expression is finite in three dimensions and reads
\begin{eqnarray}
\Pi_{00}^{\mathrm{S}}(\bk)&=&
\frac{g^2C_AT}{4\pi}\left[2\frac{m_D^2-\bk^2}{\mbk}\arctan\frac{\mbk}{m_D} -m_D+(\xi-1)m_D\right].
\label{staticvacpol2}
\end{eqnarray}
Finally for the static part $\mbk\gg T$ implies $\mbk\gg m_D$ and
\begin{equation}
\Pi_{00}^\mathrm{S}(\mbk\gg m_D)=-g^2C_A\left\{
\frac{T\mbk^{1-2\epsilon} \m2 }{4}\left[1+\epsilon(-\gamma_E + \ln (16\pi))\right]
+\mathcal{O}(m_DT)\right\},
\label{staticcontrhighk}
\end{equation}
where again we have kept up to order $\epsilon$ terms proportional to $T\mbk^{1-2\epsilon}$.

\item{\emph{$\Pi_{00}(\bk)$}}\\
$\Pi_{00}(\bk)$ is obtained by summing \eqref{pi00vacuum}, \eqref{pi00matter}, \eqref{pi00zero}, 
\eqref{pi00sing}, \eqref{matterfermion}, \eqref{vacuumfermion} and \eqref{staticvacpol} (or \eqref{staticvacpol2}).
In particular, the asymptotic expression for the gluon polarization at high momenta is 
\begin{eqnarray}
\nonumber\Pi_{00}(\mbk\gg T)&=&
-\frac{g^2\bk^2}{(4\pi)^2}\left(\beta_0\ln\frac{\mu^2}{\bk^2}+\frac{31}{9}\ca-\frac{10}{9}n_f\right)
+g^2C_A\left(-\frac{T^2}{18}-\frac{\mbk^3}{192T}\right)\\
&&
-\epsilon g^2\ca\frac{T\mbk^{1-2\epsilon} \m2 }{4}
+\mathcal{O}\left(g^2\frac{T^4}{\bk^2},g^2m_DT\right).
\label{kggtfull}
\end{eqnarray}
Note that the term proportional to $T\mbk^{1-2\epsilon}\epsilon^0$ in Eq. \eqref{staticcontrhighk} 
has canceled against the term proportional to $T\mbk^{1-2\epsilon}\epsilon^0$ in Eq. \eqref{kggt}.
\end{enumerate}

\section{The Polyakov loop\label{sec_ploop}}
The quantity we are interested in computing is the trace of the
Polyakov line $L \equiv L_R$ in a representation $R$ of dimension $d(R)$, 
where $R$ is either the fundamental representation ($R=F$, $d(F)=\nc$) 
or the adjoint representation ($R=A$, $d(A)=\nc^2-1$): 
\begin{equation}
\langle L_R\rangle\equiv\langle\trt L_R\rangle,\quad\trt \equiv \frac{\mathrm{Tr}}{d(R)}.
\label{defploopvev}
\end{equation}
The brackets stand for the average in a thermal ensemble at a temperature $T$.
Expanding the Polyakov line in the static gauge up to order $g^4$ yields
\begin{equation}
\langle L_R\rangle=
1-\frac{g^2}{2!} \frac{\langle\trt A_0^2\rangle}{T^2}
-i\frac{g^3}{3!} \frac{\langle\trt A_0^3\rangle}{T^3}
+\frac{g^4}{4!} \frac{\langle\trt A_0^4\rangle}{T^4}
+\ldots \,.
\label{expandedlooptraces}
\end{equation}
In computing Eq. \eqref{expandedlooptraces} perturbatively,  
each diagram can receive contributions from both scales $T$ and $m_D$, 
for which we assume a weak-coupling hierarchy:\footnote{
When discussing energy scales, we will consider $T$ and multiple of $\pi T$ to be 
parametrically of the same order.
}
\begin{equation}
T\gg m_D.	
\label{hierarchy}
\end{equation}	
In the weak-coupling regime, the calculation of $\langle L_R\rangle$ may be organized in an 
expansion in the coupling $g$; our aim is to compute $\langle L_R\rangle$ up to order 
$g^4$. Sometimes, we will find it useful to keep $m_D/T$ as a separate expansion 
parameter with respect to $g$, in order to identify more easily
the origin of the various terms. We will call the $g^3$ term the NLO correction to the 
Polyakov loop and the $g^4$ term the NNLO correction.
We will also identify the source of some higher-order corrections of order $g^5$ and
$g^4\times(m_D/T)^2$ that will play a role in Sec.~\ref{sec_EFT}.

\begin{figure}[ht]
\begin{center}
\includegraphics{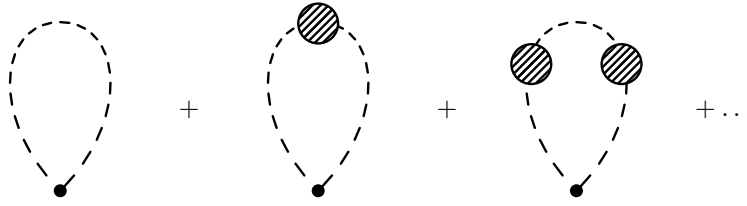}
\end{center}
\caption{Diagrams contributing to the perturbative expansion of $g^2\langle\trt A_0^2\rangle$. 
The dashed line is a temporal-gluon propagator, the dot is the point $\bx$ where the loop originates. 
The blob stands for the gluon self energy.}
\label{fig:square}
\end{figure}

\subsection{The order $g^3$ contribution\label{sub_lo}}
Let us start examining $g^2\langle\trt A_0^2\rangle$. Diagrams 
contributing to  $g^2\langle\trt A_0^2\rangle$ are shown in Fig.~\ref{fig:square}.
Summing up all these diagrams, $g^2\langle\trt A_0^2\rangle$ can be written as 
\begin{equation}
\delta\langle L_R\rangle
=-\frac{g^2}{2!} \frac{\langle\trt A_0^2\rangle}{T^2}
=-\frac{g^2C_R}{2T}\m2  \int \frac{d^dk}{(2\pi)^d}\frac{1}{\bk^2+\Pi_{00}(\bk)},
\label{loopb}
\end{equation}
where $C_R$ is the quadratic Casimir operator of the representation $R$ 
($C_A=N$, $C_F = (N^2-1)/(2N)$). We observe that the integral
receives contributions from the scales $T$ and $m_D$. We set out to
separate the contributions from these two scales assuming the hierarchy \eqref{hierarchy}. 

\begin{enumerate}
\item{\emph{Modes at the scale $T$}} \\
We evaluate the integral \eqref{loopb} for $\mbk\sim T\gg m_D$. 
In this momentum region, $\Pi_{00}(\mbk\sim T\gg m_D)\ll\bk^2$ and we may
expand the gluon propagator in $\Pi_{00}$. 
The LO term yields a scaleless integral
\begin{equation}
\delta\langle L_R\rangle=-\frac{g^2}{2T}C_R \m2  \int\frac{d^dk}{(2\pi)^d}\frac{1}{\bk^2}=0,
\label{tcontribnull}
\end{equation}
whereas the following term gives	
\begin{equation}
\delta\langle L_R\rangle_{T}=\frac{g^2C_R}{2T}\m2  \int \frac{d^dk}{(2\pi)^d}\frac{\Pi_{00}(\mbk\sim T\gg m_D)}{\bk^4}.
\label{loopbT}
\end{equation}
This term is of order $g^4$.

\item{\emph{Modes at the scale $m_D$}} \\
We evaluate now the contribution from the scale $m_D$. 
We recall from Eqs. \eqref{kllt} and \eqref{staticvacpol2} that, 
for $\mbk\ll T$, $\Pi_{00}(\bk)=m_D^2(1 +\mathcal{O}(g))$. 
We then rewrite the propagator in Eq. \eqref{loopb} as 
$1/(\bk^2+\Pi_{00}(\mbk\ll T))=1/(\bk^2+m_D^2+(\Pi_{00}(\mbk\ll T)-m_D^2))$ 
and expand in $\Pi_{00}(\mbk\ll T)-m_D^2$. The LO term yields 
\begin{equation}
\delta\langle L_R\rangle_{{\rm LO}\, m_D}
= -\frac{g^2C_R}{2T} \m2  \int \frac{d^dk}{(2\pi)^d}\frac{1}{\bk^2+m_D^2} 
=  \frac{\crr \als}{2} \frac{m_D}{T},
\label{loopbm0}
\end{equation}
whereas the following one gives 
\begin{equation}
\delta\langle L_R\rangle_{{\rm NLO}\,m_D}
= \frac{g^2C_R}{2T} \m2  \int \frac{d^dk}{(2\pi)^d}
\frac{\Pi_{00}(\mbk\sim m_D\ll T)-m_D^2}{(\bk^2+m_D^2)^2},
\label{loopbm}
\end{equation}
which is at least of order $g^4$.
\end{enumerate}

Up to order $g^3$, we then have
\begin{equation}
\langle L_R\rangle=1+\frac{\crr \als}{2}\frac{m_D}{T}+\mathcal{O}\left(g^4\right).
\label{fleading}
\end{equation}

\begin{figure}[ht]
\begin{center}
\includegraphics[width=10cm]{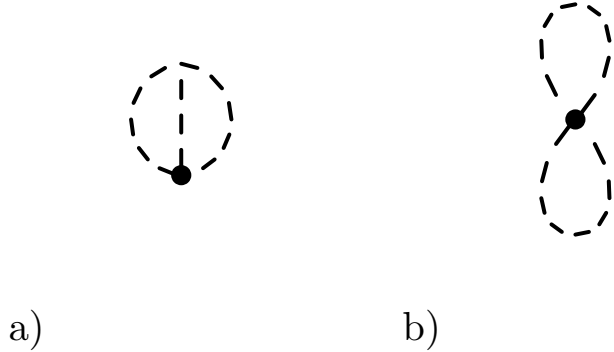}
\end{center}
\caption{Diagram a) is the leading-order contribution to $\langle\trt A_0^3\rangle$: 
it vanishes because of the three-gluon vertex involving only temporal gluons.
Diagram b) is the LO term of $g^4\langle\trt A_0^4\rangle$:  
it vanishes because scaleless.}
\label{fig:cubefourth}
\end{figure}

The LO contribution to the cubic term $g^3\langle\trt A_0^3\rangle$
is shown in Fig.~\ref{fig:cubefourth} a). 
It vanishes due to the structure of the three-gluon vertex.
This is just a LO manifestation of the charge-conjugation symmetry;
in fact, due to this symmetry, $g^3\langle\trt A_0^3\rangle$ vanishes
to all orders.
The quartic term $g^4\langle\trt A_0^4\rangle$ gets its LO
contribution from the diagram shown in Fig.~\ref{fig:cubefourth} b), which 
vanishes because scaleless. At higher order, a comparison with the analysis 
we have just performed for $\langle\trt A_0^2\rangle$ 
makes it clear that $g^4\langle\trt A_0^4\rangle$ 
starts to contribute at order $g^4\times(m_D/T)^2$, 
which is again beyond the accuracy of this analysis.
We can therefore identify as the only contributions to the Polyakov
loop at order $g^4$ the ones of Eqs. \eqref{loopbT} and \eqref{loopbm}. 
In Sec.~\ref{sub_nlo}, we will compute these contributions 
and, in Sec.~\ref{sub_nnlo}, we will analyze some sub-leading terms.

\subsection{The order $g^4$ contribution\label{sub_nlo}}
We now set out to compute Eqs. \eqref{loopbT} and \eqref{loopbm}. 
Following the discussion in Sec.~\ref{self_energy}, we
separate the non-static from the static modes in $\Pi_{00}(\bk)$.  We then
have four sources of contributions: non-static modes at the scale $T$,
non-static modes at the scale $m_D$, static modes at the scale $T$ and 
static modes at the scale $m_D$.  

\begin{enumerate}
\item{\emph{Non-static modes at the scale $T$}} \\
The non-static contribution to Eq. \eqref{loopbT} reads
\begin{equation}
\delta\langle L_R\rangle_{\mathrm{NS},\,T}=
\frac{g^2C_R}{2T}\m2  \int
\frac{d^dk}{(2\pi)^d}\frac{\Pi_{00}^{\mathrm{NS}}(\mbk\sim T)+\Pi_{00}^{\mathrm{F}}(\mbk\sim T)}{\bk^4},
\label{tcontribnlons}
\end{equation}
where $\Pi_{00}^{\mathrm{NS}}(\mbk\sim T)$ is the full
non-static contribution as defined in Eq. \eqref{pi00nonstatic} and
similarly $\Pi_{00}^{\mathrm{F}}(\mbk\sim T)$ is the full fermionic
contribution as defined in Eqs. \eqref{matterfermion} and \eqref{vacuumfermion}. 
We can rewrite Eq. \eqref{tcontribnlons} as
\begin{eqnarray}
\nonumber\delta\langle L_R\rangle_{\mathrm{NS},\,T}&=&
\frac{g^4C_R}{T}\left[-\ca\left(\frac{d-1}{2}J_0 - (d-1) J_1 +
  J_2+\frac{1}{2}J_3 +\frac{1}{4}J_4\right)\right.
\\
&&\left.+ n_f\left(-\tilde{J}_0+2\tilde{J}_1+\frac{1}{2}\tilde{J}_2\right)\right],
\label{tcontribnlonsj}
\end{eqnarray}
where we have defined the two-loop master sum-integrals $J_i$ and $\tilde{J}_i$ as
\begin{equation}
\label{defji}
J_i=\m2  \int \frac{d^dk}{(2\pi)^d} \frac{1}{\bk^4} I_i,\qquad 
\tilde{J}_i=\m2  \int \frac{d^dk}{(2\pi)^d} \frac{1}{\bk^4} \tilde{I}_i.
\end{equation}
These integrals are evaluated in appendix \ref{app_integrals} and their sum yields
\begin{equation}
\delta\langle L_R\rangle_{\mathrm{NS},\,T}=
\frac{g^4\crr}{2(4\pi)^2}\left[\ca\left(\frac{1}{2\epsilon}-\ln\frac{4T^2}{\mu^2}+1-\gamma_E
+\ln (4\pi) \right)- n_f\ln2\right]. 
\label{finalcontribtnlons}
\end{equation}
The divergence stems from the $J_2$ integral and is expected 
to cancel against an opposite divergence coming from the scale $m_D$.

\item{\emph{Non-static modes at the scale $m_D$}}\\
The non-static contribution to Eq. \eqref{loopbm} reads 
\begin{equation}
\delta\langle L_R\rangle_{\mathrm{NS},\,m_D}=\frac{g^2C_R}{2T}\m2  
\int \frac{d^dk}{(2\pi)^d}\frac{\Pi_{00}^{\mathrm{NS}}(\mbk\sim m_D)
+\Pi_{00}^{\mathrm{F}}(\mbk\sim m_D)-m_D^2}{(\bk^2+m_D^2)^2}.
\label{mcontribnlons}
\end{equation} 
For $\mbk$ much smaller than the temperature, Eq. \eqref{kllt} applies
and thus $\Pi_{00}^{\mathrm{NS}}(\bk) + \Pi_{00}^{\mathrm{F}}(\bk) =m_D^2+\mathcal{O}(g^2\bk^2)$.
Therefore, the contribution of Eq. \eqref{mcontribnlons} is of order $g^4 \times (m_D/T) \sim g^5$.
More explicitly, plugging Eq. \eqref{kllt} into Eq. \eqref{mcontribnlons} gives
\begin{equation}
\delta\langle L_R\rangle_{\mathrm{NS},\,m_D}=
\frac{3g^4C_R}{4(4\pi)^3}\frac{m_D}{T}\left[\beta_0\ln\left(\frac{\mu }{4\pi T}\right)^2
+2 \beta_0\gamma_E +\frac{11}{3}\ca-\frac{2}{3}n_f\left(4\ln 2-1\right)\right].
\label{mcontribnlonsfinal}
\end{equation}
Although a term of order $g^5$ is beyond our accuracy, the contribution
\eqref{mcontribnlonsfinal} is of interest because 
it fixes the renormalization scale of $g^3$ in the LO term \eqref{fleading} ($\als m_D/T \sim g^3$)
to $\mu = 4\pi T$. 

\item{\emph{Static modes at the scale $T$}}\\
The static contribution at the scale $T$ to Eq. \eqref{loopbT} reads
\begin{equation}
\delta\langle L_R\rangle_{\mathrm{S}\,T}=
\frac{g^2C_R}{2T}\m2  \int \frac{d^dk}{(2\pi)^d}\frac{\Pi_{00}^{\mathrm{S}}(\mbk\sim T)}{\bk^4}=0.
\label{tcontribnlos}
\end{equation}
It vanishes because $\Pi_{00}^{\mathrm{S}}(\mbk\sim T\gg m_D)\sim g^2T \mbk$ 
(see Eq. \eqref{staticcontrhighk}) and thus the resulting integration over $\bk$ is scaleless. 

\item{\emph{Static modes at the scale $m_D$}}\\
The static contribution to Eq. \eqref{loopbm} is
\begin{equation}
\delta\langle L_R\rangle_{\mathrm{S}\,m_D}=
\frac{g^2C_R}{2T}\m2  \int \frac{d^dk}{(2\pi)^d}\frac{\Pi_{00}^{\mathrm{S}}(\mbk)}{(\bk^2+m_D^2)^2},
\label{mcontribnlos}
\end{equation}
where $\Pi_{00}^{\mathrm{S}}(\mbk)$ is the full static
contribution of Eq. \eqref{staticvacpol}.  The computation
is carried out in detail in appendix \ref{app_3d}; the result reads 
\begin{equation}
\delta\langle L_R\rangle_{\mathrm{S}\,m_D}=
\frac{g^4\crr\ca}{2(4\pi)^2}\left(-\frac{1}{2\epsilon}-\ln\frac{\mu^2}{4m_D^2}-\frac{1}{2}+\gamma_E -\ln (4\pi)\right).
\label{finalcontribm}
\end{equation}
The divergence cancels against the one of Eq. \eqref{finalcontribtnlons} 
coming from non-static modes at the scale $T$.\footnote{
Both divergences in Eqs. \eqref{finalcontribtnlons} and \eqref{finalcontribm}
are of ultraviolet origin. This seems to contradict the expectation according to which 
infrared divergences from higher scales should cancel against ultraviolet divergences 
from lower scales. The contradiction is only apparent.
The static modes at the scale $T$ develop both an ultraviolet 
and an infrared divergence that cancel against each other if regularized by the same 
cut off in dimensional regularization as assumed in Eq. \eqref{tcontribnlos}.  
In general, however, the ultraviolet divergence of the static modes at the
scale $T$ cancels against the ultraviolet divergence of the non-static modes,
such that the sum of static and non-static modes at the scale $T$ ends up 
having only a residual infrared divergence. It is precisely this infrared
divergence coming from the scale $T$, formally identical to the divergence in Eq. \eqref{finalcontribtnlons}, 
that cancels against the ultraviolet divergence in (\ref{finalcontribm}) coming from the scale $m_D$.} 
Note that the gauge-dependent part of Eq. \eqref{staticvacpol} gives a vanishing integral, 
thus yielding the expected gauge-independent result.
\end{enumerate}

Summing all contributions (static and the non-static) 
from the scales $T$ and $m_D$ up to order $g^4$ thus gives
\begin{equation}
\langle L_R\rangle=1+\frac{\crr \als }{2}\frac{m_D}{T}+\frac{\crr\alpha^2_s}{2}
\left[\ca\left(\ln\frac{m_D^2}{T^2}+\frac{1}{2}\right)-n_f\ln2\right]+\mathcal{O}(g^5).
\label{finalg4loop}
\end{equation}

\subsection{Comparison with the literature} 
At order $g^4$, the Polyakov loop was first calculated in the pure gauge 
case ($n_f=0$) and in Feynman gauge, by Gava and Jengo (GJ) \cite{Gava:1981qd}, 
who find  
\begin{equation}
\langle L_R\rangle_{\mathrm{GJ}}=
1 +\frac{\crr \als }{2}\frac{m_D}{T}+\frac{C_RC_A\als^2}{2}
\left(\ln \frac{m_D^2}{T^2 }- 2\ln 2+\frac{3}{2}\right)+\mathcal{O}(g^5)\,.
\label{gavajengo}
\end{equation}
Their result disagrees with ours, given in Eq. \eqref{finalg4loop}.

The disagreement may be traced back to an incorrect treatment of the 
static modes at the scale $m_D$ in \cite{Gava:1981qd}. In Feynman gauge, 
at order $g^4$, three terms contribute to the Polyakov loop: 
the non-static gluon self energy, whose dominant contribution comes from the scale $T$, the
static gluon self energy, getting contributions from the scale
$m_D$ only, and a third term coming from the fourth-order expansion of
the Polyakov line. The computation of Gava and Jengo correctly
reproduces the first and the third term. We show this with some detail 
in  appendix \ref{app_feynman}. However, in the evaluation of the static 
gluon self energy, the Debye mass is not resummed in the temporal 
gluons, leading to an inconsistent treatment of the scale $m_D$.\footnote{
In \cite{Gava:1981qd}, some contributions coming from the resummation of the Debye mass 
seem to have been included in $\delta W(0)$.
} 
Indeed, they have 
\begin{equation}
\Pi_{00}^{\mathrm{S}}(\bk)_\mathrm{GJ}= g^2C_AT \m2  
\int\frac{d^dp}{(2\pi)^d}\left(\frac{d-1}{\bp^2}-\frac{2\bk^2}{\bp^2\bq^2}\right),
\label{staticvacpolgj}
\end{equation}
which is the static self energy in Feynman gauge but \emph{without} resumming the 
Debye mass in the internal propagators. If, instead, the Debye mass is resummed, the expression 
of the static self energy changes to Eq. \eqref{staticvacpol} with $\xi=1$.
In this case, the calculation of the Polyakov loop in Feynman gauge 
leads to exactly the same result as in Eq. \eqref{finalg4loop}.

While the last part of this paper was being completed, 
Burnier, Laine and Veps\"al\"ainen \cite{Burnier:2009bk} 
published a perturbative analysis of the singlet quark-antiquark free energy.
In the first part of their work, they consider also the Polyakov loop at order $g^4$
within a dimensionally reduced effective field theory framework in a covariant 
or Coulomb gauge. Our result \eqref{finalg4loop} agrees with theirs.

\subsection{Higher-order contributions\label{sub_nnlo}}
In Sec.~\ref{sub_nlo}, we obtained in Eq. \eqref{mcontribnlonsfinal} 
a term that is of order $g^4\times(m_D/T)\sim g^5$. Other contributions of order $g^5$ 
can only come from $\langle\trt A_0^2\rangle$.  Hence, they are encoded in the 
two-loop expression of the gluon self energy.

At order $g^6$, we can expect other contributions from the two-loop self energy 
and contributions coming from the diagram in Fig.~\ref{fig:cubefourth} b). 
We explicitly calculate these last ones due to their relevance for  Sec.~\ref{sec_EFT}.
The computation is carried out by evaluating the colour trace of the
diagram in the representation $R$, whereas the loop integrations are
easily obtained by comparison with Eq. \eqref{fleading}. Thus we obtain 
\begin{equation} 
\delta\langle L_\mathrm{R}\rangle=
\left(3\crr^2-\frac{\crr\ca}{2}\right)\frac{\als^2}{24}\left(\frac{m_D}{T}\right)^2.
\label{als2m2}
\end{equation} 
The colour structure of this quartic term is not linear in $C_R$, 
a fact that will play a role in Sec.~\ref{sec_EFT}.
We recall here that the linear dependence of $\ln \langle L_\mathrm{R}\rangle$ 
on the Casimir operator $C_R$ is called \emph{Casimir scaling} of the Polyakov loop.
Equation \eqref{als2m2} provides the leading perturbative correction that breaks the 
Casimir scaling. It is a tiny correction of order $g^6$, 
which may explain, at least in the weak-coupling regime, the approximate Casimir scaling 
observed in lattice calculations \cite{Gupta:2007ax}.

\section{The Polyakov-loop correlator at order $g^6$ for $rT\ll 1$\label{sec_corr}}
The spatial correlator of Polyakov loops in the fundamental representation 
is defined as~\cite{McLerran:1981pb}
\begin{equation}
\label{defcorr}
\langle\tilde{\mathrm{Tr}}L_F^\dagger({\bf0})\tilde{\mathrm{Tr}}L_F(\br)\rangle.
\end{equation}
Following the notation of \cite{Nadkarni:1986cz}, we define 
$C_{\mathrm{PL}}(r,T)$ as the connected part of the correlator 
\begin{equation}
\label{defcpl}
C_{\mathrm{PL}}(r,T)\equiv
\langle\tilde{\mathrm{Tr}}L_F^\dagger({\bf0})\tilde{\mathrm{Tr}}L_F(\br)\rangle_\mathrm{c}
= \langle\tilde{\mathrm{Tr}}L_F^\dagger({\bf0})\tilde{\mathrm{Tr}}L_F(\br)\rangle-\langle L_F\rangle^2.
\end{equation}
Expanding Eq. \eqref{defcpl} up to order $g^6$ yields\footnote{
We adopt a slightly different definition of
$C_\mathrm{PL}(r,T)$ with respect to \cite{Nadkarni:1986cz}, in that we
consider the zeroth-order term in the perturbative expansion, i.e $1$,
as part of $\langle L_F\rangle^2$ rather than of $C_\mathrm{PL}$.}
\begin{eqnarray}
C_{\mathrm{PL}}(r,T)&=&
\frac{g^4}{(2!)^2}\frac{\langle\trt A_0^2({\bf0})\trt A_0^2(\br)\rangle_\mathrm{c}}{T^4}
+\frac{g^6}{(3!)^2}\frac{\langle \trt A_0^3({\bf0})\trt A_0^3(\br)\rangle_\mathrm{c}}{T^6}
\nonumber\\
&&
-\frac{2g^6}{2!\,4!}\frac{\langle \trt A_0^4({\bf0})\trt A_0^2(\br)\rangle_\mathrm{c}}{T^6}
+\mathcal{O}(g^8).
\label{pertcpl}
\end{eqnarray}
Since the generators of SU$(\nc)$ are traceless, the first term in the
expansion, which is $g^2\langle\trt A_0({\bf0})\trt A_0(\br)\rangle_\mathrm{c}$, 
vanishes and thus the correlator starts in perturbation theory with
a two-gluon exchange term. Terms with an odd number of gauge fields have been omitted from Eq. \eqref{pertcpl} since they vanish for charge-conjugation symmetry.

We will perform a complete calculation of the Polyakov-loop correlator 
for distances $r T \ll 1$. This situation corresponds to temperatures lower 
than the inverse distance of the quark-antiquark pair, hence it is the right one to make 
contact with known zero-temperature results. We assume the following hierarchy: 
\begin{equation}
\frac{1}{r}\gg T\gg m_D\gg \frac{g^2}{r}.
\label{defhierarchy}
\end{equation}
The scales $1/r$ and $g^2/r$ are the typical scales appearing in any perturbative 
static quark-antiquark correlator calculation \cite{Brambilla:1999qa,Brambilla:1999xf}.
The scales $T$ and $m_D$ are associated to the thermodynamics of the system.
We assume that they are smaller than $1/r$, because we are interested in short 
distances. We assume that they are larger than $g^2/r$, because 
we would like to study a situation where both thermodynamical scales affect the 
quark-antiquark potential \cite{Brambilla:2008cx}. In the weak-coupling regime, 
as discussed above, $T \gg m_D$, where $m_D$ is given by Eq. \eqref{fulldebye}. 
Equation \eqref{defhierarchy} amounts 
to having two largely unrelated small parameters, $g$ and $rT$, the hierarchy 
only requiring $rT \gg g$. Differently from the Polyakov-loop calculation where we had only $g$, 
the perturbative expansion of the Polyakov-loop correlator is, therefore, organized as a double 
expansion in $g$ and $rT$. We will stop the expansion for the Polyakov-loop correlator 
at order $g^6(rT)^0$, meaning that, given a term of order  $g^k(rT)^n$, 
we will display it only if $k<6$, for any (positive or negative) $n$, or if $k=6$, for $n\le 0$; 
we will not display it elsewhere. We should note here that,  
as in any double expansion whose expansion parameters are unrelated, 
undisplayed terms may, under some circumstances, turn out to be numerically 
as large as or larger than some of the displayed ones.\footnote{ 
A posteriori (see the final result in Eq.~\eqref{finalcpltot}), this may be avoided, in our case,  
by further requiring that $rT \gg \sqrt{g}$.}

In \cite{Nadkarni:1986cz}, Nadkarni computed the Polyakov-loop correlator
up to order $g^6$ using resummed temporal-gluon propagators throughout the
computation, which amounts to calculating the Polyakov-loop correlator for
distances $rm_D\sim 1$. Our calculation will differ from Nadkarni's one in that 
we adopt the different hierarchy \eqref{defhierarchy}. Nevertheless, some of our results can be obtained 
by expanding Nadkarni's result for $rm_D\ll 1$; we refer to Sec.~\ref{sub_nadkarni} 
for a detailed comparison between the two results.

The calculation of the different contributions to Eq. \eqref{pertcpl} will proceed 
similarly to the calculation of the Polyakov loop performed in the previous 
section. We will consider the different Feynman diagrams contributing to each of the terms 
in \eqref{pertcpl}, separate the contributions from the different energy scales and, in case, 
distinguish between static and non-static modes.

\begin{figure}[ht]
\begin{center}
\includegraphics[width=16cm]{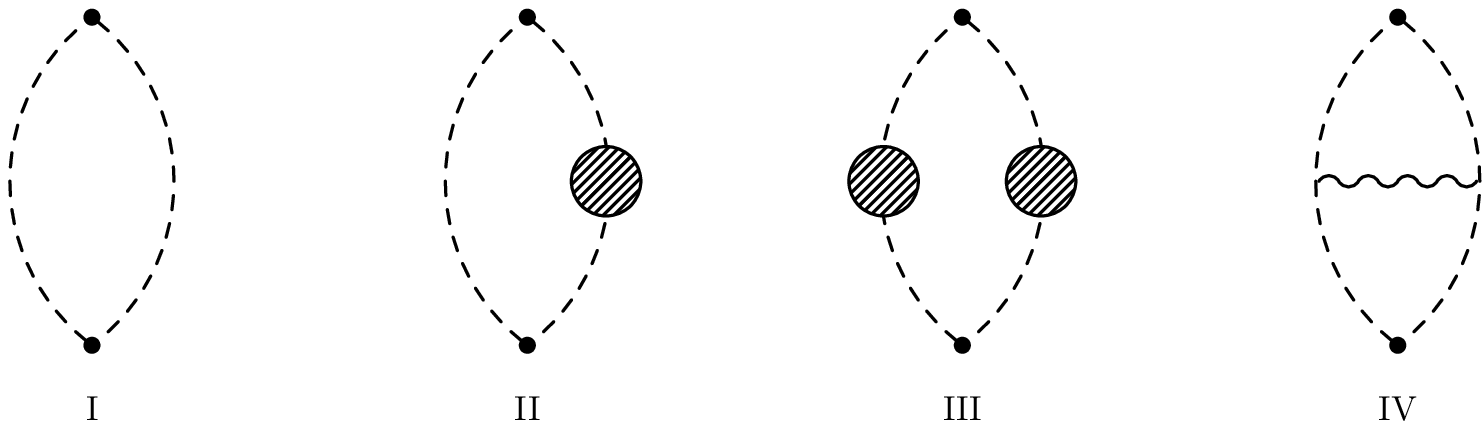}
\end{center}
\caption{Diagrams contributing to $\langle\trt A_0^2({\bf0})\trt A_0^2(\br)\rangle_\mathrm{c}$.}
\label{fig:corr-lo}
\end{figure}

\subsection{The leading-order contribution: diagram I}
We start by evaluating the four-field correlation function: its
leading-order contribution is given by diagram I in Fig.~\ref{fig:corr-lo}. 
It does not vanish only for momenta of order $1/r$, giving 
\begin{equation}
\delta C_\mathrm{PL}(r,T)_\mathrm{I}
=\frac{\nc^2-1}{8\nc^2}\frac{g^4}{T^2}\left(\m2
\int\frac{d^dk}{(2\pi)^d}\frac{e^{-i\bk\cdot\br}}{\bk^2}\right)^2
=\frac{\nc^2-1}{8\nc^2}\frac{\als^2}{ (rT)^2}.
\label{rg4contrib}
\end{equation}

\subsection{The contribution from diagrams of type II}
As we go beyond leading order, the first class of diagrams that we consider are 
those with gluon self-energy insertions in one temporal-gluon line, whose 
first example is diagram II in Fig.~\ref{fig:corr-lo}.
They give 
\begin{equation}
\delta C_\mathrm{PL}(r,T)_\mathrm{II} =
2\frac{\nc^2-1}{8\nc^2}\frac{g^4}{T^2}\frac{1}{4\pi r}\m2 
\int\frac{d^dk}{(2\pi)^d}e^{-i\bk\cdot\br}
\left(\frac{1}{\bk^2+\Pi_{00}(\bk)} - \frac{1}{\bk^2}\right),
\label{defselfenergyinsert}
\end{equation}
where the factor $2$ comes from the symmetric diagrams and 
$\Pi_{00}$ is the sum of bosonic and fermionic contributions to the gluon self
energy, as in the Polyakov-loop case. This diagram receives contributions from all scales and
depends on the gauge parameter $\xi$. However it can be shown that the gauge
dependence cancels with diagram IV \cite{Nadkarni:1986cz}, so, for
simplicity, here we write our results in static Feynman gauge, $\xi=1$.

\begin{enumerate}
\item{\emph{Contribution from the scale $1/r$}} \\
We start by evaluating the contribution from the scale $1/r$ in the integral.
If $\mbk\sim 1/r\gg T$, then we have 
\begin{equation}
\delta C_\mathrm{PL}(r,T)_{\mathrm{II}\,1/r} =
-\frac{\nc^2-1}{4\nc^2}\frac{g^4}{T^2}\frac{1}{4\pi r}\m2 
\int\frac{d^dk}{(2\pi)^d} e^{-i\bk\cdot\br} \frac{\Pi_{00}(\mbk\gg T)}{\bk^4}
\left[1 + {\cal O}\left( \frac{g^2}{rT} \right) \right],
\label{defselfenergyinsertr}
\end{equation}
where $\Pi_{00}(\mbk\gg T)$ is given by Eq. \eqref{kggtfull}. 
The Fourier transform of the vacuum part corresponds to the one-loop
static QCD potential and can be read from \cite{Fischler:1977yf,Billoire:1979ih}. 
Using that in dimensional regularization the Fourier transform of $1/\mbk^n$
becomes \cite{Gelfand:1964}
\begin{equation}
\int\frac{d^dk}{(2\pi)^d}\frac{e^{-i\bk\cdot\br}}{\mbk^n}=
\frac{2^{-n}\pi^{-d/2}}{r^{d-n}}\frac{\Gamma\left(d/2-n/2\right)}{\Gamma\left(n/2\right)},
\label{ftregdim}
\end{equation}
we have
\begin{eqnarray}
\delta C_\mathrm{PL}(r,T)_{\mathrm{II}\,1/r}&=&
\frac{\nc^2-1}{8\nc^2}\frac{\als^3}{ (rT)^2}
\left\{\frac{1}{2\pi}
\left[2\beta_0(\ln(\mu r)+\gamma_E)+ \frac{31}{9}\ca-\frac{10}{9}n_f \right]\right.
\nonumber\\
&&\left.+\ca\left(\frac{1}{12rT} -rT -\frac{2 }{9}\pi(rT)^2\right)\right\}
+\mathcal{O}\left(g^6(rT)^2,g^7\right).
\label{selfenergyinsertr}
\end{eqnarray}
The term in the first line comes from the Fourier transform of the vacuum
contribution, whereas the terms in the second line come respectively from the
singular part, the (zero mode) order $\epsilon$ term\footnote{
The dimensionally-regularized Fourier transform of the order $\epsilon$ term 
in Eq. \eqref{kggtfull}  yields a $1/\epsilon$ pole, eventually leading to a finite contribution.}  
and the $T^2$ term in Eq. \eqref{kggtfull}. 
Higher-order corrections to Eq. \eqref{kggtfull} contribute at order 
$g^6(rT)^2$ or $g^7$. Higher order radiative corrections to the gluon
self energy contribute at order $g^8$. Note that the $(\als^3/\pi)  \beta_0 \ln(\mu r)$ term in 
Eq. \eqref{selfenergyinsertr} fixes the natural scale of  $\als^2$ in the LO 
term $\delta C_\mathrm{PL}(r,T)_\mathrm{I}$ to be $1/r$.

\item{\emph{Contributions from the scales $T$ and $m_D$}} \\
We now consider the contributions from the thermal scales. For what
concerns the temperature, $\mbk\sim T$ translates into $r \mbk\ll 1$
and $m_D \ll \mbk$. Integrating out the temperature leads to the following 
contribution 
\begin{eqnarray}
\delta C_\mathrm{PL}(r,T)_{\mathrm{II}\,T}&=&
-\frac{\nc^2-1}{4\nc^2}\frac{g^4}{T^2}\frac{1}{4\pi r}
\m2 \int\!\frac{d^dk}{(2\pi)^d}
\left[1+  {\cal O}((\bk \cdot \br)^2) \right]
\frac{\Pi_{00}(\mbk\sim T)}{\bk^4}
\nonumber\\
&& \hspace{5cm}
\times
\left[ 1 +  {\cal O}(g^2)\right],
\label{defselfenergyinsertt}
\end{eqnarray}
where we have implemented the condition  $r\mbk\ll 1$ by expanding the Fourier exponent.
Integrating out the Debye-mass scale leads to the following contribution
\begin{eqnarray}
\nonumber\delta C_\mathrm{PL}(r,T)_{\mathrm{II}\,m_D}&=&
\frac{\nc^2-1}{4\nc^2}\frac{g^4}{T^2}\frac{1}{4\pi r}\m2 \int\frac{d^dk}{(2\pi)^d}
\left[1+  {\cal O}((\bk \cdot \br)^2) \right]
\left[\frac{1}{\bk^2+m_D^2}   \right.
\\
&& \hspace{3cm} \left.-\frac{\Pi_{00}(\mbk\sim m_D)-m_D^2}{(\bk^2+m_D^2)^2}
+ {\cal O}\left( \frac{g^4}{m_D^2}\right) 
\right].
\label{defselfenergyinsertm}
\end{eqnarray}
The integrals to be evaluated are the same needed to evaluate 
Eqs. \eqref{loopbT}, \eqref{loopbm0} and \eqref{loopbm}.  
Thus, summing the $T$ and $m_D$ contributions, we obtain
\begin{eqnarray}
\nonumber\delta C_\mathrm{PL}(r,T)_{\mathrm{II}\,T+m_D}&=&
-\frac{\nc^2-1}{4\nc^2}\frac{\als^2}{rT}\left\{\frac{m_D}{T}
+\als\left[\ca\left(\ln\frac{m_D^2}{T^2}+\frac12\right)-n_f\ln2\right]\right\}
\\
&&+\mathcal{O}\left(\frac{g^7}{rT},g^6(rT)\right).
\label{selfenergyinserttm}
\end{eqnarray}
The term of order $g^5/(rT)$ comes from the first term in
\eqref{defselfenergyinsertm}, the terms of order $g^6/(rT)$ come
from the non-static modes in \eqref{defselfenergyinsertt} and from the 
static ones in the second term of Eq. \eqref{defselfenergyinsertm}, the 
appearance of the logarithm $\ln{m_D^2}/{T^2}$ signals the cancellation between 
divergences at the scale $T$ and $m_D$, the suppressed term $g^7/(rT)$ 
comes from the non-static modes in the second term of
Eq. \eqref{defselfenergyinsertm} (see Eqs. \eqref{mcontribnlons} and 
\eqref{mcontribnlonsfinal} for the analogous case in the Polyakov-loop
calculation), whereas the suppressed term $g^6(rT)$ comes from the
$(\bk\cdot\br)^2$ term in Eq. \eqref{defselfenergyinsertt}. 
\end{enumerate}

\subsection{The contribution from diagrams of type III}
Diagram III in Fig.~\ref{fig:corr-lo} is the first example of the class of
diagrams with gluon self-energy insertions in both temporal-gluon lines.
They may be evaluated from the diagrams of type II:
\begin{eqnarray}
\delta C_\mathrm{PL}(r,T)_\mathrm{III} 
&=&\frac{\nc^2-1}{8\nc^2}\frac{g^4}{T^2}\left[\m2
\int\frac{d^dk}{(2\pi)^d} e^{-i\bk\cdot\br}
\left( \frac{1}{\bk^2+\Pi_{00}(\bk)} - \frac{1}{\bk^2} \right) \right]^2
\nonumber\\
&=&\frac{8N^2}{N^2-1}\frac{T^2}{g^4}\left(4\pi r\frac{\delta C_\mathrm{PL}(r,T)_\mathrm{II}}{2}\right)^2 .
\label{defIII}
\end{eqnarray}
The leading-order term in \eqref{selfenergyinserttm} gives a $g^6$ contribution 
to $\delta C_\mathrm{PL}(r,T)_\mathrm{III}$, all other contributions being at
least of order $g^7/(rT)^2$, 
\begin{equation}
\delta C_\mathrm{PL}(r,T)_{\mathrm{III}}=
\frac{\nc^2-1}{8\nc^2}\als^2\frac{m_D^2}{T^2}+\mathcal{O}\left(\frac{g^7}{(rT)^2}\right).
\label{finalIII}
\end{equation}

\subsection{The contribution from diagrams of type IV}
The transverse static-gluon exchange between the two temporal-gluon lines
(diagram IV and the diagrams derived from IV by inserting gluon self energies 
in each of the gluon lines) receives the following contributions. 

\begin{enumerate}
\item{\emph{Contribution from the scale $1/r$}} \\
The contribution from the scale $1/r$ reads at leading order (with $\xi=1$) 
\begin{eqnarray}
\delta C_\mathrm{PL}(r,T)_\mathrm{IV\,1/r}
&=& \frac{g^6}{4T}\frac{\nc^2-1}{2\nc^2}\ca 
\mu^{6\epsilon} \int \frac{d^dk_1}{(2\pi)^d} \int \frac{d^dk_2}{(2\pi)^d} \int \frac{d^dp}{(2\pi)^d}
e^{-i(\bk_1-\bk_2)\cdot\br}
\nonumber\\
&&
\hspace{5cm} \times 
\frac{(2\bk_1+\bp)\cdot(2\bk_2+\bp)}{\bk_1^2\bk^2_2(\bk_1+\bp)^2(\bk_2+\bp)^2\bp^2}.
\label{nadkarnicontrib}
\end{eqnarray}
Gluon self-energy insertions are suppressed by $g^2$.

\item{\emph{Contribution from the scale $T$}} \\
The contribution from the scale $T$ vanishes, because scaleless, if no 
self-energy insertions are considered. Hence, the leading contribution from the
scale $T$ is of order $g^6/T \times g^2T \sim g^8$. 

\item{\emph{Contribution from the scale $m_D$}} \\
The contribution from the scale $m_D$ reads 
\begin{eqnarray}
&&\hspace{-10mm}
\delta C_\mathrm{PL}(r,T)_\mathrm{IV\,m_D} =
\nonumber\\
&& 
\frac{g^6}{4T}\frac{\nc^2-1}{2\nc^2}\ca
\mu^{6\epsilon} \int \frac{d^dk_1}{(2\pi)^d} \int\frac{d^dk_2}{(2\pi)^d}  \int\frac{d^dp}{(2\pi)^d}
\left[1+  {\cal O}(((\bk_1-\bk_2) \cdot \br)^2) \right]
\nonumber\\
&&
\hspace{0.5cm} 
\times 
\frac{(2\bk_1+\bp)\cdot(2\bk_2+\bp)}{(\bk_1^2+m_D^2)(\bk^2_2+m_D^2)((\bk_1+\bp)^2+m_D^2)((\bk_2+\bp)^2+m_D^2)\bp^2}
\left[1+  {\cal O}(g) \right].
\label{nadkarnicontribm}
\end{eqnarray}
This corresponds to a contribution  of order $g^6(m_D/T)\sim g^7$, which is beyond our accuracy. 
\end{enumerate}

The leading contribution to $\delta C_\mathrm{PL}(r,T)_\mathrm{IV}$ comes, therefore, from  $\delta
C_\mathrm{PL}(r,T)_\mathrm{IV\,1/r}$, which can be computed in dimensional
regularization with the help of Eq. \eqref{ftregdim}. Our final result reads
\begin{equation}
\delta C_\mathrm{PL}(r,T)_\mathrm{IV}
=\frac{\nc^2-1}{2\nc}\frac{\als^3 }{rT}\left(1-\frac{\pi^2}{16}\right)+\mathcal{O}\left(g^7\right).
\label{corr-IV}
\end{equation}
The same result follows from \cite{Nadkarni:1986cz} by expanding in $rm_D\ll 1$.

\begin{figure}[ht]
\begin{center}
\includegraphics{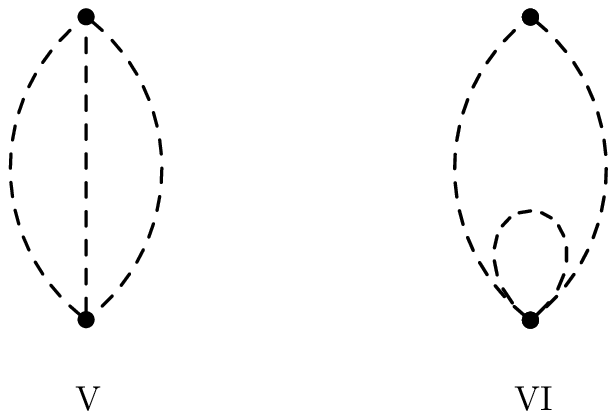}
\end{center}
\caption{Diagram V is the first contribution to $\langle \trt
  A_0^3({\bf0})\trt A_0^3(\br)\rangle_\mathrm{c}$, whereas diagram VI is the
  first contribution to $\langle \trt A_0^4({\bf0})\trt A_0^2(\br)\rangle_\mathrm{c}$.}
\label{fig:corr-nlo}
\end{figure}

\subsection{The contribution from diagrams of type V}
Diagrams contributing to the correlators of six $A_0$ fields in
Eq. \eqref{pertcpl} are shown in Fig.~\ref{fig:corr-nlo}. 
The LO diagram  contributing to $\langle \trt A_0^3({\bf0})\trt
A_0^3(\br)\rangle$ is diagram V, which gives 
\begin{equation}
\delta C_\mathrm{PL}(r,T)_\mathrm{V}
=\frac{(\nc^2-4)(\nc^2-1)}{96\nc^3}\frac{\als^3}{(rT)^3}.
\label{corr-33}
\end{equation}
If we consider diagram V with gluon self-energy insertions in one of the
temporal lines, in analogy to \eqref{defselfenergyinsertm},  then this starts 
contributing at order $g^7/(rT)^2$, which is beyond our accuracy.

\subsection{The contribution from diagrams of type VI}
Diagrams contributing to $\langle \trt A_0^4({\bf0})\trt
A_0^2(\br)\rangle$ are like diagram VI in Fig.~\ref{fig:corr-nlo}
and diagrams derived from VI by inserting gluon self energies 
and other radiative corrections. 
Colour factors aside, their leading contribution may be estimated 
by simply multiplying the contribution of the diagrams of
Fig.~\ref{fig:corr-lo} to the Polyakov-loop correlator 
with the contribution of the diagrams of Fig.~\ref{fig:square}
to the Polyakov loop. Hence, diagrams of type VI
contribute at LO to order $g^4/(rT)^2 \times g^2m_D/T \sim g^7/(rT)^2$, 
which is beyond our accuracy.

\subsection{The Polyakov-loop correlator up to order $g^6$}
Summing up all contributions, we then have 
\begin{eqnarray}
C_{\mathrm{PL}}(r,T)&=&
\frac{\nc^2-1}{8\nc^2}\left\{ 
\frac{\als(1/r)^2}{(rT)^2}
-2\frac{\als^2}{rT}\frac{m_D}{T} \right.
\nonumber\\
&&
\hspace{1.5cm}
+\frac{\als^3}{(rT)^3}\frac{\nc^2-2}{6\nc}
+ \frac{1}{2\pi }\frac{\als^3}{(rT)^2}\left(\frac{31}{9}\ca-\frac{10}{9}n_f +2\gamma_E\beta_0\right)
\nonumber\\
&&
\hspace{1.5cm}
+ \frac{\als^3}{rT}\left[
\ca\left(-2 \ln\frac{m_D^2}{T^2} + 2-\frac{\pi^2}{4}\right) + 2n_f\ln 2\right]
\nonumber\\
&&
\hspace{1.5cm}
\left.
+\als^2\frac{m_D^2}{T^2} -\frac{2}{9}\pi \als^3 \ca
\right\}
+\mathcal{O}\left(g^6(rT),\frac{g^7}{(rT)^2}\right),
\label{finalcpltot}
\end{eqnarray}
where we have made explicit the scale dependence of $\als$ in the leading term. 
Note that the $r$, $T$ and $m_D$ independent term proportional to 
$-2\pi \als^3 \ca/9$ comes from Eq. \eqref{selfenergyinsertr}, so it is 
actually a contribution from the scale $1/r$ that accounts for the matter part of the 
gluon self energy. The term proportional 
to $\als^3/(rT)^3$ comes from diagram V, Eq. \eqref{corr-33}, 
and from the singular part of the gluon self energy in the static gauge, 
Eq. \eqref{selfenergyinsertr}.

\subsection{Comparison with the result of Nadkarni\label{sub_nadkarni}}
We compare here with Nadkarni's (N) computation of the Polyakov-loop correlator \cite{Nadkarni:1986cz}. 
The regime of validity of Nadkarni's computation is $T \gg 1/r \sim  m_D$, while ours is 
$1/r \gg T \gg m_D$. Therefore, we may only compare results obtained here that 
do not involve the hierarchy $rT \ll 1$, with Nadkarni's results that do not 
involve the hierarchy $rT \gg 1$, expanded for $rm_D \ll 1$.

In \cite{Nadkarni:1986cz}, the tree-level expression of 
$g^4\langle\trt A_0^2({\bf0})\trt A_0^2(\br)\rangle_\mathrm{c}/(4T^4)$ reads 
$(\nc^2-1)/(8\nc^2)\als^2$ $\exp(-2rm_D)/(rT)^2$, which expanded for $rm_D \ll 1$ 
gives $\delta C_\mathrm{PL}(r,T)_\mathrm{I}$, the LO of 
$\delta C_\mathrm{PL}(r,$ $T)_{\mathrm{II}\,m_D}$ (to be read from Eq. \eqref{selfenergyinserttm})
and $\delta C_\mathrm{PL}(r,T)_{\mathrm{III}}$. Also, the tree-level expression of 
$g^6\langle \trt A_0^3({\bf0}) \trt A_0^3(\br)\rangle_\mathrm{c}/(36T^6)$
in \cite{Nadkarni:1986cz} agrees with $\delta C_\mathrm{PL}(r,T)_\mathrm{V}$
once expanded for $rm_D \ll 1$.

Diagram IV in Fig.~\ref{fig:corr-lo} also contributes 
to Nadkarni's calculation. The diagram does not involve gluon self-energy 
insertions and therefore its calculation does not rely on the hierarchy 
between $1/r$ and $T$. As already remarked, $\delta C_\mathrm{PL}(r,T)_\mathrm{IV}$ 
agrees with  Nadkarni's result once expanded for $rm_D\ll 1$.\footnote{
In Nadkarni's paper this contribution is called  $f_{II}$.}

Let's now consider the NLO contribution to $\delta C_\mathrm{PL}(r,T)_{\mathrm{II}\,m_D}$.
This contribution is given by the static part of Eq. \eqref{defselfenergyinsertm}:
\begin{equation}
\delta C_\mathrm{PL}(r,T)_{\mathrm{II}\,{\rm N}\,m_D}=
-\frac{\nc^2-1}{4\nc^2}\frac{g^4}{T^2}\frac{1}{4\pi r}\m2
\int\frac{d^dk}{(2\pi)^d}\frac{\Pi_{00}^\mathrm{S}(\mbk)}{(\bk^2+m_D^2)^2}.
\label{defselfenergyinsertmnad}
\end{equation}
The integral is divergent. In our case, i.e. assuming $1/r \gg T \gg m_D$, 
the divergence cancels against $\delta C_\mathrm{PL}(r,T)_{\mathrm{II}\,T}$, eventually 
leading to a finite result in $\delta C_\mathrm{PL}(r,T)_{\mathrm{II}\,T+m_D}$.
The $\ln m_D/T$ term in Eq. \eqref{selfenergyinserttm} signals precisely that 
a divergence at the scale $m_D$ has canceled against a divergence at the scale $T$.
In  Nadkarni's case,  i.e. assuming $T \gg 1/r \gg m_D$, we get, along with 
$\delta C_\mathrm{PL}(r,T)_{\mathrm{II}\,{\rm N}\,m_D}$, a contribution from the scale $1/r$, which is 
\begin{equation}
\delta C_\mathrm{PL}(r,T)_{\mathrm{II}\,{\rm N}\,1/r} =
-\frac{\nc^2-1}{4\nc^2}\frac{g^4}{T^2}\frac{1}{4\pi r}\m2 
\int\frac{d^dk}{(2\pi)^d}e^{-i\bk\cdot\br}\frac{\Pi_{00}^\mathrm{S}(\mbk\gg m_D)}{\bk^4}.
\label{defselfenergyinsertrnad}
\end{equation}
This is like Eq. \eqref{defselfenergyinsertr},   
but involves only the static part of the self energy \eqref{staticcontrhighk},
since non-static modes have been already integrated out at the larger scale $T$.
According to Eq. \eqref{staticcontrhighk}, we have $\Pi_{00}^\mathrm{S}(\mbk\gg m_D)\sim T\mbk^{1-2\epsilon}$.
The Fourier transform of $1/\mbk^{3+2\epsilon}$ originates a $1/\epsilon$ pole.
It is this divergence that in Nadkarni's hierarchy cancels against the divergence in 
$\delta C_\mathrm{PL}(r,T)_{\mathrm{II}\,{\rm N}\,m_D}$ leading to the finite result
\begin{equation}
\delta C_\mathrm{PL}(r,T)_{\mathrm{II}\,{\rm N}\,m_D} + \delta C_\mathrm{PL}(r,T)_{\mathrm{II}\,{\rm N}\,1/r} =
-\frac{\nc^2-1}{2\nc}\frac{\als^3}{rT}\left[\ln (2 m_D r)
  +\gamma_E-\frac{3}{4} +\mathcal{O}(rm_D)\right],
\label{nadkarniII}
\end{equation}
which agrees with the result in  \cite{Nadkarni:1986cz}.\footnote{
In Nadkarni's paper this contribution is called  $f_I$.} 
In this case, the $\ln m_D r$ term signals that a divergence at the scale $m_D$ 
has canceled against a divergence at the scale $r$.

\section{The Polyakov-loop correlator in an EFT language\label{sec_EFT}}
The calculation of the Polyakov loop discussed in the previous section can be 
conveniently rephrased in an effective field theory (EFT) language that exploits 
at the Lagrangian level the hierarchy of energy scales in Eq.~\eqref{defhierarchy}.
The EFT framework has the advantage to allow more easily for systematic improvements of the 
calculation and to make more transparent its physical meaning.

Our starting point is QCD with a static quark and a static antiquark, denoted 
in the following as {\em static QCD}. Its action in Euclidean space-time reads
\begin{equation}
{\cal S}_{\rm QCD} = \int_0^{1/T} \!\! d\tau \int d^3x \; \left( \psi^\dagger D_0 \psi + \chi^\dagger D_0 \chi 
+ \frac{1}{4}F^a_{\mu\nu}F^a_{\mu\nu} + \sum_{l=1}^{n_f} \bar{q}_lD\!\!\!\!/
\, q_l \right),
\label{QCD}
\end{equation}
where $D_0 = \partial_0 - igA_0$, $\psi$ is the Pauli spinor field that annihilates 
a static quark, $\chi$ is the Pauli spinor field that creates a static antiquark, 
and $q_1$, ..., $q_{n_f}$ are the light quark fields, which are 
assumed to be massless in this study. 

The Polyakov-loop correlator may be expressed in static QCD as 
\begin{equation}
\langle\tilde{\mathrm{Tr}}L_F^\dagger({\bf 0})\tilde{\mathrm{Tr}}L_F(\br)\rangle 
= \frac{1}{\nc^2}\frac{1}{\cal N} 
\langle \chi^\dagger_j({\bf 0},1/T)\psi_i({\bf r},1/T)\psi^\dagger_i({\bf r},0)\chi_j({\bf 0},0)\rangle,
\label{PLC-QCD}
\end{equation}
where ${\cal N} = [\delta^3({\bf 0})]^2$ and we have written explicitly the colour indices.
The thermal average on the right-hand side reduces to the Polyakov-loop correlator on 
the left-hand side after integrating out the fields $\psi$ and $\chi$ \cite{McLerran:1981pb}.
On general grounds, one also expects that \cite{Luscher:2002qv,Jahn:2004qr}
\begin{equation}
\langle\tilde{\mathrm{Tr}}L_F^\dagger({\bf 0})\tilde{\mathrm{Tr}}L_F(\br)\rangle 
= \frac{1}{\nc^2} \sum_n e^{-E_n/T},
\label{PLC-spectrum}
\end{equation}
where $E_n$ are the eigenvalues of the Hamiltonian associated to the static QCD Lagrangian.

\subsection{pNRQCD\label{secpNRQCD}}
Potential non-relativistic QCD (pNRQCD) is the EFT that follows from QCD 
by integrating out from the static quark-antiquark sector
gluons of energy or momentum that scale like 
the inverse of the distance $r$ between the quark and the antiquark.
Since $1/r$ is the largest scale, the matching of the pNRQCD Lagrangian may be done 
by setting to zero all other scales and, in particular, the thermal ones; 
as a consequence, the Lagrangian is identical to the one derived at zero 
temperature \cite{Pineda:1997bj,Brambilla:1999xf,Brambilla:2004jw,Brambilla:2002nu,Brambilla:2003nt}.
In Euclidean space-time, the action reads
\begin{eqnarray}
{\cal S}_{\rm pNRQCD} &=&  
\int_0^{1/T} \!\! d\tau \int d^3x \int d^3r\, {\rm Tr} \Bigg\{ {\rm S}^\dagger (\partial_0+V_s){\rm S}
+ {\rm O}^\dagger (D_0+V_o){\rm O}
\nonumber\\
&& \hspace{1.5cm}
- iV_A \left( {\rm S}^\dagger {\bf r}\cdot g{\bf E} {\rm O} + {\rm O}^\dagger {\bf r}\cdot g{\bf E} {\rm S}\right)
- \frac{i}{2}V_B \left({\rm O}^\dagger {\bf r}\cdot g{\bf E} {\rm O} + {\rm O}^\dagger{\rm O} {\bf r}\cdot g{\bf E} \right) 
\nonumber\\
&&  
\hspace{1.5cm} 
+ \frac{i}{8} V_C \left(r^i r^j  {\rm O}^\dagger D^igE^j {\rm O} - r^i r^j  {\rm O}^\dagger {\rm O} D^igE^j \right)
+ \delta {\cal L}_{\rm pNRQCD}
\Bigg\}
\nonumber\\
&&
+ \int_0^{1/T} \!\! d\tau \int d^3x \; \left(
\frac{1}{4}F^a_{\mu\nu}F^a_{\mu\nu} + \sum_{l=1}^{n_f} \bar{q}_lD\!\!\!\!/\,q_l
\right),
\label{pNRQCD}
\end{eqnarray}
where the trace is over the colour indices, ${\rm S} = 1\hspace{-1.1mm}{\rm l}_{\nc \times \nc} S/\sqrt{N}$ is  
a quark-antiquark field in a colour-singlet configuration, ${\rm O} = \sqrt{2} T^a O^a$ 
is a quark-antiquark field in a colour-octet configuration, $D_0{\rm O} = \partial_0 - ig[A_0,{\rm O}]$, 
${\bf D} = \bfnabla -ig{\bf A}$ and $E^i = F_{i0}$ is the chromoelectric field. 
The fields $S$ and $O^a$ depend on the continuous parameter $\br$ that labels the distance between the quark and the antiquark, the centre-of-mass coordinate ${\bf x}$ and the Euclidean time $\tau$; 
the gluon fields have been multipole expanded and, therefore, depend on
${\bf x}$ and $\tau$ only. The quantities $V_s$, $V_o$, $V_A$, $V_B$ and $V_C$ are 
the matching coefficients of the EFT. These are non-analytic functions of $r$. Since 
$V_A(r) = 1 + {\cal O}(\als^2)$ \cite{Brambilla:2006wp}, 
$V_B(r) = 1 + {\cal O}(\als^2)$ and  $V_C(r) = 1 + {\cal O}(\als)$ it will suffice to our 
purposes to put $V_A(r)=V_B(r)=V_C(r)=1$ from now on. $V_s$ and $V_o$ are the singlet and octet 
potentials in pNRQCD: $V_s$ is known up to three loops \cite{Anzai:2009tm} and 
$V_o$ is known up to two loops \cite{Kniehl:2004rk}. For the purpose of obtaining the Polyakov-loop 
correlator at NNLO accuracy it is sufficient to know $V_s$ and $V_o$ at one-loop accuracy and their 
difference at two-loop accuracy:
\begin{eqnarray}
&& V_s(r) = -C_F\frac{\als(1/r)}{r}\left[ 1 
+ \left( \frac{31}{9}\ca-\frac{10}{9}n_f +2\gamma_E\beta_0 \right)\frac{\als}{4\pi} + {\cal O}(\als^2)\right],
\label{Vspnrqcd}\\
&& V_o(r) = \frac{1}{2\nc} \frac{\als(1/r)}{r}\left[ 1 
+ \left( \frac{31}{9}\ca-\frac{10}{9}n_f +2\gamma_E\beta_0 \right)\frac{\als}{4\pi} + {\cal O}(\als^2)\right],
\label{Vopnrqcd}\\
&& (\nc^2-1)V_o(r) + V_s(r) =
\frac{\nc(\nc^2-1)}{8}\frac{\als^3}{r}\left(\frac{\pi^2}{4}-3\right)\left[ 1 + {\cal O}(\als)\right].
\label{Vo-Vs}
\end{eqnarray}
Finally, $\delta {\cal L}_{\rm pNRQCD}$ includes all operators that are of
order $r^3$ or smaller. At tree-level, they may be read from the multipole expansion
of the quark and antiquark coupling to the temporal gluon in the static QCD
Lagrangian \eqref{QCD}, hence they just involve covariant derivatives acting on a chromoelectric field: 
the leading-order operator being
$- i r^i r^j r^k {\rm Tr} \{ {\rm O}^\dagger D^iD^jgE^k {\rm S}$ $+{\rm S}^\dagger D^iD^jgE^k {\rm O} \}/24$ 
\cite{Brambilla:2002nu}. As we will argue in the next section, these terms 
are of order $g^4$, however, their contribution eventually cancels 
in the Polyakov-loop correlator up to order $g^6(rT)^0$. For this reason, 
we do not need to specify them further here.

Matching the connected Polyakov-loop correlator to pNRQCD gives 
\begin{eqnarray}
C_\mathrm{PL}(r,T)
&=& 
\frac{1}{\nc^2}\Bigg[
Z_s \frac{\langle S(\br,{\bf 0},1/T)S^\dagger(\br,{\bf 0},0)\rangle}{\cal N}
+
Z_o \frac{\langle O^a(\br,{\bf 0},1/T)O^{a \, \dagger}(\br,{\bf 0},0)\rangle}{\cal N}
\nonumber\\
&& \hspace{0.8cm}
+ {\cal O}\left(\als^3(rT)^4\right)\Bigg] - \langle L_F \rangle^2.
\label{PLC-pNRQCD}
\end{eqnarray}
The right-hand side is the pNRQCD part of the matching.
It contains the singlet and octet correlators, 
$\langle S(\br,{\bf 0},1/T)S^\dagger(\br,{\bf 0},0)\rangle$ and 
$\langle O^a(\br,{\bf 0},1/T)O^{a \, \dagger}(\br,{\bf 0},0)\rangle$, not surprisingly 
because in the $r\to 0$ limit the tensor fields $\chi^\dagger_j({\bf 0},1/T)\psi_i({\bf r},1/T)$ 
and $\psi^\dagger_i({\bf r},0)\chi_j({\bf 0},0)$,  
appearing in the right-hand side of Eq. \eqref{PLC-QCD}, decompose into the direct sum 
of a colour-singlet and a colour-octet component.
The colour-singlet and colour-octet correlators may be read from the Lagrangian \eqref{pNRQCD}:
\begin{eqnarray}
\frac{\langle S(\br,{\bf 0},1/T)S^\dagger(\br,{\bf 0},0)\rangle}{\cal N} &=& e^{-V_s(r)/T}(1+ \delta_s),
\label{SSpNRQCD}
\\
\frac{\langle O^a(\br,{\bf 0},1/T)O^{a\,\dagger}(\br,{\bf 0},0)\rangle}{\cal  N} &=&
e^{-V_o(r)/T}\left[(\nc^2-1)\, \langle L_A \rangle  +  \delta_o\right],
\label{OOpNRQCD}
\end{eqnarray}
where $\delta_s$ and $\delta_o$ stand for loop corrections to the singlet and octet correlators respectively.
The factor $\langle L_A \rangle$
comes from the covariant derivative $D_0$ acting on the octet field in \eqref{pNRQCD}.\footnote{
The adjoint Polyakov loop $\langle L_A \rangle$ factorizes the contribution
coming from the gluons in the thermal bath that bind with the colour-octet quark-antiquark states to
form part of the spectrum appearing in the right-hand side of Eq. \eqref{PLC-spectrum}.
In pNRQCD at zero temperature, a similar expression
factorizes the non-perturbative gluonic contribution to the gluelumps masses \cite {Brambilla:1999xf}.
}
Note that at finite temperature, for $T\simg g^2/r$, 
the octet correlator is not suppressed with respect to the singlet one, while 
in the opposite limit, $T \ll g^2/r$, the Polyakov-loop correlator 
is dominated by the singlet contribution.
Higher-dimensional operators have not been
displayed, because they are negligible with respect to our present accuracy, which is 
of order $g^6(rT)^0$.  The reason is that higher-dimensional operators involve the coupling with 
at least two field-strength tensors, hence the corresponding matrix elements
are at least of order $(rT)^4$; moreover, as can be seen by adding two
external gluons to diagram I of Fig.~\ref{fig:corr-lo}, the matrix element of an operator coupled
with two external gluons is at least of order $g^6$.
The normalization factors $Z_s$ and $Z_o$ have to be determined from the
matching condition \eqref{PLC-pNRQCD}. While $V_s$ and $V_o$ 
are the same at zero and finite temperature, the normalization
factors are not for they depend on the boundary conditions. 

In order to determine the normalization factors $Z_s$ and $Z_o$, let us consider in Eq. \eqref{PLC-pNRQCD}
only contributions coming from the scale $1/r$. 
In dimensional regularization, all loop corrections vanish in the pNRQCD part
of the matching and the Polyakov loops $\langle L_F \rangle$ and $\langle L_A \rangle$ 
reduce to one; therefore, the matching condition reads
\begin{equation}
C_\mathrm{PL}(r,T)_{1/r} = 
\langle\tilde{\mathrm{Tr}}L_F^\dagger({\bf 0})\tilde{\mathrm{Tr}}L_F(\br)\rangle_{1/r} - 1= 
\frac{1}{\nc^2}\left[ Z_se^{-V_s(r)/T} + Z_o(\nc^2-1)e^{-V_o(r)/T} \right] -1.
\label{pNRQCDrmatching}
\end{equation}
We may now proceed in different ways. A way consists in matching 
with the spectral decomposition \eqref{PLC-spectrum}. 
By noting that at the scale $1/r$ the spectrum is just given by a singlet state 
of energy $V_s(r)$ and $\nc^2-1$ degenerate octet states of energy $V_o(r)$,
the matching condition  implies that $Z_s = Z_o = 1$.
Another way consists in taking advantage of the Polyakov-loop correlator 
calculation done in Sec.~\ref{sec_corr} and matching to it.
$C_\mathrm{PL}(r,T)_{1/r}$ is the sum of Eq. \eqref{rg4contrib}, 
Eq. \eqref{selfenergyinsertr}  without the contribution from the matter part 
of the gluon self energy, Eq. \eqref{corr-IV} and Eq. \eqref{corr-33}; it reads
\begin{eqnarray}
C_\mathrm{PL}(r,T)_{1/r} &=& 
\frac{\nc^2-1}{8\nc^2}\left\{ \frac{\als(1/r)^2}{(rT)^2} 
+ \frac{\als^3}{(rT)^3}\frac{\nc^2-2}{6\nc} 
\right.
+ \frac{1}{2\pi }\frac{\als^3}{(rT)^2}\left(\frac{31}{9}\ca-\frac{10}{9}n_f +2\gamma_E\beta_0\right)
\nonumber\\
&& \hspace{1.7cm}
\left.
+ \frac{\als^3}{rT}\ca \left(3-\frac{\pi^2}{4}\right)
+{\cal O}\left(\frac{\als^4}{(rT)^4}\right)
\right\}.
\label{CPLr}
\end{eqnarray}
A direct inspection shows that this expression satisfies 
\begin{equation}
C_\mathrm{PL}(r,T)_{1/r} = 
\frac{1}{\nc^2}\left[ e^{-V_s(r)/T} + (\nc^2-1)e^{-V_o(r)/T} \right] -1,
\label{pNRQCDrverification}
\end{equation}
up to order $\als^3$, for $V_s(r)$ and $V_o(r)$ given by
Eqs. \eqref{Vspnrqcd}-\eqref{Vo-Vs}.\footnote{
More precisely, the matching to \eqref{CPLr} fixes $Z_s=Z_o=1$ up to order
$\als^2$ and $Z_s + (\nc^2-1)Z_o= \nc^2$ up to order $\als^3$.
}
We note that Eqs. (\ref{CPLr}) and (\ref{pNRQCDrverification}) are equivalent for $rT \gg g^2$,  
however, in Eq. \eqref{pNRQCDrverification}, we resum some contributions
that would become large for $r T \siml g^2$. Equation \eqref{pNRQCDrverification} is therefore valid also in that regime.
Finally, we observe that the combination of the two procedures provides a
non-trivial verification of Eq. \eqref{Vo-Vs}, i.e. of the two-loop difference between the octet 
and the singlet potentials, known, so far, only from the direct calculation 
of the two-loop octet potential in a covariant gauge, done in Ref.~\cite{Kniehl:2004rk}.

Loop corrections to the singlet and octet correlators in Eqs. \eqref{SSpNRQCD} and \eqref{OOpNRQCD} 
get contributions from the scales $T$, $m_D$ and lower ones. 
We now proceed to evaluate these corrections, separating the contributions 
of the temperature from the ones of the Debye mass.

\subsection{The temperature scale\label{secpNRQCDT}}
In the hierarchy \eqref{defhierarchy}, the next scale after the inverse
distance is the temperature. Our aim is thus to compute the
temperature contributions to loop corrections in pNRQCD. These loop
corrections are the terms $\delta_s$ and $\delta_o$ 
that were introduced in Eqs.~\eqref{SSpNRQCD} and \eqref{OOpNRQCD}. 
We call $\delta_{s,T}$ and $\delta_{o,T}$ the parts of $\delta_{s}$ and
$\delta_{o}$ respectively that encode the contributions coming from the scale $T$; they
may be obtained by expanding $\delta_{s}$ and $\delta_{o}$ in $m_D$,
$V_s$, $V_o$ and in any lower energy scale. Similarly, 
$\delta \langle L_R \rangle_T$ is the part of $\langle L_R \rangle$ that encodes 
the contributions coming from the scale $T$.
Different terms contribute to  $\delta_{s,T}$, $\delta_{o,T}$ and $\delta \langle L_R \rangle_T$; 
we examine them in the following.

\begin{figure}[ht]
\begin{center}
\includegraphics[width=7cm]{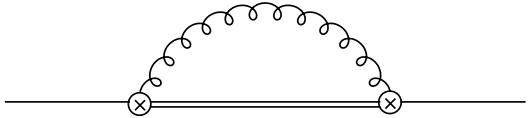}
\end{center}
\caption{The pNRQCD Feynman diagram giving the leading-order correction to $\delta_s$.
The single continuous line stands for a singlet propagator, 
the double line for an octet propagator, the circle with a cross for the 
chromoelectric dipole vertex proportional to $V_A$ in the Lagrangian \eqref{pNRQCD}
and the curly line connecting the two circles with a cross for a chromoelectric correlator.
}
\label{fig:singEEV}
\end{figure}

\begin{enumerate}
\item{\emph{The singlet $r^2$ contributions}} \\
We start considering the one-loop, order $r^2$ in the multipole expansion, correction 
to the singlet correlator induced by the diagram shown in Fig.~\ref{fig:singEEV}; it reads
\begin{eqnarray}
\delta_{s}^{\,{\cal O}(r^2)} &=& \left( ig\sqrt{\frac{1}{2\nc}} \right)^2 r^ir^j T\sum_n
\int \frac{d^dk}{(2\pi)^d} \int_0^{1/T} \!\! d\tau \int_0^{\tau} \!\! d\tau'
e^{\tau V_s} \, e^{-(\tau-\tau')V_o} \, e^{-\tau' V_s}
\nonumber\\
&& \hspace{4cm} \times e^{-i(\tau-\tau')\omega_n}\,
\langle E^{i\,a} U_{ab}E^{j\,b}\rangle(\omega_n,\bk).
\label{EEVa}
\end{eqnarray}
In the sum integral, we may distinguish between contributions coming from the non-zero modes and from the zero modes.

For the contribution coming from the non-zero modes, only 
the leading-order chromoelectric correlator in momentum space $\langle E^{i\,a}
U_{ab}E^{j\,b}\rangle(\omega_n,\bk)$ ($U_{ab}$ stands for a Wilson straight line in the adjoint representation
connecting $E^{i\,a}$ with $E^{j\,b}$; at leading order $U_{ab} = \delta_{ab}$) is relevant at our accuracy:
\begin{equation}
\langle E^{i\,a} U_{ab}E^{j\,b}\rangle(\omega_n,\bk) = 
(\nc^2-1)\left[ \frac{k^ik^j}{\bk^2} + (\delta_{ij}-\hat{k}^i\hat{k}^j)\frac{\omega_n^2}{\omega_n^2+\bk^2}\right].
\label{EEleading}
\end{equation}
Loop corrections to the chromoelectric correlator contribute to the Polyakov-loop correlator 
at order $g^6(rT)$ or smaller.
Because of the hierarchy \eqref{defhierarchy}, we  can  expand the
right-hand side of \eqref{EEVa} in $V_o-V_s$. The longitudinal part of
the chromoelectric correlator, i.e. the first term in square brackets,
vanishes in dimensional regularization, whereas the transverse part is
sensitive to the scale $T$ through the Matsubara frequencies. After
performing the sum integral over the non-zero modes, we obtain
\begin{eqnarray}
\delta_{s,T}^{\, {\cal O}(r^2)\, {\rm NS}} &=& 
- g^2C_F \frac{r^2T}{9}(V_o-V_s) + g^2C_F \frac{r^2}{36}(V_o-V_s)^2 +  {\cal  O}\left(g^6(rT),\frac{g^8}{rT}\right)
\nonumber\\
&=& - \frac{2}{9} \pi \nc C_F\als^2rT + \frac{\pi}{36}\nc^2 C_F\als^3 
+  {\cal  O}\left(g^6(rT),\frac{g^8}{rT}\right).
\label{deltasEa}
\end{eqnarray}

The contribution coming from the zero modes reads
\begin{eqnarray}
\delta_{s,T}^{\, {\cal O}(r^2)\, {\rm S}} &=& 
\left( ig\sqrt{\frac{1}{2\nc}} \right)^2  \frac{r^ir^j}{2T}
\int \frac{d^dk}{(2\pi)^d} \langle E^{i\,a} U_{ab}E^{j\,b}\rangle(0,\bk)\vert_{|\bk|\sim T}
 + {\cal  O}\left(g^6(rT)\right).
\label{EEVb}
\end{eqnarray}
Here, the first non-vanishing contribution in dimensional regularization
comes from the one-loop correction to the chromoelectric correlator.
The integral with $\langle E^{i\,a} $ $U_{ab}E^{j\,b}\rangle(0,\bk)$ at one loop 
has been calculated in \cite{Brambilla:2008cx}. Using that result we obtain\footnote{\label{footEE}
The chromoelectric correlator is gauge invariant.
In static gauge, thermal corrections arise from the non-static part of the spatial 
gluon propagator. Hence, at one loop, only gluon self-energy diagrams may provide 
thermal corrections; we have 
$$
\langle E^{i\,a} U_{ab}E^{j\,b}\rangle(0,\bk)\vert_{|\bk|\sim T} = 
\langle \partial_i A_0^{a} \, \partial_j A_0^{a}\rangle(0,\bk)\vert_{|\bk|\sim T} = 
(\nc^2-1)\frac{k^ik^j}{\bk^2+\Pi_{00}^{\rm NS}(\bk)_{\rm mat}},
$$
where $\Pi_{00}^{\rm NS}(\bk)_{\rm mat}$ is the matter part of the gluon self-energy's temporal component  
calculated in static gauge, which can be read from Eq. \eqref{pi00matter}.  
Finally, we recall that $\Pi_{00}^{\rm NS}(\bk)_{\rm mat}$ is the same in static gauge and in Coulomb gauge.
} 
 \begin{eqnarray}
\delta_{s,T}^{\, {\cal O}(r^2)\, {\rm S}} &=& 
\frac{3}{2}\zeta(3)C_F\frac{\als}{\pi}(rm_D)^2 
- \frac{2}{3}\zeta(3)\nc C_F \als^2(rT)^2 
 + {\cal  O}\left(g^6(rT)\right).
\label{deltasEb}
\end{eqnarray}

\item{\emph{Higher multipole terms}} \\
Our aim is to calculate in the EFT the Polyakov-loop correlator at order $g^6$, neglecting terms of order 
$g^6(rT)$ or smaller. Contributions coming from the $\delta {\cal L}_{\rm pNRQCD}$ part of the pNRQCD Lagrangian, 
which includes terms of order $r^3$ or smaller coming from the multipole expansion, share, at leading order,
the same colour structure and the same order in $\als$ as Eqs. \eqref{deltasEa} and \eqref{deltasEb} but are suppressed 
by powers of $rT$. We may write these contributions as
\begin{eqnarray}
\delta_{s,T}^{\,\delta {\cal L}_{\rm pNRQCD}} 
&=& \delta_{s,T}^{\, {\cal O}(r^2)\, {\rm NS}}
\sum_{n=0}^\infty c_n^{\rm NS}(rT)^{2n+2}  
+
\delta_{s,T}^{\, {\cal O}(r^2) \, {\rm S}}
\sum_{n=0}^\infty c_n^{\rm S}(rT)^{2n+2}  
+  {\cal  O}\left(g^6(rT)^3\right),
\nonumber\\
\label{Vsmult}
\end{eqnarray}
where the unknown coefficients $c_n^{\rm NS}$ and $c_n^{\rm S}$ are, as we will see, 
irrelevant for the purpose of calculating the Polyakov-loop correlator at order 
$g^6(rT)^0$.

\item{\emph{The octet contributions}} \\
As in the singlet case, one loop-corrections to the octet correlator may be divided into 
order $r^2$ non-zero mode contributions ($\delta_{o,T}^{\, {\cal O}(r^2)\, {\rm NS}}$), 
order $r^2$ zero-mode contributions ($\delta_{o,T}^{\, {\cal O}(r^2)\, {\rm S}}$), 
and higher multipole terms ($\delta_{o,T}^{\,\delta {\cal L}_{\rm pNRQCD}}$). 
It turns out that 
\begin{eqnarray}
\delta_{o,T}^{\, {\cal O}(r^2)\, {\rm NS}}   &=&  \delta_{s,T}^{\, {\cal O}(r^2)\, {\rm NS}}\vert_{V_s\leftrightarrow V_o},
\label{deltaoEa}
\end{eqnarray}
and, up to order $g^6(rT)^0$, 
\begin{eqnarray}
\delta_{o,T}^{\, {\cal O}(r^2)\, {\rm S}}   &=&  - \delta_{s,T}^{\, {\cal O}(r^2)\, {\rm S}},
\label{deltaoEb}\\
\delta_{o,T}^{\,\delta {\cal L}_{\rm pNRQCD}} &=& -\delta_{s,T}^{\,\delta {\cal L}_{\rm pNRQCD}}.
\label{Vomult}
\end{eqnarray}
These equalities are proved in appendix \ref{app_octet}.

\item{\emph{$\delta \langle L_R \rangle_T$}} \\
Finally, we need to calculate the contributions to the Polyakov loop coming from the scale $T$.
The order $g^4$ contribution may be read from Eq. \eqref{finalcontribtnlons}. 
Since we do not know the order $\crr \, g^6$ contribution, we write $\delta \langle L_R \rangle_T$ as 
\begin{equation}
\delta\langle L_R\rangle_{T}=
\frac{\crr\als^2}{2}\left[\ca\left(\frac{1}{2\epsilon}-\ln\frac{4T^2}{\mu^2}+1-\gamma_E
+\ln (4\pi) \right)- n_f\ln2 + a\,\als \right]
+ {\cal  O}\left(\als^4\right),
\label{LRT}
\end{equation}
where the explicit value of the coefficient $a$ does not matter.
Instead, what matters here is that this coefficient is common to all colour representations.
The first correction from the scale $T$ not of the type $\crr \, \als^n$ 
appears at order $\als^4$ and comes from diagram~b) in Fig.~\ref{fig:cubefourth} 
with two self-energy insertions, one in each temporal gluon.
Note that Eq.~\eqref{als2m2} provides the first correction not of the type $\crr \, \als^n$ 
coming from the scale $m_D$.
\end{enumerate}
In summary, we obtain the contribution of the scale $T$ to the singlet and octet correlators:
\begin{eqnarray}
&&
e^{-V_s(r)/T} \delta_{s,T} =
\nonumber\\
&&\hspace{1cm}
 e^{-V_s(r)/T}\Bigg\{  
- \frac{2}{9} \pi \nc C_F\als^2rT\left[1 + \sum_{n=0}^\infty c_n^{\rm NS}(rT)^{2n+2}\right]
+ \frac{\pi}{36}\nc^2 C_F\als^3 
\nonumber\\
&&\hspace{25mm}
+ \left(\frac{3}{2}\zeta(3)C_F\frac{\als}{\pi}(rm_D)^2 
- \frac{2}{3}\zeta(3)\nc C_F \als^2(rT)^2 \right)
\left[1 + \sum_{n=0}^\infty c_n^{\rm S}(rT)^{2n+2}\right]
\nonumber\\
&&\hspace{25mm}
+ {\cal  O}\left(g^6(rT),\frac{g^8}{rT}\right) 
\Bigg\},
\label{VsE}
\\
&&
 e^{-V_o(r)/T}\left[(\nc^2-1)\, \delta \langle L_A \rangle_T + \delta_{o,T}\right] =
\nonumber\\
&&\hspace{1cm}
(\nc^2-1)e^{-V_o(r)/T}\Bigg\{
\frac{\ca}{2}  \als^2 \left[\ca\left(\frac{1}{2\epsilon}-\ln\frac{4T^2}{\mu^2}+1-\gamma_E
+\ln (4\pi) \right)- n_f\ln2 + a\,\als \right]
\nonumber\\
&&\hspace{25mm}
+ \frac{1}{9} \pi \als^2rT\left[1 + \sum_{n=0}^\infty c_n^{\rm NS}(rT)^{2n+2}\right]
+ \frac{\pi}{72}\nc \als^3 
\nonumber\\
&&\hspace{25mm}
- \left(\frac{3}{4}\zeta(3)\frac{1}{\nc}\frac{\als}{\pi}(rm_D)^2 
- \frac{1}{3}\zeta(3) \als^2(rT)^2 \right)
\left[1 + \sum_{n=0}^\infty c_n^{\rm S}(rT)^{2n+2}\right]
\nonumber\\
&&\hspace{25mm}
+ {\cal  O}\left(g^6(rT),\frac{g^8}{rT}\right) 
\Bigg\}.
\label{VoE}
\end{eqnarray}

Inserting Eqs.~\eqref{LRT}-\eqref{VoE} into Eq.~\eqref{PLC-pNRQCD} and expanding, 
we obtain that the connected Polyakov-loop correlator is given by
\begin{eqnarray}
C_\mathrm{PL}(r,T) &=& C_\mathrm{PL}(r,T)_{1/r}
\nonumber\\
&&
-\frac{\pi}{18} C_F\als^3 + \frac{\nc^2-1}{8\nc^2}\frac{\als^3}{rT}
\bigg[
\ca\left( - \frac{1}{\epsilon}- 2\ln\frac{\mu^2}{4T^2} -2 +2\gamma_E
-2\ln (4\pi) \right) 
\nonumber\\
&&
\hspace{6.2cm}
+2 n_f\ln 2 \bigg]
+{\cal  O}\left(g^6(rT),\frac{g^8}{(rT)^4}\right)
\nonumber\\
&& + \;\hbox{loop corrections at the scale }m_D\hbox{ or lower}\,,
\label{CPLT}
\end{eqnarray}
where  $C_\mathrm{PL}(r,T)_{1/r}$ may be read from Eq. \eqref{CPLr}.
We observe that, in the connected Polyakov-loop correlator, terms proportional to the 
unknown coefficients  $c_n^{\rm NS}$, $c_n^{\rm S}$ and $a$ have canceled.
The thermal corrections in \eqref{CPLT} agree with those calculated 
in Sec.~\ref{sec_corr}; in particular, they correspond to the sum of the 
gluon self-energy matter-part contribution in Eq. \eqref{selfenergyinsertr} with 
Eq. \eqref{defselfenergyinsertt}. The result in Eq. \eqref{CPLT} has an infrared 
divergence that originates at the scale $T$. This divergence shall cancel against 
an opposite ultraviolet one at the scale $m_D$, which will be the subject of the next section.

\subsection{The Debye mass scale\label{secpNRQCDmD}}
Here we compute the contributions to the singlet correlator, 
the octet correlator and the Polyakov loop coming from loop momenta sensitive 
to the Debye mass scale. We call these contributions  
$\delta_{s,m_D}$, $\delta_{o,m_D}$ and $\delta \langle L_R\rangle_{m_D}$
respectively. They may be computed by evaluating the loop integrals in $\delta_{s}$, 
$\delta_{o}$ and $\delta \langle L_R\rangle$ over momenta of the order 
$m_D$ and expanding with respect to any other scale.
The Debye mass scale is the lowest scale we need to consider here; 
contributions coming from scales lower than $m_D$ are beyond our accuracy.
Different terms contribute to  $\delta_{s,m_D}$, $\delta_{o,m_D}$ and $\delta \langle L_R \rangle_{m_D}$; 
we examine them in the following.

\begin{enumerate}
\item{\emph{The singlet and octet contributions}} \\
The leading-order contribution to $\delta_{s,m_D}$ comes from the self-energy diagram 
shown in Fig.~\ref{fig:singEEV} when evaluated over loop momenta of order $m_D$.
The contribution reads
\begin{eqnarray}
\delta_{s,m_D} &=&
\left( ig\sqrt{\frac{1}{2\nc}} \right)^2 r^ir^j T\sum_n
\int \frac{d^dk}{(2\pi)^d} \int_0^{1/T} \!\! d\tau \int_0^{\tau} \!\! d\tau'
e^{\tau V_s} \, e^{-(\tau-\tau')V_o} \, e^{-\tau' V_s}
\nonumber\\
&& \hspace{4cm} \times e^{-i(\tau-\tau')\omega_n}\,
\langle E^{i\,a} U_{ab}E^{j\,b}\rangle(\omega_n,\bk)\vert_{\mbk\sim m_D}.
\label{defEEVsmE}
\end{eqnarray}
The chromoelectric correlator evaluated over the region 
$\mbk\sim m_D$ gives rise to scaleless momentum integrals 
unless for the temporal part of the zero mode, $n=0$, which is at leading order 
$\langle E^{i\,a} U_{ab}E^{j\,b}\rangle(0,\bk)\vert_{|\bk|\sim m_D} = (\nc^2-1)\,k^ik^j\,/(\bk^2+m_D^2)$.
We obtain
\begin{equation}
\delta_{s,m_D}=-g^2 \cf\frac{r^ir^j}{2T}\,\int \frac{d^dk}{(2\pi)^d} \frac{k^ik^j}{\bk^2+m_D^2}
\left[1 + {\cal  O}\left(\frac{g^2}{rT}\right)\right] =
- \cf \frac{\als}{6} r^2\frac{m_D^3}{T} + {\cal  O}\left(g^7(rT)\right).
\label{EEVsmE}
\end{equation}

The leading-order contribution to $\delta_{o,m_D}$ comes from the octet self-energy diagrams 
shown in Fig.~\ref{fig:octetEEV}, when evaluated over the region $\mbk\sim m_D$.
Also in this case, the only non-vanishing contribution comes from the zero mode
of the temporal gluon propagator, which is $1/(\bk^2+m_D^2)$ (see Eq.~\eqref{propscreened}). 
For the same argument developed in appendix \ref{app_octet}, we find that 
\begin{equation}
\delta_{o,m_D} = -\delta_{s,m_D}.
\label{EEVomE}
\end{equation}
Higher multipole terms are of order $\displaystyle \als r^2\frac{m_D^3}{T} (rm_D)^2 \sim g^7 (rT)^4$ or smaller 
and, therefore, beyond our accuracy.

\item{\emph{$\delta\langle L_{R} \rangle_{m_D}$}} \\
We need to calculate the contribution to the Polyakov loop coming from the scale $m_D$.
It may be read from Eqs. \eqref{loopbm0}, \eqref{finalcontribm} and \eqref{als2m2}.
Since we do not know the order $\crr \,g^5$ and $\crr \,g^6$ contributions, 
we write $\langle L_{R} \rangle_{m_D}$ as 
\begin{eqnarray}
\delta\langle L_{R}\rangle_{m_D} &=& 
\frac{\crr\als}{2}\frac{m_D}{T} 
\nonumber\\ 
&& 
+ \frac{\crr\als^2}{2}
\left[\ca\left(- \frac{1}{2\epsilon}-\ln\frac{\mu^2}{4m_D^2}
-\frac{1}{2}+\gamma_E - \ln (4\pi) \right) + b_1\,g + b_2\,g^2 \right]
\nonumber\\
&& 
+ \left(3\crr^2-\frac{\crr\ca}{2}\right)\frac{\als^2}{24}\left(\frac{m_D}{T}\right)^2
+ {\cal  O}\left(g^7\right),
\label{LRmE}
\end{eqnarray}
where the explicit values of the coefficients $b_1$  and $b_2$ do not matter.
Instead, what matters here is that these coefficients are common to all colour representations.
\end{enumerate}
In summary, we obtain the contribution of the scale $m_D$ to the singlet and octet correlators:
\begin{eqnarray}
&& e^{-V_s(r)/T} \delta_{s,m_D} = 
e^{-V_s(r)/T} \Bigg\{
- \cf \frac{\als}{6} r^2 \frac{m_D^3}{T} 
+   {\cal  O}\left(g^7(rT)\right) \Bigg\},
\label{VsM}
\\
&& 
e^{-V_o(r)/T}\left[(\nc^2-1)\, \delta \langle L_A \rangle_{m_D} + \delta_{o,m_D}\right] =
(\nc^2-1)e^{-V_o(r)/T}\Bigg\{
\frac{\ca\als}{2}\frac{m_D}{T} 
+ \frac{5}{48} \ca^2\als^2 \left(\frac{m_D}{T}\right)^2
\nonumber\\
&&
\hspace{25mm}
+ \frac{\ca\als^2}{2}
\left[\ca\left(- \frac{1}{2\epsilon}-\ln\frac{\mu^2}{4m_D^2}
-\frac{1}{2}+\gamma_E - \ln (4\pi)\right) 
+ b_1\,g + b_2\,g^2 \right]
\nonumber\\
&& 
\hspace{25mm}
+ \frac{1}{\nc} \frac{\als}{12} r^2 \frac{m_D^3}{T} 
+  {\cal  O}\left(g^7\right) \Bigg\}.
\label{VoM}
\end{eqnarray}

Inserting Eqs.~\eqref{LRmE}-\eqref{VoM} into Eq.~\eqref{CPLT} and expanding,\footnote{
In terms of $\delta_{s,T}$, $\delta_{s,m_D}$, $\delta_{o,T}$, $\delta_{o,m_D}$, 
$\delta \langle L_F \rangle_T$, $\delta \langle L_F \rangle_{m_D}$, 
$\delta \langle L_A \rangle_T$ and $\delta \langle L_A \rangle_{m_D}$,  
$C_\mathrm{PL}(r,T)$ reads 
\begin{eqnarray*}
C_\mathrm{PL}(r,T) &=& 
\frac{1}{\nc^2}\left\{ e^{-V_s(r)/T}\left(1+\delta_{s,T}+\delta_{s,m_D}\right) \right.
\\
&& \hspace{5mm} 
\left.
+ e^{-V_o(r)/T} \left[ (\nc^2-1)\left(1 + \delta \langle L_A \rangle_T + \delta \langle L_A \rangle_{m_D}\right)
+\delta_{o,T}+\delta_{o,m_D} \right]
\right\} 
\\
&& - \left(1 + \delta \langle L_F \rangle_T + \delta \langle L_F \rangle_{m_D}\right)^2.
\end{eqnarray*}
} 
we obtain that the connected Polyakov-loop correlator is given by
\begin{eqnarray}
C_\mathrm{PL}(r,T) &=& C_\mathrm{PL}(r,T)_{1/r}
\nonumber\\
&&
-\frac{C_F}{18}\pi \als^3 + \frac{\nc^2-1}{8\nc^2} \als^2 \left(\frac{m_D}{T}\right)^2
\nonumber\\
&&
+ \frac{\nc^2-1}{\nc^2}\frac{\als}{rT}\left\{
-\frac{\als}{4}\frac{m_D}{T} - \frac{\als^2}{4}\left[
\ca\left( - \ln\frac{T^2}{m_D^2} +\frac{1}{2}\right) 
- n_f\ln 2 \right]\right\}
\nonumber\\
&&
+{\cal  O}\left(g^6(rT),\frac{g^7}{(rT)^2}\right),
\label{CPLmE}
\end{eqnarray}
where  $C_\mathrm{PL}(r,T)_{1/r}$ may be read from Eq. \eqref{CPLr}.
We observe that, in the Polyakov-loop correlator, terms proportional to the 
unknown coefficients $b_1$ and $b_2$, as well as the divergences, have canceled.
The origin of the thermal corrections to the Polyakov-loop correlator 
in the situation $1/r \gg T \gg m_D \gg g^2/r$ is clear. 
The term $-C_F\pi \als^3/18$ arises from the dipole interaction contributions and from 
their interference with the zero-temperature potentials.
The other thermal corrections arise from the interference of the 
adjoint Polyakov loop with the zero-temperature potentials.

The result coincides with Eq.~\eqref{finalcpltot}, obtained in
Sec.~\ref{sec_corr} after a direct calculation.
The differences in the way the two results were achieved illustrate 
well the typical differences between a direct computation and a computation 
in an EFT framework. In the EFT framework, some more conceptual work was
necessary in order to identify the relevant contributions.
Once this was done, we could take advantage of previously done calculations 
(in particular for $V_s(r)$ and $V_o(r)$) and reduce the calculation to essentially 
one diagram, shown in Fig.~\ref{fig:singEEV}, evaluated in different momentum regions. 
In the EFT framework, we could also gain some new insight by reconstructing 
the spectral decomposition of the Polyakov-loop correlator and 
by providing two new quantities: the colour-singlet and the colour-octet quark-antiquark correlators.

\subsection{Singlet and octet free energies}
Potential NRQCD at finite temperature allows to define 
a colour-singlet correlator, $\langle S(\br,{\bf 0},1/T)S^\dagger(\br,{\bf 0},0)\rangle$,  
and a colour-octet correlator, 
$\langle O^a(\br,{\bf 0},1/T)O^{a \, \dagger}(\br,{\bf 0},0)\rangle$, which 
are both gauge-invariant quantities.
We may associate to them 
a  \emph{colour-singlet free energy}, 
$f_{s}(r,T,m_D)$, and a \emph{colour-octet free energy}, 
$f_{o}(r,T,m_D)$, such that 
\begin{eqnarray}
\langle S(\br,{\bf 0},1/T)S^\dagger(\br,{\bf 0},0)\rangle 
&=& e^{-V_s(r)/T}\left(1+\delta_{s,T}+\delta_{s,m_D}\right) 
\nonumber\\
&\equiv&e^{-f_{s}(r,T,m_D)/T},\\
\langle O^a(\br,{\bf 0},1/T)O^{a \, \dagger}(\br,{\bf 0},0)\rangle 
&=& e^{-V_o(r)/T} \left[(\nc^2-1)\left(1 + \delta \langle L_A \rangle_T + \delta \langle L_A \rangle_{m_D}\right)
+\delta_{o,T}+\delta_{o,m_D}\right]
\nonumber\\
&\equiv&(\nc^2-1)e^{-f_{o}(r,T,m_D)/T}.
\end{eqnarray}
Using the results of the previous sections, we have that 
\begin{eqnarray}
f_{s}(r,T,m_D) 
&=& 
V_s(r) 
\nonumber\\
&&
+ \frac{2}{9} \pi \nc C_F\als^2rT^2\left[1 + \sum_{n=0}^\infty c_n^{\rm NS}(rT)^{2n+2}\right]
- \frac{\pi}{36}\nc^2 C_F\als^3T 
\nonumber\\
&&
- \left(\frac{3}{2}\zeta(3)C_F\frac{\als}{\pi}(rm_D)^2T 
- \frac{2}{3}\zeta(3)\nc C_F \als^2r^2T^3 \right)
\left[1 + \sum_{n=0}^\infty c_n^{\rm S}(rT)^{2n+2}\right]
\nonumber\\
&&
+ \cf \frac{\als}{6} r^2 m_D^3 
+ T {\cal  O}\left(g^6(rT),\frac{g^8}{rT}\right),
\label{freesinglet}
\end{eqnarray}
and
\begin{eqnarray}
f_{o}(r,T,m_D) 
&=&
V_o(r) 
\nonumber\\
&&
- \frac{\ca\als}{2}m_D 
+ \frac{1}{48} \ca^2\als^2 \frac{m_D^2}{T}
\nonumber\\
&&
-\frac{\ca\als^2}{2}T\left[\ca\left(-\ln\frac{T^2}{m_D^2}+\frac{1}{2}\right)-
  n_f\ln2 + b_1\,g + b_2\,g^2 + a\,\als \right]
\nonumber\\
&&
- \frac{\pi}{9}\als^2rT^2\left[1 + \sum_{n=0}^\infty c_n^{\rm NS}(rT)^{2n+2}\right]
- \frac{\pi}{72}\nc \als^3T 
\nonumber\\
&&
+ \left(\frac{3}{4\nc}\zeta(3)\frac{\als}{\pi}(rm_D)^2T - \frac{1}{3}\zeta(3) \als^2r^2T^3 \right)
\left[1 + \sum_{n=0}^\infty c_n^{\rm S}(rT)^{2n+2}\right]
\nonumber\\
&&
- \frac{1}{\nc} \frac{\als}{12} r^2 m_D^3 
+ T {\cal  O}\left(g^6(rT),\frac{g^8}{rT}\right).
\label{freeoctet}
\end{eqnarray}
We note that $f_{s}(r,T,m_D)$ and $f_{o}(r,T,m_D)$ are both finite and gauge invariant.
They also do not depend on some special choice of Wilson lines connecting 
the initial and final quark and antiquark states.

In \cite{Brambilla:2008cx}, the colour-singlet quark-antiquark potential was calculated 
in real-time formalism in the same thermodynamical situation considered 
here and specified by Eq.~\eqref{defhierarchy}. The result may be found 
in Eq.~(92) of Ref.~\cite{Brambilla:2008cx}. 
Comparing terms of the same order, the real part of the real-time potential 
differs from $f_{s}(r,T,m_D)$ by  
$\displaystyle \frac{1}{9} \pi \nc C_F\als^2rT^2 - \frac{\pi}{36}\nc^2 C_F\als^3T$.
The origin of the difference may be traced back to terms in Eq.~\eqref{EEVa} that 
would vanish for large real times. Indeed, performing the calculation of 
$\langle S(\br,{\bf 0},\tau)S^\dagger(\br,{\bf 0},0)\rangle$ for an imaginary time $\tau \le 1/T$,  
along the lines of Secs.~\ref{secpNRQCDT} and \ref{secpNRQCDmD}, 
and then continuing analytically $\tau$ to large real times, one gets 
back exactly both the real and the imaginary parts of the real-time colour-singlet potential 
derived in \cite{Brambilla:2008cx}. 
The difference between the singlet free energy and the real part of the real-time 
colour-singlet potential appears to be a relevant finding to be considered when using free-energy 
lattice data for the quarkonium in media phenomenology.

\subsection{Comparison with the literature} 
An EFT approach for the calculation of the correlator of Polyakov loops 
was developed in \cite{Braaten:1994qx} for the situation $m_D  \simg 1/r$ and 
in \cite{Nadkarni:1986cz} for $T \gg 1/r$. In neither of the two cases, the
scale $1/r$ was integrated out: the Polyakov-loop correlator was
described in terms of dimensionally reduced effective field theories of QCD, 
while the complexity of the bound-state dynamics remained implicit in the correlator. 
The description developed in  \cite{Nadkarni:1986cz,Braaten:1994qx} is valid for 
largely separated Polyakov loops. Under that condition, the correlator turns out 
to be screened either by the Debye mass, for $rm_D \sim 1$, or by the mass of the
lowest-lying glueball, for $rm_D \gg 1$.

In \cite{Jahn:2004qr}, the spectral decomposition of the Polyakov-loop correlator was analyzed. 
It was concluded that the quark-antiquark component of an allowed intermediate state, 
i.e. a field $\varphi$ describing a quark located in $\bx_1$  and an antiquark located in $\bx_2$,
should transform as $\varphi(\bx_1,\bx_2) \to g(\bx_1) \varphi(\bx_1,\bx_2) g^\dagger(\bx_2)$
under a gauge transformation $g$. 
Equation \eqref{PLC-pNRQCD}  is in accordance with that result for, 
in pNRQCD, both the singlet field S and the octet field O transform in that way  \cite{Pineda:1997bj}. 
We remark, however, a difference in language:
in our work, singlet and octet refer to the gauge transformation properties of 
the quark-antiquark fields, while, in \cite{Jahn:2004qr}, they refer to the gauge 
transformation properties of the physical states.

In \cite{Burnier:2009bk}, a weak-coupling calculation of the untraced
Polyakov-loop correlator in Coulomb gauge and of the cyclic Wilson loop was
performed up to order $g^4$. Each of these objects contributes to
the correlator of two Polyakov loops through a Fierz transformation that also generates some 
octet counterparts. It is expected that large cancellations occur between 
those correlators and their octet counterparts in
order to reproduce the Polyakov-loop correlator given in Eq. \eqref{finalcpltot}.
Such large cancellations should occur at the level of the scales $1/r$, 
$T$ and $m_D$ as we have already experienced in this work.
Note that in the case of the untraced Polyakov-loop correlator, 
the octet contribution shall also restore gauge invariance.

\section{Summary and outlook\label{sec_concl}}
In the weak-coupling regime, we have calculated the Polyakov loop up to order
$g^4$ and the correlator of two Polyakov loops up to order $g^6(rT)^0$, 
assuming the hierarchy of scales $\displaystyle \frac{1}{r}\gg T\gg m_D\gg \frac{g^2}{r}$.

The Polyakov-loop calculation differs from the result of Gava and Jengo
\cite{Gava:1981qd} by a finite contribution at order $g^4$. We have analyzed
in detail the origin of the difference and shown in an appendix that our
result may be reproduced also performing the calculation in Feynman gauge.
Our calculation agrees with the recent finding of Ref.~\cite{Burnier:2009bk}.

The calculation of the Polyakov-loop correlator is new in the considered
regime, although some partial results may be deduced from a previous work of
Nadkarni, who studied distances $r \sim 1/m_D$ \cite{Nadkarni:1986cz}. 
We have performed the calculation in two different approaches:  
by a direct computation in static gauge and by calculating the Polyakov-loop correlator 
in a suitable EFT that exploits the hierarchy of scales in the problem. In this second
approach, we have used pNRQCD at finite temperature and subsequently integrated 
out lower momentum regions. The advantages of this second approach are that the 
calculations do not rely on any specific choice of gauge 
and the systematics is clearer. Moreover, it makes explicit the quark-antiquark colour-singlet and 
colour-octet contributions to the Polyakov-loop correlator.
In particular, we have shown that at leading order in the multipole expansion
the Polyakov-loop correlator can be written as the colour average of a colour-singlet 
correlator, which defines a gauge-invariant colour-singlet free energy, and a 
colour-octet correlator, which defines a gauge-invariant colour-octet free energy.
This is in line with some early intuitive arguments given in 
\cite{McLerran:1981pb,Gross:1980br,Nadkarni:1986cz}. 
In general, however, such a decomposition does not hold.

In the weak-coupling regime, the degrees of freedom of pNRQCD are quark-antiquark 
colour-singlet fields, quark-antiquark colour-octet fields, gluons and light quarks.
The obtained result for the Polyakov-loop correlator 
is consistent with its spectral decomposition.
In the strong-coupling regime, the degrees of freedom are expected to change 
when the typical energy of the bound state is smaller than the 
confinement scale $\Lambda_{\rm QCD}$. In that situation, the bound state 
would become sensitive to confinement and give rise to a new 
spectrum of gluonic excitations (hybrids, glueballs). 
In the present work, we have not discussed this situation, which 
surely deserves investigation.

Possible further extensions of this work also include the study of the
Polyakov-loop correlator in different scale hierarchies, in particular at
temperatures of the same order as or higher than $1/r$, where the present
analysis should smoothly go over the ones performed in \cite{Nadkarni:1986cz,Braaten:1994qx}. 
As mentioned above, also analyses that involve the strong-coupling scale
should be addressed.

Finally, the present study should be completed by the study of correlators 
different from the Polyakov-loop one. Among these, the most studied in lattice 
gauge theories are the untraced Polyakov-loop correlator and the cyclic Wilson
loop. Also the octet Wilson loop should be included for its role in the
Polyakov-loop correlator. Since some partial perturbative results are already available for some 
of these correlators, it would be interesting to see how they can be reproduced 
in the EFT framework introduced here and how they combine to give back the Polyakov-loop 
correlator.

\begin{acknowledgments}
We thank Mikko Laine for correspondence and the authors of Ref.~\cite{Burnier:2009bk} 
for acknowledging some of the results presented here prior to publication. 
N.B., J.G. and A.V. thank Owe Philipsen for discussions.
A part of this work was done at the Kavli Institute for Theoretical Physics
China (KITPC), CAS, Beijing. P.P. and J.G. thank the KITPC for hospitality and support.
Part of this work was also carried out during the
CATHIE-INT mini program ``Quarkonia in hot matter: from QCD to experiment''
held at the Institute for Nuclear Theory (INT).
N.B., P.P. and A.V. thank the Institute for Nuclear Theory at the University of Washington for its hospitality and the Department of Energy for partial support.
J.G. thanks for hospitality Brookhaven National Laboratory, 
where this work was started. 
The work of P.P. was supported by U.S. Department of Energy under
Contract No.~DE-AC02-98CH10886. 
N.B., J.G. and A.V. acknowledge financial support from the RTN Flavianet MRTN-CT-2006-035482 (EU)
and from the DFG cluster of excellence ``Origin and structure of the universe''
(\href{http://www.universe-cluster.de}{www.universe-cluster.de}).
\end{acknowledgments}

\appendix

\section{Feynman rules in the static gauge\label{app_rules}}
\begin{fmffile}{foo}
\setlength{\unitlength}{1mm}
In the following, we list the Feynman rules in Euclidean space-time  
under the gauge condition $\partial_0A^0=0$.
The temporal propagator reads (dropping colour indices)  
\begin{equation}
\label{temporal}
D_{00}(\omega_n,\bk)=\quad\parbox{30mm}{
\begin{fmfchar*}(30,10)
\fmfleft{in}
\fmfright{out}
\fmf{dashes}{in,out}
\end{fmfchar*}}\;=\; \frac{\delta_{n0}}{\bk^2},
\end{equation}
where, as usual, $\omega_n=2\pi nT$ and the Kronecker delta fixes
$n=0$, making this propagator purely static. The spatial propagator
can be divided into a non-static ($n\ne0$) and a static ($n=0$)
part. The former reads
\begin{equation}
\label{nonstatic}
D_{ij}(\omega_n\ne0,\bk)=\quad\parbox{30mm}{
\begin{fmfchar*}(30,10)
\fmfleft{in}
\fmfright{out}
\fmf{curly}{in,out}
\end{fmfchar*}}\;=\; 
\frac{1}{\omega_n^2+\bk^2}\left(\delta_{ij}+\frac{k_ik_j}{\omega_n^2}\right)(1-\delta_{n0}),
\end{equation}
and thus mixes longitudinal and transverse components. The static part
has a residual gauge dependence on the parameter $\xi$; it reads
\begin{equation}
\label{magnetostatic}
D_{ij}(\omega_n=0,\bk)=\quad\parbox{30mm}{
\begin{fmfchar*}(30,10)
\fmfleft{in}
\fmfright{out}
\fmf{photon}{in,out}
\end{fmfchar*}}\;=\; \frac{1}{\bk^2}\left(\delta_{ij}-(1-\xi)\frac{k_ik_j}{\bk^2}\right)\delta_{n0}.
\end{equation}
Finally the ghost propagator reads
\begin{equation}
\label{ghost}
D_{\mathrm{ghost}}(\omega_n,\bk)=\quad\parbox{30mm}{
\begin{fmfchar*}(30,10)
\fmfleft{in}
\fmfright{out}
\fmf{ghost}{in,out}
\end{fmfchar*}}\;=\; \frac{\delta_{n0}}{\bk^2},
\end{equation}
and is thus purely static\footnote{The non-static ghost can be shown
to decouple \cite{D'Hoker:1981us}.}. The interaction vertices
(gluon-gluon and gluon-ghost) are the usual ones.
\end{fmffile}

\section{The gluon self energy in the static gauge\label{app_pi00}}
We proceed to the computation of the Matsubara sums in
Eq. \eqref{defmaster} in order to obtain Eqs. \eqref{pi00vacuum},
\eqref{pi00matter}, \eqref{pi00zero} and \eqref{pi00sing}. 
We recall the two basic bosonic Matsubara sums \cite{Kapusta:2006pm}
\begin{eqnarray}
T\sum_{n=-\infty}^{+\infty}\frac{1}{\bp^2+\omega_n^2}&=&\frac{1+2n_\mathrm{B}(\mbp)}{2\mbp},
\label{matsup2}
\\
T\sum_{n=-\infty}^{+\infty}\frac{1}{(\bp^2+\omega_n^2)(\bq^2+\omega_n^2)}&=&\frac{1}{2
  \mbp \mbq}\left(	
\frac{1+n_{\mathrm{B}}(\mbp)+n_{\mathrm{B}}(\mbq)}{\mbp+\mbq}
+\frac{n_{\mathrm{B}}(\mbq)-n_{\mathrm{B}}(\mbp)}{\mbp-\mbq}\right),
\nonumber\\
\label{matsup2q2}
\end{eqnarray}
where $n_\mathrm{B}$ is the Bose--Einstein distribution. 
Since the sums include also the zero mode, in 
evaluating the master sum integrals defined in
Eqs. \eqref{defmaster} and \eqref{shorthand} 
we will have to subtract it. 
Furthermore, we identify the temperature-independent part (the unity) in the numerators
on the r.h.s of Eqs. \eqref{matsup2} and \eqref{matsup2q2}
as the vacuum part and the part proportional to the thermal distributions 
as the matter part.

For $I_0$, we have
\begin{equation}
\label{sumi0}
I_0=\int_p^\prime\frac{1}{p^2}=\m2 \int\frac{d^dp}{(2\pi)^d}\left(\frac{1+2n_\mathrm{B}(\mbp)}{2\mbp}-\frac{T}{\bp^2}\right)
=\frac{T^2}{12};
\end{equation}
the subtracted zero mode along with the vacuum part vanish in dimensional regularization. 

For $I_1$, we have ($\bq = \bk -\bp$)
\begin{eqnarray*}
I_1&=&\m2 \int\frac{d^dp}{(2\pi)^d}\left[\frac{\mbp}{2\mbq}\left(	
\frac{1+n_{\mathrm{B}}(\mbp)+n_{\mathrm{B}}(\mbq)}{\mbp+\mbq}
+\frac{n_{\mathrm{B}}(\mbq)-n_{\mathrm{B}}(\mbp)}{\mbp-\mbq}\right)-\frac{T}{\bq^2}\right]
\\
&=&\m2 \int\frac{d^dp}{(2\pi)^d}\left[\frac{\bp^2}{2\mbp\mbq(\mbp+\mbq)}
+\frac{\mbp n_{\mathrm{B}}(\mbp)}{2  \mbq}\left(\frac{-2\mbq}{\bp^2-\bq^2}\right)
\right.
\\
&&
\hspace{5.6cm}
\left. +\frac{|\bq'|n_{\mathrm{B}}(\mbp)}{2 \mbp }\left(\frac{-2|\bq'|}{\bp^2-\bq'^2}\right)-\frac{T}{\bq^2}\right],
\end{eqnarray*}
where we have operated a shift $\bp\to \bq'= \bp + \bk$, $\bq \to -\bp$ in some terms of the matter part. 
The vacuum part can be brought into a more standard form by noting that
\begin{equation}
\label{3to4}
\int_{-\infty}^{+\infty}\frac{dp_0}{2\pi}\frac{1}{(\bp^2+p_0^2)(\bq^2+p_0^2)}=\frac{1}{2\mbp\mbq(\mbp+\mbq)}.
\end{equation}
This allows to write the three-dimensional integral as a standard Euclidean four-dimensional integral,  
which can be computed with the formulas listed in appendix \ref{sub_oneloop} setting  
$d+1=4-2\epsilon$. We thus have
\begin{equation}
\label{i1vacuum}
(I_1)_\mathrm{vac}=
\m2 \int\frac{d^{d+1}p}{(2\pi)^{d+1}}\frac{p^\mu p^\nu (\delta_{\mu\nu}-\delta_{\mu0}\delta_{\nu0})}{p^2q^2}
=(\delta_{\mu\nu}-\delta_{\mu0}\delta_{\nu0})\m2 L_{d+1}^{\mu\nu}(k,1,1)\vert_{k^0=0}.
\end{equation}
The zero-mode integral vanishes in dimensional regularization, 
whereas the remaining matter part is finite and gives 
\begin{eqnarray}
(I_1)_{\mathrm{mat}}
&=&\frac{1}{2\pi^2}\int_0^\infty d\mbp\,\mbp n_{\mathrm{B}}(\mbp)\left(1+\frac{\mbp}{2\mbk}
\ln\left\vert\frac{\mbk+2\mbp}{\mbk-2\mbp}\right\vert\right).
\label{i1matter}
\end{eqnarray}

Analogously, we have for $I_2$
\begin{equation*}
I_2=\bk^2\m2 \int\frac{d^dp}{(2\pi)^d}\left[\frac{1}{2\mbp\mbq}\left(	
\frac{1+n_{\mathrm{B}}(\mbp)+n_{\mathrm{B}}(\mbq)}{\mbp+\mbq}
+\frac{n_{\mathrm{B}}(\mbq)-n_{\mathrm{B}}(\mbp)}{\mbp-\mbq}\right)-\frac{T}{\bp^2\bq^2}\right].
\end{equation*}
The vacuum part is 
\begin{equation}
(I_2)_\mathrm{vac}=\m2 \int\frac{d^{d+1}p}{(2\pi)^{d+1}}\frac{\bk^2}{p^2q^2}=\bk^2 \m2 L_{d+1}(k,1,1)\vert_{k^0=0},
\label{i2vacuum}
\end{equation}
the matter part is
\begin{equation}
(I_2)_\mathrm{mat}=
\frac{1}{2\pi^2}\left(\int_0^\infty d\mbp\, n_{\mathrm{B}}(\mbp)
\frac{\mbk}{2}\ln\left\vert\frac{\mbk+2\mbp}{\mbk-2\mbp}\right\vert\right), 
\label{i2mat}
\end{equation}
and the subtracted zero-mode part is 
\begin{equation}
(I_2)_\mathrm{zero}= - \m2 \int\frac{d^dp}{(2\pi)^d}\frac{T\bk^2}{\bp^2\bq^2},
\label{i2zero}
\end{equation}
which has been kept in dimensional regularization.

We consider now $I_3$:
\begin{eqnarray}
I_3&=&\bk^2\m2 \int\frac{d^dp}{(2\pi)^d}\left[\frac{1+2n_{\mathrm{B}}(\mbp)}{2\mbp^3}-\frac{T}{\bp^4}\right],
\\
\label{i3vac}
(I_3)_\mathrm{vac}&=&\bk^2\m2 \int\frac{d^dp}{(2\pi)^d}\frac{1}{2\mbp^3}=0,
\\
\label{i3mat}
(I_3)_\mathrm{mat}&=& \frac{1}{2\pi^2}\int_0^\infty d\mbp\,\mbp n_{\mathrm{B}}(\mbp)\frac{\bk^2}{\bp^2}.
\end{eqnarray}
In dimensional regularization the subtracted zero mode vanishes. 
The matter part is infrared divergent.
Since this divergence will cancel against terms from $I_4$ in the sum \eqref{defp00}, 
we present the result directly in the three-dimensional limit.

$I_4$ is given by
\begin{equation}
I_4
=I_4^a-I_4^b-I_4^c
=\int_p^\prime\frac{\bk^4}{\bp^2\bq^2\omega_n^2}-\int_p^\prime\frac{\bk^4}{p^2q^2\bp^2}-\int_p^\prime\frac{\bk^4}{\bp^2\bq^2q^2}.
\label{i4rewrite}
\end{equation}
$I_4^a$ is
\begin{equation}
\label{i4a}
I_4^a=\frac{2T}{(2\pi T)^2}\frac{\bk^4}{8\mbk}\sum_{n=1}^\infty\frac{1}{n^2}=\frac{\mbk^3}{96T},
\end{equation}
which is a term peculiar to this gauge; it is singular in the $T\to0$ limit and
constitutes $\Pi_{00}^\mathrm{NS}(\bk)_{\mathrm{sing}}$. 
$I_4^b$ is
\begin{equation}
\label{i4b}
I_4^b=\bk^4\m2 \int\frac{d^dp}{(2\pi)^d}\left[\frac{1}{2\mbp^3\mbq}\left(
\frac{1+n_{\mathrm{B}}(\mbp)+n_{\mathrm{B}}(\mbq)}{\mbp+\mbq}+\frac{n_{\mathrm{B}}(\mbq)
-n_{\mathrm{B}}(\mbp)}{\mbp-\mbq}\right)-\frac{T}{\bp^4\bq^2}\right].
\end{equation}
The vacuum part can be brought into a more familiar form by adding and subtracting $1/(2\mbp^3\bq^2)$
\begin{eqnarray}
(I_4^b)_{\mathrm{vac}}&=&
\bk^4\m2 \int\frac{d^dp}{(2\pi)^d}\left[\frac{1}{2\mbp^3\mbq(\mbp+\mbq)}
-\frac{1}{2\mbp^3\mbq^2}\right]+\bk^4\m2 \int\frac{d^dp}{(2\pi)^d}\frac{1}{2\mbp^3\mbq^2}
\nonumber\\
&=&-\bk^2\frac{\ln2}{2\pi^2}+\frac{\bk^4}{2} \m2 L_d(\bk,3/2,1).
\label{i4bvac}
\end{eqnarray}
Although the matter part of $I_4$ is infrared divergent, its infrared divergence cancels against 
the matter part of $I_3$, i.e. Eq. \eqref{i3mat},  in the sum \eqref{defp00}.
Hence, we may evaluate it directly in three dimensions. In contrast, we will keep regularized 
the subtracted zero modes. As discussed in the main text, 
these subtracted zero modes behave like $\epsilon \mbk^{1-2\epsilon}$ and are going to contribute 
when evaluating the Fourier transform of $\mbk^{1-2\epsilon}/\mbk^4$ in the Polyakov-loop 
correlator calculation, like in Eq. \eqref{defselfenergyinsertr}.
Therefore, $(I_4^b)_{\mathrm{mat}}$ and $(I_4^b)_{\mathrm{zero}}$ read
\begin{eqnarray}
(I_4^b)_{\mathrm{mat}}&=&
\frac{1}{2\pi^2}\int_0^\infty d\mbp\,\mbp n_{\mathrm{B}}(\mbp)
\frac{\mbk^3 }{2\mbp^3}\left[\ln\left\vert\frac{\mbk+2\mbp}{\mbk-2\mbp}\right\vert 
+\ln\left\vert\frac{\mbk-\mbp}{\mbk+\mbp}\right\vert\right],
\label{i4bmat}\\
(I_4^b)_{\mathrm{zero}}&=&
-\m2 \int\frac{d^dp}{(2\pi)^d}\frac{T\bk^4}{\bp^4\bq^2}.
\label{i4bzero}
\end{eqnarray}
Similarly $I_4^c$ reads
\begin{eqnarray}
I_4^c&=&\bk^4\m2 \int\frac{d^dp}{(2\pi)^d}\left[\frac{1+2n_{\mathrm{B}}(\mbq)}{2\bp^2\mbq^3}-\frac{T}{\bp^2\bq^4}\right],
\label{i4c}\\
(I_4^c)_{\mathrm{vac}}&=&\bk^4\m2 \int\frac{d^dp}{(2\pi)^d}\frac{1}{2\mbp^2\mbq^3}=\frac{\bk^4}{2} \m2 L_d(\bk,1,3/2),
\label{i4cvac}\\
(I_4^c)_{\mathrm{mat}}&=&
\frac{1}{2\pi^2}\int_0^\infty d\mbp\,\mbp n_{\mathrm{B}}(\mbp)\frac{\mbk^3}{2\mbp^3}
\ln\left\vert\frac{\mbk+\mbp}{\mbk-\mbp}\right\vert, 
\label{i4cmat}
\\
(I_4^c)_{\mathrm{zero}}&=& -\m2 \int\frac{d^dp}{(2\pi)^d}\frac{T\bk^4}{\bp^2\bq^4}.
\label{i4zero}
\end{eqnarray}
Notice that, as we anticipated, 
the sum $(I_3)_\mathrm{mat}/2-(I_4^b)_\mathrm{mat}/4-(I_4^c)_\mathrm{mat}/4$, 
which is the combination appearing in $\Pi_{00}^{\mathrm{NS}}(\bk)$, is infrared finite.
It is also worthwhile noticing that the vacuum parts $(I_4^b)_{\mathrm{vac}}$ and 
$(I_4^c)_{\mathrm{vac}}$ are infrared divergent, but that in the sum 
$(I_3)_\mathrm{vac}/2-(I_4^b)_\mathrm{vac}/4-(I_4^c)_\mathrm{vac}/4$,
these infrared divergences are canceled and replaced by an ultraviolet 
divergence eventually removed by renormalization. The canceling infrared divergence and the 
remaining ultraviolet one come from $(I_3)_\mathrm{vac}$, 
which vanishes, like in Eq. \eqref{i3vac}, if the two are set equal, 
as usually done in dimensional regularization.

Putting all pieces together in Eq. \eqref{defp00} and using  
\begin{eqnarray*}
\m2 \int\frac{d^dp}{(2\pi)^d}\frac{\bk^4}{\bp^4\bq^2} &=& 
\mbk^{1-2\epsilon}\m2 (4\pi)^{-3/2+\epsilon}
\frac{\Gamma(3/2+\epsilon)\Gamma(1/2-\epsilon)\Gamma(-1/2-\epsilon)}{\Gamma(-2\epsilon)}
\\
&=& \epsilon\frac{\mbk^{1-2\epsilon} \m2 }{4}\left[1 + {\cal O}(\epsilon)\right],
\\
\m2 \int\frac{d^dp}{(2\pi)^d}\frac{\bk^2}{\bp^2\bq^2} &=& 
\mbk^{1-2\epsilon}\m2 (4\pi)^{-3/2+\epsilon}\frac{\Gamma(1/2+\epsilon)\Gamma(1/2-\epsilon)^2}{\Gamma(1-2\epsilon)}
\\
&=& \frac{\mbk^{1-2\epsilon} \m2 }{8}\left[1+\epsilon(-\gamma_E + \ln (16\pi)) + {\cal O}(\epsilon^2)\right],
\end{eqnarray*}
we obtain Eqs. \eqref{pi00vacuum}, \eqref{pi00matter}, \eqref{pi00zero} and \eqref{pi00sing}.

\subsection{One-loop integrals\label{sub_oneloop}}
We list here the loop integrals $L_d$, $L_d^{\mu}$ and $L_d^{\mu\nu}$, 
obtained with the Gegenbauer polynomials technique \cite{Pascual:1984zb}:
\begin{eqnarray}
L_d(k,r,s)&=&\int\frac{d^d p}{(2\pi)^d}\frac{1}{(p+k)^{2r}p^{2s}}\nonumber\\
\label{iscalar}
&=&\frac{k^{d-2(r+s)}}{(4\pi)^{d/2}}\frac{\Gamma\left(r+s-{d}/{2}\right)}
{\Gamma(r)\Gamma(s)}\,\frac{\Gamma\left({d}/{2}-s\right)
\Gamma\left({d}/{2}-r\right)}{\Gamma(d-s-r)},
\\
L_d^{\mu}(k,r,s)&=&\int\frac{d^d p}{(2\pi)^d}\frac{p_{\mu}}{(p+k)^{2r}p^{2s}}
\nonumber\\
\label{ivector}
&=&-k^{\mu}\,\frac{k^{d-2(r+s)}}{(4\pi)^{d/2}}
\frac{\Gamma\left(r+s-{d}/{2}\right)}{\Gamma(r)\Gamma(s)}\,\frac{\Gamma\left({d}/{2}+1-s\right)
\Gamma\left({d}/{2}-r\right)}{\Gamma(d+1-s-r)},
\\
L_d^{\mu\nu}(k,r,s)&=&\int\frac{d^dp}{(2\pi)^d}\frac{p^{\mu}p^{\nu}}{(p+k)^{2r}p^{2s}}
\nonumber\\
&=&\frac{k^{d-2(r+s)}}{(4\pi)^{d/2}}\biggl[\frac{k^2}{2}
\frac{\Gamma\left(r+s-1-{d}/{2}\right)}{\Gamma(r)\Gamma(s)}\,\frac{\Gamma\left({d}/{2}+1-s\right)
\Gamma\left({d}/{2}+1-r\right)}{\Gamma(d+2-s-r)}\enspace\delta^{\mu\nu}
\nonumber\\
\label{itensor}
&&\qquad\qquad+\frac{\Gamma\left(r+s-{d}/{2}\right)}{\Gamma(r)\Gamma(s)}\,
\frac{\Gamma\left({d}/{2}+2-s\right)\Gamma\left({d}/{2}-r\right)}
{\Gamma(d+2-s-r)}\enspace k^{\mu}k^{\nu}
\biggr].
\end{eqnarray}

\section{Expansions\label{app_exp}}
In this appendix, we list the expansions of the gluon self energy for
temperatures much greater or smaller than the momentum $k$.

We start with $T\gg\mbk$. In the non-static sector, 
$I_0$ gives its exact result \eqref{sumi0} and $I_3$ reads in dimensional regularization
\begin{equation}
I_3= - \frac{2 T \bk^2\Gamma(1-d/2)(2\pi T)^{d-4}\mu^{2\epsilon}}{(4\pi)^{d/2}}\zeta(4-d).
\label{i3exact}
\end{equation}
For the other integrals, we first carry out the integral, then
Taylor expand the result in ${\bk^2}/{\omega_n^2}$ and finally perform the
sums with the zeta function, thus obtaining 
\begin{eqnarray}
I_1&=&
\frac{T^2}{12}-\frac{\Gamma(2-d/2)\mu^{2\epsilon}\left(\sqrt{\pi}T\right)^d}{2\pi^2 T}
\sum_{l=0}^\infty\frac{\Gamma(d/2-1)\Gamma(l+1)}{\Gamma(d/2-1-l)\Gamma(2l+2)}\zeta(2l+2-d)
\left(\frac{k}{2\pi T}\right)^{2l},
\nonumber\\
\\
I_2&=&
\frac{\bk^2\Gamma(2-d/2)\mu^{2\epsilon}\left(\sqrt{\pi}T\right)^{d}}{8\pi^4T^3}
\sum_{l=0}^\infty\frac{\Gamma(d/2-1)\Gamma(l+1)}{\Gamma(d/2-1-l)\Gamma(2l+2)}\zeta(2l+4-d)
\left(\frac{k}{2\pi T}\right)^{2l},
\\
I_4&=&
\frac{\bk^4 \Gamma(2-d/2)\mu^{2\epsilon}\left(\sqrt{\pi} T\right)^{d}}{32\pi^6T^5}
\sum_{l=0}^\infty\frac{\Gamma(d/2-1)\Gamma(l+1)}{\Gamma(d/2-1-l)\Gamma(2l+2)}\zeta(2l+6-d)
\left(\frac{k}{2\pi T}\right)^{2l}.
\end{eqnarray}
In the fermionic, sector we have
\begin{equation}
\tilde{I}_0=-\frac{T^2}{24},
\label{i0f}
\end{equation}
and we can derive the expansions for $\tilde{I}_1$ and $\tilde{I}_2$ following the 
same procedure used for the bosonic integrals, but ending up with the Hurwitz zeta
function as a result of the odd frequency sums. 
Thus we have
\begin{eqnarray}
\tilde{I}_1&=&\frac{\Gamma(2-d/2)\mu^{2\epsilon}\left(\sqrt{\pi} T\right)^{d}}{2\pi^2T}
\sum_{l=0}^\infty\frac{\Gamma(d/2-1)\Gamma(l+1)}{\Gamma(d/2-1-l)\Gamma(2l+2)}
\zeta(2l+2-d,1/2)\left(\frac{k}{2\pi T}\right)^{2l},
\\
\tilde{I}_2&=&\frac{\bk^2\Gamma(2-d/2)\mu^{2\epsilon}\left(\sqrt{\pi} T\right)^{d}}{8\pi^4T^3}
\sum_{l=0}^\infty\frac{\Gamma(d/2-1)\Gamma(l+1)}{\Gamma(d/2-1-l)\Gamma(2l+2)}\zeta(2l+4-d,1/2)
\left(\frac{k}{2\pi T}\right)^{2l}.
\nonumber\\
\end{eqnarray}
Plugging these expressions in Eqs. \eqref{defp00} and \eqref{pi00fermion} we
obtain the high-temperature expansion \eqref{kllt}.

We consider now the low-temperature expansion.
The vacuum part gives the order $\bk^2$ term in the expansion,
whereas, for the matter part, the condition $\mbk\gg T$ translates 
in Eq. \eqref{pi00matter} into $\mbk\gg \mbp$, since the internal
momentum $\mbp$ is of order $T$. Expanding this expression in
$\mbp/\mbk\ll 1$ yields
\begin{eqnarray} 
\Pi_{00}^{\mathrm{NS}}(\mbk\gg T)_{\mathrm{mat}}&=&
- g^2C_A\frac{T^2}{18}
+g^2T^2\mathcal{O}\left(\frac{T^2}{\bk^2}\right).
\label{nonstaticcontrlowt}
\end{eqnarray}
The singular term ($\propto \mbk^3/T$) and the subtracted zero-mode part 
also contribute in this region. 
The sum of Eq. \eqref{nonstaticcontrlowt} with the vacuum, subtracted zero-mode and
singular parts yields Eq. \eqref{kggt}.
For what concerns the static modes, the only scales are $\mbk$ and $m_D$, 
thus the condition $\mbk\gg T$ becomes $\mbk\gg m_D$ and we end up with 
Eq. \eqref{staticcontrhighk}.
Finally, the fermionic contribution is suppressed in this region, i.e. the first 
nonzero term in the expansion of Eq. \eqref{matterfermion} is of order 
$g^2T^4/{\bk^2}$.

\section{Non-static two-loop sum-integrals\label{app_integrals}}
We set on the evaluation of the two-loop sum-integrals defined by Eq. \eqref{defji}.
$J_0$ does not contribute in dimensional regularization because the integral over $\bk$ has no scale.
$J_1$ can be rewritten as
\begin{equation}
\label{rewritej1}
J_1=\m2 \int \frac{d^dk}{(2\pi)^d}\int_p^\prime\frac{\bp^2}{\bk^4p^2q^2}
=\m2 \int \frac{d^dk}{(2\pi)^d}\frac{1}{\bk^4}\int_p^\prime\frac{1}{q^2}
-\m2 \int \frac{d^dk}{(2\pi)^d}\int_p^\prime\frac{\omega_n^2}{\bk^4p^2q^2}.
\end{equation}
The first term vanishes in dimensional regularization, whereas the second one
yields\footnote{
A convenient way to proceed is by performing first the momentum integrations,
by means of two Feynman parameters, and then the frequencies sum, which gives
$\zeta(0) = -1/2$.}
\begin{equation}
J_1=-\m2 \int
\frac{d^dk}{(2\pi)^d}\int_p^\prime\frac{\omega_n^2}{\bk^4p^2q^2}
=-\frac{T}{8(4\pi)^2}.
\label{finalj1}
\end{equation}
$J_2$ can be read from \cite{Arnold:1994ps},
\begin{equation}
J_2 = \m2 \int \frac{d^dk}{(2\pi)^d}\int_p^\prime\frac{1}{\bk^2p^2q^2}
=\frac{T}{(4\pi)^2}\left(-\frac{1}{4\epsilon}+\ln\frac{2T}{\mu}
-\frac{1}{2}+\frac{\gamma_E}{2}-\frac{\ln (4\pi)}{2}\right).
\end{equation}
$J_3$ vanishes in dimensional regularization because the $\bk$ integral has no scale and finally $J_4$ yields
\begin{equation}
J_4=\m2 \int\frac{d^dk}{(2\pi)^d}\int_p^\prime\frac{1}{p^2q^2\omega_n^2}
=T\sum_{n\ne0}\frac{1}{\omega_n^2}\left(-\frac{\vert\omega_n\vert}{4\pi}\right)^2
=-\frac{T}{(4\pi)^2}.
\end{equation}

We consider now the fermionic integrals. $\tilde{J}_0$ vanishes because it has
a scaleless $\bk$ integration, whereas $\tilde{J}_1$ can be computed along
the lines of its bosonic counterpart, performing the sum over odd frequencies
by means of the the generalized (Hurwitz) zeta function, 
\begin{equation}
\tilde{J}_1=\m2 \int \frac{d^dk}{(2\pi)^d}T\sum_{n}\m2 
\int\frac{d^dp}{(2\pi)^d}\frac{\tilde{\omega}_n^2}{\bk^4p^2q^2}=-\frac{T\zeta(0,1/2)}{4(4\pi)^2}=0.
\label{finalj1f}
\end{equation}
$\tilde{J}_2$ can be read from \cite{Arnold:1994eb},
\begin{equation}
\tilde{J}_2=
\m2 \int \frac{d^dk}{(2\pi)^d}T\sum_{n}\m2 \int\frac{d^dp}{(2\pi)^d}\frac{1}{\bk^2p^2q^2}=-\frac{T}{(4\pi)^2}\ln2.
\end{equation}

\section{Static-modes contribution to the Polyakov loop\label{app_3d}}
In this appendix, we evaluate the 6-dimensional two-loop
integral entering Eq. \eqref{mcontribnlos}. 
We will perform the calculation modifying the magnetostatic propagator in Eq. \eqref{propnonstatic} into
\begin{equation}
\frac{1}{\bk^2}\left(\delta_{ij}-(1-\xi)\frac{k_ik_j}{\bk^2}\right)\delta_{n0}
\to\frac{1}{\bk^2+m_m^2}\left(\delta_{ij}-(1-\xi)\frac{k_ik_j}{\bk^2}\right)\delta_{n0}, 
\label{magnprop}
\end{equation}
where $m_m$ may be interpreted as a small magnetic mass to be put to zero at the end of the calculation.
The magnetic mass modifies the static gluon self-energy expression with resummed gluon propagators 
from Eq. \eqref{staticvacpol} to 
\begin{equation}
\Pi_{00}^{\mathrm{S}}(\bk)=g^2\ca T\m2 \int\frac{d^dp}{(2\pi)^d}
\left[
\frac{d-(1-\xi)}{\bp^2+m_m^2} 
-\frac{(\bk+\bq)^2-(1-\xi)\displaystyle\frac{((\bk+\bq)\cdot\bp)^2}{\bp^2}}{(\bp^2+m_m^2)(\bq^2+m_D^2)}
\right],
\label{magnmass}
\end{equation}
where $q=k-p$.\footnote{
We use here a different parameterization of the integrand  with respect to  Eq. \eqref{staticvacpol}.
} 

In  Eq. \eqref{mcontribnlos}, the integral over the first term in
Eq. \eqref{magnmass}, i.e. the tadpole contribution, gives 
\begin{equation}
-\frac{d-(1-\xi)}{4\pi}\frac{g^4C_R\ca m_m}{2}\m2 \int \frac{d^dk}{(2\pi)^d}\frac{1}{(\bk^2+m_D^2)^2}
=-\left[d-(1-\xi)\right]\frac{g^4C_R\ca}{4(4\pi)^2}\frac{m_m}{m_D}.
\label{tadpolecontrib}
\end{equation}
For the second term, we start by considering the term proportional to $(\bk+\bq)^2$.
We rewrite 
\begin{equation}
(\bk+\bq)^2=2(\bk^2+m_D^2)+2(\bq^2+m_D^2)-(\bp^2+m_m^2)+(m_m^2-4m_D^2),
\label{algebraicrewrite}
\end{equation}
and consider the contributions given by each of the four terms in brackets. 
The first one gives
\begin{eqnarray}
&& 2\mu^{4\epsilon} \int \frac{d^dp}{(2\pi)^{d}}\int \frac{d^dk}{(2\pi)^{d}}
\frac{1}{(\bk^2+m_D^2)(\bp^2+m_m^2)(\bq^2+m_D^2)}
\nonumber\\
&=&\frac{2}{(4\pi)^2}\left[\frac{1}{4\epsilon}+\ln\frac{\mu}{2m_D+m_m}+\frac{1}{2}
-\frac{\gamma_E}{2}+\frac{\ln (4\pi)}{2}+\mathcal{O}(\epsilon)\right],
\label{staticozero}
\end{eqnarray}	
the second one gives 
\begin{equation}
2\mu^{4\epsilon}\int \frac{d^dp}{(2\pi)^{d}}\int \frac{d^dk}{(2\pi)^{d}}
\frac{1}{(\bk^2+m_D^2)^2(\bp^2+m_m^2)}= -\frac{1}{(4\pi)^2}\frac{m_m}{m_D},
\label{noq}
\end{equation}
the third one gives 
\begin{equation}
-\mu^{4\epsilon}\int \frac{d^dp}{(2\pi)^{d}}\int \frac{d^dk}{(2\pi)^{d}}
\frac{1}{(\bk^2+m_D^2)^2(\bq^2+m_D^2)}=\frac{1}{2(4\pi)^2},
\end{equation}
and the last one 
\begin{eqnarray*} 
&&(m_m^2-4m_D^2)\m2 \int \frac{d^dk}{(2\pi)^d}\m2 \int \frac{d^dp}{(2\pi)^d}
\frac{1}{(\bk^2+m_D^2)^2(\bp^2+m_m^2)(\bq^2+m_D^2)}
\\
=&&\frac{m_m^2-4m_D^2}{-2m}\frac{\partial}{\partial m}\m2 
\left. 
\int \frac{d^dk}{(2\pi)^d}\m2 
\int\frac{d^dp}{(2\pi)^d}\frac{1}{(\bk^2+m^2)(\bp^2+m_m^2)(\bq^2+m_D^2)}\right|_{m=m_D}
\\
=&& \frac{1}{(4\pi)^2}\frac{m_m^2-4m_D^2}{2m_D(2m_D+m_m)}.
\end{eqnarray*}
Finally, we consider the term proportional to ${((\bk+\bq)\cdot\bp)^2}/{\bp^2}$
in Eq. \eqref{magnmass}. We rewrite the numerator as
\begin{equation}
(1-\xi)\frac{((\bk+\bq)\cdot\bp)^2}{\bp^2}=
\frac{1-\xi}{\bp^2}[(\bk^2+m_D^2)^2+(\bq^2+m_D^2)^2-2(\bk^2+m_D^2)(\bq^2+m_D^2)].
\end{equation}
The first term gives
\begin{equation}
\mu^{4\epsilon} \int \frac{d^dp}{(2\pi)^{d}}\int \frac{d^dk}{(2\pi)^{d}}
\frac{1}{\bp^2(\bp^2+m_m^2)(\bq^2+m_D^2)}=
-\frac{1}{(4\pi)^2}\frac{m_D}{m_m},
\end{equation}
the third term is $-2$ times this one and the second term gives
\begin{eqnarray*}
&&\m2 \int \frac{d^dp}{(2\pi)^{d}}\frac{1}{\bp^2(\bp^2+m_m^2)}\m2 
\int\frac{d^dk}{(2\pi)^{d}}\frac{\bk^2+\bp^2-2\bp\cdot\bk+m_D^2}{(\bk^2+m_D^2)^2}
\\
&=&\m2 \int \frac{d^dp}{(2\pi)^{d}}\frac{1}{\bp^2(\bp^2+m_m^2)}
\left[\frac{\bp^2}{8\pi m_D}-\frac{m_D}{4\pi}\right]
=-\frac{1}{(4\pi)^2}\left[\frac{m_D}{m_m}+\frac{m_m}{2m_D}\right].
\end{eqnarray*}
The static contribution is thus
\begin{eqnarray}
\nonumber\delta\langle L_R\rangle_{\mathrm{S}\,m_D}
&=&\frac{g^4C_AC_R}{2(4\pi^2)}\left[
-\frac{1}{2\epsilon}-\ln\frac{\mu^2}{(2m_D+m_m)^2}+\gamma_E -\ln (4\pi)
+\frac{2-d}{2}\frac{m_m}{m_D}
\right.\\
&&\hspace{2cm} 
\left.-\frac{3}{2}-\frac{m_m^2-4m_D^2}{2m_D(2m_D+m_m)}\right].
\label{finalcontribmmagn}
\end{eqnarray}
The final result is independent of the gauge parameter $\xi$. 
The expression is well behaved for $m_m\to0$ and yields Eq. \eqref{finalcontribm}.

\section{The Polyakov loop in Feynman gauge\label{app_feynman}}
In this section, we sketch the computation of the vacuum expectation
value of the Polyakov loop in Feynman gauge. We restrict ourselves to  
the fundamental representation ($L\equiv L_F$). 
Since the fermionic contribution, evaluated in
Sec.~\ref{sec_ploop}, is to that order gauge-invariant, 
we do not need to compute it here again.

The perturbative expansion of the Polyakov line through the Baker--Campbell--Hausdorff formula is, 
following \cite{Curci:1984rd} and up to order $g^4$, 
\begin{eqnarray}
\nonumber\langle \trt L\rangle &=& 
\frac{1}{N}\left\langle{\rm Tr} {\rm P} \exp\left(ig\int_0^{1/T} d\tau A^0(\bx,\tau)\right)\right\rangle=
\frac{1}{N}\left\langle{\rm Tr}\left(1+\frac{g^2}{2}(H_0^2+g^2H_1^2\right.\right.\\
&&\left.\left.+2gH_0H_1+2g^2H_0H_2)+\frac{1}{3!}g^3(H_0^3+3gH_0^2H_1)+\frac{1}{4!}g^4H_0^4\right)\right\rangle+\ldots,
\label{bch}
\end{eqnarray}
where
\begin{eqnarray}
 H_0&=&i\int_0^{1/T} d\tau A^0(\tau),
\nonumber\\
 H_1&=&-\frac{1}{2}\int_0^{1/T} d\tau_1 \int_0^{\tau_1} d\tau_2 \left[A^0(\tau_2),A^0(\tau_1)\right],
\nonumber \\
H_2 &=&-\frac{1}{6}\left[H_0,H_1\right]
-\frac{i}{6}\int_0^{1/T} d\tau_1 \int_0^{\tau_1} d\tau_2\left[A^0(\tau_2),\left[A^0(\tau_2),A^0(\tau_1)\right]\right]
\nonumber\\
&&-\frac{i}{3}\int_0^{1/T} d\tau_1 \int_0^{\tau_1} d\tau_2\int_0^{\tau_2} d\tau_3
\left[A^0(\tau_3),\left[A^0(\tau_2),A^0(\tau_1)\right]\right],
\label{defh2}
\end{eqnarray}
and $A^0(\tau,\bx)\equiv A^0(\tau)$. 
We recall that 
\begin{eqnarray*}
D_{00}(\tau)&\equiv& 
\theta(\tau)\langle A_0(\tau) A_0(0) \rangle + \theta(-\tau)\langle A_0(0) A_0(\tau) \rangle 
= T\sum_{n}e^{i\omega_n\tau}\m2 \int\frac{d^dk}{(2\pi)^d}D_{00}(\omega_n,\bk),
\end{eqnarray*}
where in Feynman gauge the free temporal-gluon propagator is 
\begin{equation}
D_{00}^{(0)}(\omega_n,\bk)=\frac{1}{\omega_n^2+\bk^2}.
\end{equation}

\begin{figure}[ht]
\begin{center}
\includegraphics[width=14cm]{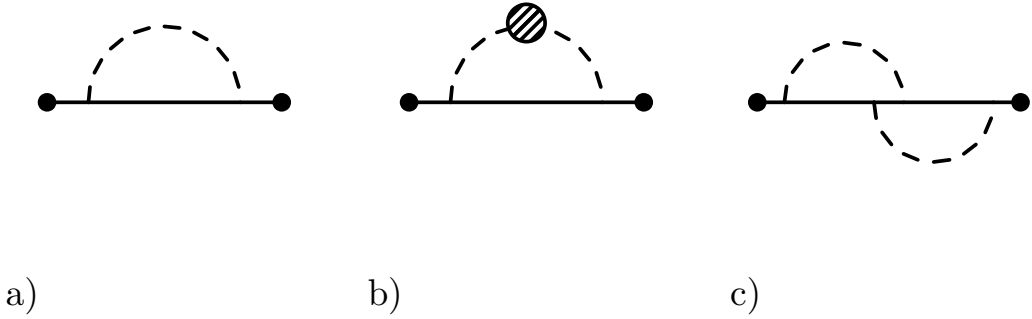}
\end{center}
\caption{Diagrams contributing to the Polyakov loop up to order $g^4$
in Feynman gauge. The blob stands for the one-loop gluon self energy,
the solid line for the Polyakov line and the dots at its beginning/end
represent the points $(0,\bx)$ and $(1/T,\bx)$, which are
compactified by the periodic boundary conditions. 
When integrating over loop momenta of order $m_D$, the dashed lines 
stand for resummed temporal propagators, elsewhere for free ones.}
\label{fig:feynman}
\end{figure}

We can now start working on the different terms in Eq. \eqref{bch}. 
The first one gives 
\begin{equation}
\frac{1}{N}\left\langle{\rm Tr}\frac{g^2}{2}H_0^2\right\rangle
=-\frac{1}{2}\frac{g^2\cf}{T}\m2 \int\frac{d^dk}{(2\pi)^d}D_{00}(0,\bk).
\label{lofeynman}
\end{equation}
Following the same approach as in Sec.~\ref{sec_ploop}, at order $g^4$, 
the relevant diagrams contributing to \eqref{lofeynman} are shown in 
Fig.~\ref{fig:feynman} a) and b). At leading order, the Debye mass is gauge invariant, 
whereas the full one-loop gluon self-energy is not. It is
convenient to separate non-static from static modes. The former
yield \cite{Arnold:1994ps}
\begin{equation}
\Pi_{00}^{\mathrm{NS}}(0,\bk)=-2g^2\ca\left(\frac{d-1}{2}I_0 - (d-1) I_1 + I_2\right),
\label{pi00nsfeynman}
\end{equation}
where the master integrals $I_j$ are those defined in Eq. \eqref{defmaster}, 
hence Eq. \eqref{pi00nsfeynman} equals the first three terms
of the static-gauge expression \eqref{defp00}. The static mode contribution 
to the self energy is common to all gauges that share the same static propagator as the 
static gauge and the Feynman gauge do. Therefore, the static part of the 
self energy in Feynman gauge is just Eq. \eqref{staticvacpol} with $\xi=1$. 
We then have, separating the contributions coming from the scale $T$ 
from those coming from the scale $m_D$, 
\begin{equation}
\m2 \int\frac{d^dk}{(2\pi)^d}D_{00}(0,\bk)=
\m2 \int\frac{d^dk}{(2\pi)^d}
\left[\frac{1}{\bk^2+m_D^2}-\frac{\Pi_{00}^{\mathrm{NS}}(\mbk\sim T)}{\bk^4}
-\frac{\Pi_{00}^{\mathrm{S}}(\mbk)}{(\bk^2+m_D^2)^2}\right]+\ldots,
\label{feynpropg2}
\end{equation}
where the dots stand for higher orders in the perturbative expansion.  
We have omitted the non-static contribution at the scale $m_D$
(cf. Eq. \eqref{mcontribnlons}) since it can be shown that also in
Feynman gauge $\Pi_{00}^\mathrm{NS}(\mbk\sim m_D)-m_D^2
=\mathcal{O}\left(g^2\bk^2\right)$, leading to a higher-order
contribution, whereas the contribution of the static modes at the
scale $T$ leads to a scaleless integral. Plugging Eq. \eqref{feynpropg2}
into Eq. \eqref{lofeynman} and using the results of appendices
\ref{app_integrals} and \ref{app_3d} we obtain most of the final,
order $g^4$, result, except for the contribution of $J_4$ in
Eq. \eqref{tcontribnlonsj}.

We then consider the other terms in the
Baker--Campbell--Hausdorff expansion, starting from $H_1^2$:
\begin{eqnarray}
\frac{1}{N}\left\langle{\rm Tr}\,\frac{g^4}{2}H_1^2\right\rangle&=&
\frac{\cf\ca}{8}g^4\int_0^{1/T} d\tau_1 \int_0^{\tau_1} d\tau_2\int_0^{1/T} d\tau_3 
\int_0^{\tau_3} d\tau_4\left[D_{00}(\tau_2-\tau_3)D_{00}(\tau_1-\tau_4)\right.
\nonumber\\
&&\hspace{75mm}
\left.-D_{00}(\tau_2-\tau_4)D_{00}(\tau_1-\tau_3)\right]
\nonumber\\
&=&-2\zeta(0)\frac{g^4\cf\ca }{4(4\pi)^2}+\ldots=\frac{\als^2\cf\ca}{4}+\ldots,
\label{nlofeynman}
\end{eqnarray}
where we have used free propagators and the dots stand for higher orders. 
This result corresponds exactly to the contribution of $J_4$ in the static
gauge. The contribution can be traced back to diagram c) in Fig.~\ref{fig:feynman} and 
it corresponds to the term called $\mathcal{L}_4$ in Eq. (4) of \cite{Gava:1981qd}.

We now need to show that the sum of the remaining terms yields zero at order $g^4$. 
$\langle{\rm Tr}\, 2gH_0H_1\rangle$ vanishes because it involves 
a three temporal-gluon vertex.
$\langle{\rm Tr} \, 2g^2H_0H_2\rangle$ is a more complicated object,
however one can show that, working with free propagators \cite{Curci:1984rd}, 
\begin{equation}
\frac{1}{N}\langle{\rm Tr}\,g^4H_0H_2\rangle=0+\mathcal{O}(g^5,g^4\times(m_D/T)).
\label{hoh2}
\end{equation}
The $H_0^3$ term vanishes, again due to the three temporal-gluon vertex 
and the $H_0^2H_1$ term can be easily shown to be zero after performing 
the colour trace.
The $H_0^4$ term gives
\begin{equation}
\frac{1}{4!\nc}\langle{\rm Tr}\,g^4H_0^4\rangle=\frac{g^4}{4!}
\left(3\cf^2-\frac{\cf\ca}{2} \right)\frac{1}{T^2}
\left(\m2 \int\frac{d^dk}{(2\pi)^d}D_{00}(0,\bk)\right)^2,
\label{h04}
\end{equation}
which is at least of order $g^4\times(m_D/T)^2$. 
This, finally, shows that the Feynman-gauge computation of the Polyakov loop 
agrees with the static-gauge computation that led to Eq.~\eqref{finalg4loop}.

\section{Octet contributions\label{app_octet}}
In this appendix, we want to prove that, up to order $g^6 (rT)^0$, 
$\displaystyle 
\delta_{o,T}^{\, {\cal O}(r^2)\, {\rm NS}} =  \delta_{s,T}^{\, {\cal O}(r^2)\, {\rm NS}}\vert_{V_s\leftrightarrow V_o}$,
$\displaystyle 
\delta_{o,T}^{\, {\cal O}(r^2)\, {\rm S}}  =  - \delta_{s,T}^{\, {\cal O}(r^2)\, {\rm S}}$,
and $\displaystyle \delta_{o,T}^{\,\delta {\cal L}_{\rm pNRQCD}} = -\delta_{s,T}^{\,\delta {\cal L}_{\rm pNRQCD}}$, 
where the left- and right-hand sides of the equalities encode non-zero modes, zero-modes and higher-multipole 
one-loop corrections to the pNRQCD octet and singlet 
propagators respectively induced by interaction vertices of the type 
${\rm S}^\dagger r^{i_1}...r^{i_n}\partial_{i_1}...\partial_{i_{n-1}}E^{i_n} {\rm O} \, +$ Hermitian conjugate   
or ${\rm O}^\dagger r^{i_1}...r^{i_n}\partial_{i_1}...\partial_{i_{n-1}}E^{i_n} {\rm O} \, +$ charge conjugate.

\begin{figure}[ht]
\begin{center}
\includegraphics[width=14cm]{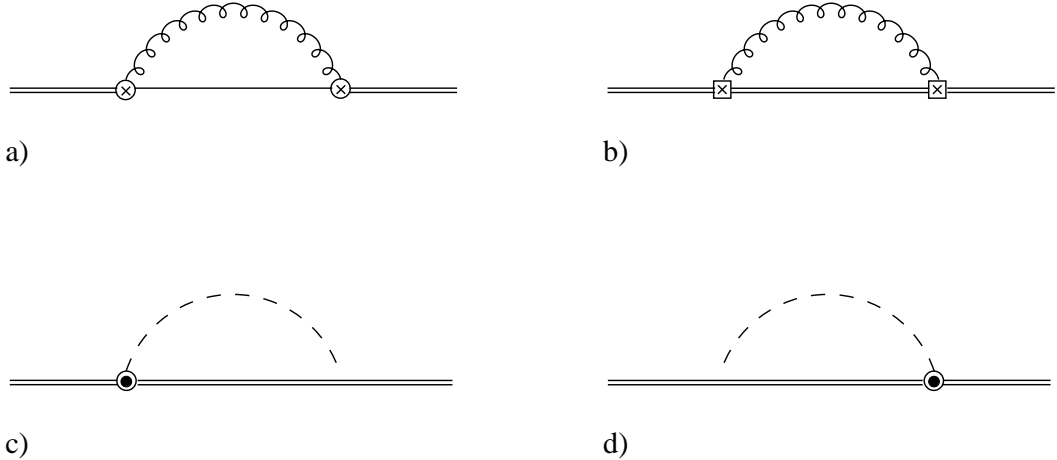}
\end{center}
\caption{The pNRQCD Feynman diagrams giving the leading-order correction to $\delta_o$.
The single continuous line stands for a singlet propagator, 
the double line for an octet propagator, the circle with a cross for the 
chromoelectric dipole vertex proportional to $V_A$ in the Lagrangian \eqref{pNRQCD}, 
the square  with a cross for the chromoelectric dipole vertex proportional to $V_B$ 
in the Lagrangian \eqref{pNRQCD},  the circle with a dot for the 
chromoelectric dipole vertex proportional to $V_C$ in the Lagrangian \eqref{pNRQCD}, 
the curly line for a chromoelectric correlator and the dashed line for a temporal-gluon propagator.
}
\label{fig:octetEEV}
\end{figure}

The general argument goes as follows. Let's first consider contributions coming from the non-zero modes
of the loop integral, Fig.~\ref{fig:singEEV}  providing the leading-order contribution to the singlet 
propagator and diagram a) in  Fig.~\ref{fig:octetEEV} providing the leading-order contribution 
to the octet propagator. As the leading-order example shows, there is a one to one correspondence between 
diagrams in the singlet and in the octet channel, to each singlet diagram corresponds an octet 
diagram whose contribution is equal to the singlet diagram contribution with $V_s$ replaced by $V_o$ and viceversa.
We note that, since at order $g^4$ these contributions are linear in $V_o-V_s$, they are at that order 
one the opposite of the other.

Let's now consider contributions coming from the zero modes of the loop integral. In order to see how things 
work, we consider, first, the order $r^2$ contribution. In the singlet channel, only one diagram, Fig.~\ref{fig:singEEV}, 
contributes; that contribution has been written in Eq. \eqref{EEVb} and evaluated in Eq. \eqref{deltasEb}.
In the octet channel, four diagrams contribute, which are shown in  Fig.~\ref{fig:octetEEV}.
Diagram a) gives the same contribution as the singlet channel:
\begin{eqnarray}
\delta_{o,T}^{\, a) \, {\rm S}} &=& 
- g^2 \frac{1}{2\nc}\frac{r^ir^j}{2T}
\int \frac{d^dk}{(2\pi)^d} \langle E^{i\,a} U_{ab}E^{j\,b}\rangle(0,\bk)\vert_{|\bk|\sim T}
 + {\cal  O}\left(g^6(rT)\right).
\label{EEVOa}
\end{eqnarray}
Diagram b) is like diagram a) with the colour factor $1/(2\nc)$ replaced by $d^{abc}d^{abc}/[4(\nc^2-1)]$:
\begin{eqnarray}
\delta_{o,T}^{\, b) \, {\rm S}} &=& 
- g^2\frac{\nc^2-4}{4\nc} \frac{r^ir^j}{2T}
\int \frac{d^dk}{(2\pi)^d} \langle E^{i\,a} U_{ab}E^{j\,b}\rangle(0,\bk)\vert_{|\bk|\sim T}
 + {\cal  O}\left(g^6(rT)\right).
\label{EEVOb}
\end{eqnarray}
Finally, diagrams c) and d) are like diagram a) 
with the colour factor $1/(2\nc)$ replaced by $f^{abc}f^{abc}/[8(\nc^2-1)]$:
\begin{eqnarray}
\delta_{o,T}^{\, c)+d) \, {\rm S}} &=& 
g^2 \frac{\nc}{4} \frac{r^ir^j}{2T}
\int \frac{d^dk}{(2\pi)^d} \langle E^{i\,a} U_{ab}E^{j\,b}\rangle(0,\bk)\vert_{|\bk|\sim T}
 + {\cal  O}\left(g^6(rT)\right),
\label{EEVOc}
\end{eqnarray}
where the positive sign comes from moving a derivative acting on the chromoelectric field in one vertex 
to the temporal gluon in the other one (see also footnote \ref{footEE}).
Summing Eqs. \eqref{EEVOa}-\eqref{EEVOc} we obtain the opposite of the singlet contribution in  Eq. \eqref{EEVb}.

This argument may be easily generalized to any order in the multipole expansion. 
Let's consider diagrams contributing to order $2n$ in the multipole expansion.
The singlet contribution is proportional to 
\begin{equation}
\delta_{s,T}^{\, {\cal O}(r^{2n}) \, {\rm S}} \propto 
r^{2n} \frac{1}{2\nc} \sum_{\ell = 0}^{n-1}\frac{1}{(2\ell+1)!}\frac{1}{(2n-(2\ell+1))!}.
\end{equation}
Again there are three classes of octet contributions that correspond to the three classes discussed 
at order $r^2$. Except for the first class, each one has a different colour factor with respect to the singlet 
contribution, but for the rest they are equal:
\begin{eqnarray}
\delta_{o,T}^{\, a) \, {\rm S}} &\propto& 
- r^{2n} \frac{1}{2\nc} \sum_{\ell = 0}^{n-1}\frac{1}{(2\ell+1)!}\frac{1}{(2n-(2\ell+1))!}, 
\\
\delta_{o,T}^{\, b) \, {\rm S}} &\propto& 
- r^{2n} \frac{\nc^2-4}{4\nc} \sum_{\ell = 0}^{n-1}\frac{1}{(2\ell+1)!}\frac{1}{(2n-(2\ell+1))!}, 
\\
\delta_{o,T}^{\, c)+d) \, {\rm S}} &\propto& 
r^{2n} \frac{\nc}{4} \sum_{\ell = 0}^{n}\frac{1}{(2\ell)!}\frac{1}{(2n-2\ell)!},
\end{eqnarray}
where the positive sign in the last expression comes from moving an odd number of derivatives 
acting on the field in one vertex to the field in the other one. 
Since $\displaystyle \sum_{\ell = 0}^{n}\frac{1}{(2\ell)!}\frac{1}{(2n-2\ell)!} = 
\sum_{\ell = 0}^{n-1}\frac{1}{(2\ell+1)!}\frac{1}{(2n-(2\ell+1))!}$, the sum of 
all octet contributions is just the opposite of the singlet contribution.


\begin{thebibliography}{99}
%\cite{Kuti:1980gh}
\bibitem{Kuti:1980gh}
  J.~Kuti, J.~Polonyi and K.~Szlachanyi,
  %``Monte Carlo Study Of SU(2) Gauge Theory At Finite Temperature,''
  Phys.\ Lett.\  B {\bf 98}, 199 (1981).
  %%CITATION = PHLTA,B98,199;%%

%\cite{McLerran:1981pb}
\bibitem{McLerran:1981pb}
  L.~D.~McLerran and B.~Svetitsky,
  %``Quark Liberation At High Temperature: A Monte Carlo Study Of SU(2) Gauge
  %Theory,''
  Phys.\ Rev.\  D {\bf 24}, 450 (1981).
  %%CITATION = PHRVA,D24,450;%%

%\cite{Rebhan:1994mx}
\bibitem{Rebhan:1994mx}
  A.~K.~Rebhan,
  %``NonAbelian Debye screening in one loop resummed perturbation theory,''
  Nucl.\ Phys.\  B {\bf 430}, 319 (1994)
  [arXiv:hep-ph/9408262].
  %%CITATION = NUPHA,B430,319;%%

%\cite{Arnold:1995bh}
\bibitem{Arnold:1995bh}
  P.~Arnold and L.~G.~Yaffe,
  %``The NonAbelian Debye screening length beyond leading order,''
  Phys.\ Rev.\  D {\bf 52}, 7208 (1995)
  [arXiv:hep-ph/9508280].
  %%CITATION = PHRVA,D52,7208;%%

%\cite{Mocsy:2007jz}
\bibitem{Mocsy:2007jz}
  A.~Mocsy and P.~Petreczky,
  %``Color Screening Melts Quarkonium,''
  Phys.\ Rev.\ Lett.\  {\bf 99}, 211602 (2007)
  [arXiv:0706.2183 [hep-ph]]; 
  %%CITATION = PRLTA,99,211602;%%
%\cite{Mocsy:2007yj}
%\bibitem{Mocsy:2007yj}
%  A.~Mocsy and P.~Petreczky,
  %``Can quarkonia survive deconfinement ?,''
  Phys.\ Rev.\  D {\bf 77}, 014501 (2008)
  [arXiv:0705.2559 [hep-ph]]; 
  %%CITATION = PHRVA,D77,014501;%%
%\cite{Mocsy:2005qw}
%\bibitem{Mocsy:2005qw}
%  A.~Mocsy and P.~Petreczky,
  %``Quarkonia correlators above deconfinement,''
  Phys.\ Rev.\  D {\bf 73}, 074007 (2006)
  [arXiv:hep-ph/0512156];
  %%CITATION = PHRVA,D73,074007;%%
%\cite{Mocsy:2004bv}
%\bibitem{Mocsy:2004bv}
%  A.~Mocsy and P.~Petreczky,
  %``Heavy quarkonia survival in potential model,''
  Eur.\ Phys.\ J.\  C {\bf 43}, 77 (2005)
  [arXiv:hep-ph/0411262].
  %%CITATION = EPHJA,C43,77;%%

%\cite{Philipsen:2008qx}
\bibitem{Philipsen:2008qx}
  O.~Philipsen,
  %``Static potentials for quarkonia at finite temperatures,''
  Nucl.\ Phys.\  A {\bf 820}, 33C (2009)
  [arXiv:0810.4685 [hep-ph]].
  %%CITATION = NUPHA,A820,33C;%%

%\cite{Laine:2006ns}
\bibitem{Laine:2006ns}
  M.~Laine, O.~Philipsen, P.~Romatschke and M.~Tassler,
  %``Real-time static potential in hot QCD,''
  JHEP {\bf 0703}, 054 (2007)
  [arXiv: hep-ph/0611300];
  %%CITATION = JHEPA,0703,054;%%
%\cite{Burnier:2007qm}
%\bibitem{Burnier:2007qm}
  Y.~Burnier, M.~Laine and M.~Veps\"al\"ainen,
  %``Heavy quarkonium in any channel in resummed hot QCD,''
  JHEP {\bf 0801}, 043 (2008)
  [arXiv:0711.1743 [hep-ph]];
  %%CITATION = JHEPA,0801,043;%%
%\cite{Laine:2008cf}
%\bibitem{Laine:2008cf}
  M.~Laine,
  %``How to compute the thermal quarkonium spectral function from first
  %principles?,''
  Nucl.\ Phys.\  A {\bf 820}, 25C (2009)
  [arXiv:0810.1112 [hep-ph]].
  %%CITATION = NUPHA,A820,25C;%%

%\cite{Beraudo:2007ky}
\bibitem{Beraudo:2007ky}
  A.~Beraudo, J.~P.~Blaizot and C.~Ratti,
  %``Real and imaginary-time $Q\bar{Q}$ correlators in a thermal medium,''
  Nucl.\ Phys.\  A {\bf 806}, 312 (2008)
  [arXiv:0712.4394 [nucl-th]].
  %%CITATION = NUPHA,A806,312;%%

%\cite{Brambilla:2008cx}
\bibitem{Brambilla:2008cx}
  N.~Brambilla, J.~Ghiglieri, A.~Vairo and P.~Petreczky,
  %``Static quark-antiquark pairs at finite temperature,''
  Phys.\ Rev.\  D {\bf 78}, 014017 (2008)
  [arXiv:0804.0993 [hep-ph]].
  %%CITATION = PHRVA,D78,014017;%%

%\cite{Escobedo:2008sy}
\bibitem{Escobedo:2008sy}
  M.~A.~Escobedo and J.~Soto,
  %``Non-relativistic bound states at finite temperature (I): the hydrogen
  %atom,''
  Phys.\ Rev.\ A {\bf 78}, 032520 (2008) 
  [arXiv:0804.0691 [hep-ph]].
  %%CITATION = ARXIV:0804.0691;%%

%\cite{Brambilla:2010vq}
\bibitem{Brambilla:2010vq}
  N.~Brambilla, M.~A.~Escobedo, J.~Ghiglieri, J.~Soto and A.~Vairo,
  %``Heavy Quarkonium in a weakly-coupled quark-gluon plasma below the melting
  %temperature,''
   JHEP {\bf 1009} (2010) 038
  [arXiv:1007.4156 [hep-ph]].
  %%CITATION = ARXIV:1007.4156;%%

%\cite{Kaczmarek:1999mm}
\bibitem{Kaczmarek:1999mm}
  O.~Kaczmarek, F.~Karsch, E.~Laermann and M.~Lutgemeier,
  %``Heavy quark potentials in quenched QCD at high temperature,''
  Phys.\ Rev.\  D {\bf 62}, 034021 (2000)
  [arXiv:hep-lat/9908010].
  %%CITATION = PHRVA,D62,034021;%%

%\cite{Petreczky:2001pd}
\bibitem{Petreczky:2001pd}
  P.~Petreczky, O.~Kaczmarek, F.~Karsch, E.~Laermann, S.~Stickan, I.~Wetzorke and F.~Zantow,
  %``Lattice calculation of medium effects at short and long distances,''
  Nucl.\ Phys.\  A {\bf 698}, 400 (2002)
  [arXiv:hep-lat/0103034].
  %%CITATION = NUPHA,A698,400;%%

%\cite{Digal:2003jc}
\bibitem{Digal:2003jc}
  S.~Digal, S.~Fortunato and P.~Petreczky,
  %``Heavy quark free energies and screening in SU(2) gauge theory,''
  Phys.\ Rev.\  D {\bf 68}, 034008 (2003)
  [arXiv:hep-lat/0304017].
  %%CITATION = PHRVA,D68,034008;%%

%\cite{Bazavov:2008rw}
\bibitem{Bazavov:2008rw}
  A.~Bazavov, P.~Petreczky and A.~Velytsky,
  %``Static quark anti-quark pair in SU(2) gauge theory,''
  Phys.\ Rev.\  D {\bf 78}, 114026 (2008)
  [arXiv:0809.2062 [hep-lat]].
  %%CITATION = PHRVA,D78,114026;%%

%\cite{Karsch:2000kv}
\bibitem{Karsch:2000kv}
  F.~Karsch, E.~Laermann and A.~Peikert,
  %``Quark mass and flavor dependence of the QCD phase transition,''
  Nucl.\ Phys.\  B {\bf 605}, 579 (2001)
  [arXiv:hep-lat/0012023].
  %%CITATION = NUPHA,B605,579;%%

%\cite{Petreczky:2004pz}
\bibitem{Petreczky:2004pz}
  P.~Petreczky and K.~Petrov,
  %``Free energy of a static quark anti-quark pair and the renormalized
  %Polyakov loop in three flavor QCD,''
  Phys.\ Rev.\  D {\bf 70}, 054503 (2004)
  [arXiv:hep-lat/0405009].
  %%CITATION = PHRVA,D70,054503;%%

%\cite{Petreczky:2005bd}
\bibitem{Petreczky:2005bd}
  P.~Petreczky,
  %``Heavy quark potentials and quarkonia binding,''
  Eur.\ Phys.\ J.\  C {\bf 43}, 51 (2005)
  [arXiv:hep-lat/0502008];
  %%CITATION = EPHJA,C43,51;%%
%\cite{Petreczky:2004xs}
%\bibitem{Petreczky:2004xs}
%  P.~Petreczky,
  %``QCD thermodynamics on lattice,''
  Nucl.\ Phys.\ Proc.\ Suppl.\  {\bf 140}, 78 (2005)
  [arXiv:hep-lat/0409139];
  %%CITATION = NUPHZ,140,78;%%
%\cite{Bazavov:2009us}
%\bibitem{Bazavov:2009us}
  A.~Bazavov, P.~Petreczky and A.~Velytsky,
  %``Quarkonium at Finite Temperature,''
  arXiv:0904.1748 [hep-ph].
  %%CITATION = ARXIV:0904.1748;%%

%\cite{Gross:1980br}
\bibitem{Gross:1980br}
  D.~J.~Gross, R.~D.~Pisarski and L.~G.~Yaffe,
  %``QCD And Instantons At Finite Temperature,''
  Rev.\ Mod.\ Phys.\  {\bf 53}, 43 (1981).
  %%CITATION = RMPHA,53,43;%%

%\cite{Nadkarni:1986cz}
\bibitem{Nadkarni:1986cz}
  S.~Nadkarni,
  %``Nonabelian Debye Screening. 1. The Color Averaged Potential,''
  Phys.\ Rev.\  D {\bf 33}, 3738 (1986).
  %%CITATION = PHRVA,D33,3738;%%

%\cite{Gava:1981qd}
\bibitem{Gava:1981qd}
  E.~Gava and R.~Jengo,
  %``Perturbative Evaluation Of The Thermal Wilson Loop,''
  Phys.\ Lett.\  B {\bf 105}, 285 (1981).
  %%CITATION = PHLTA,B105,285;%%

\bibitem{D'Hoker:1981us}
  E.~D'Hoker,
  %``Perturbative Results On QCD In Three-Dimensions At Finite Temperature,''
  Nucl.\ Phys.\  B {\bf 201}, 401 (1982).
  %%CITATION = NUPHA,B201,401;%%

\bibitem{Curci:1982fd}
  G.~Curci and P.~Menotti,
  %``Temporal Gauges And Periodic Boundary Conditions,''
  Z.\ Phys.\  C {\bf 21}, 281 (1984). 
  %%CITATION = ZEPYA,C21,281;%%
		
\bibitem{Curci:1984rd}
  G.~Curci, P.~Menotti and G.~Paffuti,
  %``Temporal Gauge On A Periodic Lattice,''
  Z.\ Phys.\  C {\bf 26}, 549 (1985). 
  %%CITATION = ZEPYA,C26,549;%%
		
%\cite{Duncan:1975kt}
\bibitem{Duncan:1975kt}
  A.~Duncan,
  %``Fine Structure In Nonabelian Gauge Theories,''
  Phys.\ Rev.\  D {\bf 13}, 2866 (1976).
  %%CITATION = PHRVA,D13,2866;%%

%\cite{Appelquist:1977es}
\bibitem{Appelquist:1977es}
  T.~Appelquist, M.~Dine and I.~J.~Muzinich,
  %``The Static Limit Of Quantum Chromodynamics,''
  Phys.\ Rev.\  D {\bf 17}, 2074 (1978).
  %%CITATION = PHRVA,D17,2074;%%

\bibitem{Kajantie:1982xx}
  K.~Kajantie and J.~I.~Kapusta,
  %``Behavior Of Gluons At High Temperature,''
  Annals Phys.\  {\bf 160}, 477 (1985). 
  %%CITATION = APNYA,160,477;%%

\bibitem{Heinz:1986kz}
  U.~W.~Heinz, K.~Kajantie and T.~Toimela,
  %``Gauge Covariant Linear Response Analysis of QCD Plasma Oscillations,''
  Annals Phys.\  {\bf 176}, 218 (1987).
  %%CITATION = APNYA,176,218;%%

\bibitem{Kapusta:2006pm}
  J.~I.~Kapusta and C.~Gale,
  ``Finite-temperature field theory: Principles and applications,'' 
  %\href{http://www.slac.stanford.edu/spires/find/hep/www?irn=7209002}{SPIRES entry}
  {\it  Cambridge, UK: Univ. Pr. (2006) 428 p}

\bibitem{Rebhan:1993az}
  A.~K.~Rebhan,
  %``The NonAbelian Debye mass at next-to-leading order,''
  Phys.\ Rev.\  D {\bf 48}, 3967 (1993)
  [arXiv:hep-ph/9308232].
  %%CITATION = PHRVA,D48,3967;%%

%\cite{Burnier:2009bk}
\bibitem{Burnier:2009bk}
  Y.~Burnier, M.~Laine and M.~Veps\"al\"ainen,
  %``Dimensionally regularized Polyakov loop correlators in hot QCD,''
  JHEP {\bf 1001}, 054 (2010)
  [arXiv:0911.3480 [hep-ph]].
  %%CITATION = JHEPA,1001,054;%%

%\cite{Gupta:2007ax}
\bibitem{Gupta:2007ax}
  S.~Gupta, K.~H\"ubner and O.~Kaczmarek,
  %``Renormalized Polyakov loops in many representations,''
  Phys.\ Rev.\  D {\bf 77}, 034503 (2008)
  [arXiv:0711.2251 [hep-lat]].
  %%CITATION = PHRVA,D77,034503;%%

%\cite{Brambilla:1999qa}
\bibitem{Brambilla:1999qa}
  N.~Brambilla, A.~Pineda, J.~Soto and A.~Vairo,
  %``The infrared behaviour of the static potential in perturbative {QCD},''
  Phys.\ Rev.\  D {\bf 60}, 091502 (1999)
  [arXiv:hep-ph/9903355].
  %%CITATION = PHRVA,D60,091502;%%

\bibitem{Brambilla:1999xf}
  N.~Brambilla, A.~Pineda, J.~Soto and A.~Vairo,
  %``Potential NRQCD: An effective theory for heavy quarkonium,''
  Nucl.\ Phys.\  B {\bf 566}, 275 (2000) 
  [arXiv:hep-ph/9907240].
  %%CITATION = NUPHA,B566,275;%%

\bibitem{Fischler:1977yf}
 W.~Fischler,
 %``Quark - Anti-Quark Potential In QCD,''
 Nucl.\ Phys.\  B {\bf 129}, 157 (1977).
 %%CITATION = NUPHA,B129,157;%%

\bibitem{Billoire:1979ih}
 A.~Billoire,
 %``How Heavy Must Be Quarks In Order To Build Coulombic Q Anti-Q Bound
 %States,''
 Phys.\ Lett.\  B {\bf 92}, 343 (1980).
 %%CITATION = PHLTA,B92,343;%%

%\cite{Gelfand:1964}
\bibitem{Gelfand:1964}
  I.~M.~Gelfand,
  ``Generalized Functions'',
{\it  New York, USA: Acad. Pr. (1964) 423 p.}

%\cite{Jahn:2004qr}
\bibitem{Jahn:2004qr}
  O.~Jahn and O.~Philipsen,
  %``The Polyakov loop and its relation to static quark potentials and free
  %energies,''
  Phys.\ Rev.\  D {\bf 70}, 074504 (2004) 
  [arXiv:hep-lat/0407042].
  %%CITATION = PHRVA,D70,074504;%%

%\cite{Luscher:2002qv}
\bibitem{Luscher:2002qv}
  M.~L\"uscher and P.~Weisz,
  %``Quark confinement and the bosonic string,''
  JHEP {\bf 0207}, 049 (2002)
  [arXiv:hep-lat/0207003].
  %%CITATION = JHEPA,0207,049;%%

%\cite{Pineda:1997bj}
\bibitem{Pineda:1997bj}
  A.~Pineda and J.~Soto,
  %``Effective field theory for ultrasoft momenta in NRQCD and NRQED,''
  Nucl.\ Phys.\ Proc.\ Suppl.\  {\bf 64}, 428 (1998)
  [arXiv:hep-ph/9707481].
  %%CITATION = NUPHZ,64,428;%%

%\cite{Brambilla:2004jw}
\bibitem{Brambilla:2004jw}
  N.~Brambilla, A.~Pineda, J.~Soto and A.~Vairo,
  %``Effective field theories for heavy quarkonium,''
  Rev.\ Mod.\ Phys.\  {\bf 77}, 1423 (2005) 
  [arXiv:hep-ph/0410047].
  %%CITATION = RMPHA,77,1423;%%

%\cite{Brambilla:2002nu}
\bibitem{Brambilla:2002nu}
  N.~Brambilla, D.~Eiras, A.~Pineda, J.~Soto and A.~Vairo,
  %``Inclusive decays of heavy quarkonium to light particles,''
  Phys.\ Rev.\  D {\bf 67}, 034018 (2003)
  [arXiv:hep-ph/0208019].
  %%CITATION = PHRVA,D67,034018;%%

%\cite{Brambilla:2003nt}
\bibitem{Brambilla:2003nt}
  N.~Brambilla, D.~Gromes and A.~Vairo,
  %``Poincare invariance constraints on NRQCD and potential NRQCD,''
  Phys.\ Lett.\  B {\bf 576}, 314 (2003)
  [arXiv:hep-ph/0306107].
  %%CITATION = PHLTA,B576,314;%%

%\cite{Brambilla:2006wp}
\bibitem{Brambilla:2006wp}
  N.~Brambilla, X.~Garcia i Tormo, J.~Soto and A.~Vairo,
  %``The logarithmic contribution to the QCD static energy at NNNNLO,''
  Phys.\ Lett.\  B {\bf 647}, 185 (2007)
  [arXiv:hep-ph/0610143].
  %%CITATION = PHLTA,B647,185;%%

%\cite{Anzai:2009tm}
\bibitem{Anzai:2009tm}
  C.~Anzai, Y.~Kiyo and Y.~Sumino,
  %``Static QCD potential at three-loop order,''
  Phys.\ Rev.\ Lett.\  {\bf 104}, 112003 (2010)
  [arXiv:0911.4335 [hep-ph]];
  %%CITATION = PRLTA,104,112003;%%
%\cite{Smirnov:2009fh}
%\bibitem{Smirnov:2009fh}
  A.~V.~Smirnov, V.~A.~Smirnov and M.~Steinhauser,
  %``Three-loop static potential,''
  Phys.\ Rev.\ Lett.\  {\bf 104}, 112002 (2010)
  [arXiv:0911.4742 [hep-ph]].
  %%CITATION = PRLTA,104,112002;%%

%\cite{Kniehl:2004rk}
\bibitem{Kniehl:2004rk}
  B.~A.~Kniehl, A.~A.~Penin, Y.~Schr\"oder, V.~A.~Smirnov and M.~Steinhauser,
  %``Two-loop static QCD potential for general colour state,''
  Phys.\ Lett.\  B {\bf 607}, 96 (2005) 
  [arXiv:hep-ph/0412083].
  %%CITATION = PHLTA,B607,96;%%

%\cite{Braaten:1994qx}
\bibitem{Braaten:1994qx}
  E.~Braaten and A.~Nieto,
  %``Asymptotic Behavior Of The Correlator For Polyakov Loops,''
  Phys.\ Rev.\ Lett.\  {\bf 74}, 3530 (1995)
  [arXiv:hep-ph/9410218].
  %%CITATION = PRLTA,74,3530;%%

%\cite{Pascual:1984zb}
\bibitem{Pascual:1984zb}
  P.~Pascual and R.~Tarrach,
  %``QCD: Renormalization For The Practitioner,''
  Lect.\ Notes Phys.\  {\bf 194}, 1 (1984).
  %%CITATION = LNPHA,194,1;%%

\bibitem{Arnold:1994ps}
  P.~Arnold and C.~x.~Zhai,
  %``The Three loop free energy for pure gauge QCD,''
  Phys.\ Rev.\  D {\bf 50},7603 (1994) 
  [arXiv:hep-ph/9408276].
 %%CITATION = PHRVA,D50,7603;%%

\bibitem{Arnold:1994eb}
  P.~Arnold and C.~x.~Zhai,
  %``The Three loop free energy for high temperature QED and QCD with
  %fermions,''
  Phys.\ Rev.\  D {\bf 51}, 1906  (1995) 
  [arXiv:hep-ph/9410360].
  %%CITATION = PHRVA,D51,1906;%%
\end{thebibliography}
\end{document}